\documentclass[preprint]{imsart}

\RequirePackage[OT1]{fontenc}
\usepackage{amsthm,amsmath}
\usepackage[english]{babel}
\usepackage[utf8]{inputenc}
\usepackage{amssymb}
\usepackage{dsfont}
\usepackage{bbm}
\usepackage{comment}
\usepackage{graphicx}
\usepackage{subfig}
\usepackage{multirow}
\usepackage{tikz}
\usepackage{color}
\usepackage[colorinlistoftodos]{todonotes}

\usepackage{algorithm}
\usepackage{algpseudocode}

\usepackage{enumitem}
\usepackage{float}
\usepackage{mathtools}
\usepackage{hyperref}
\usepackage{url}
\usepackage{appendix}

\RequirePackage[numbers]{natbib}
\RequirePackage[colorlinks,citecolor=blue,urlcolor=blue]{hyperref}

\startlocaldefs
\numberwithin{equation}{section}
\theoremstyle{plain}
\newtheorem{theorem}{Theorem}[section]

\newtheorem{corollary}{Corollary}[theorem]

\theoremstyle{definition}
\newtheorem{definition}{Definition}[section]

\theoremstyle{remark}

\endlocaldefs

\begin{document}

\begin{frontmatter}

\title{Sparse Networks with Core-Periphery Structure}
\runtitle{Sparse Networks with Core-Periphery Structure}

\author{\fnms{Cian} \snm{Naik}\ead[label=e1]{cian.naik@stats.ox.ac.uk}}
\address{Department of Statistics, University of Oxford\\ \printead{e1}}

\author{\fnms{Fran\c cois} \snm{Caron}\ead[label=e3]{caron@stats.ox.ac.uk}}
\address{Department of Statistics, University of Oxford\\\printead{e3}}

\author{\fnms{Judith} \snm{Rousseau}\ead[label=e2]{judith.rousseau@stats.ox.ac.uk}}
\address{Department of Statistics, University of Oxford\\ \printead{e2}}

\runauthor{C. Naik et al.}

\begin{abstract}
We propose a statistical model for graphs with a core-periphery structure. To do this we define a precise notion of what it means for a graph to have this structure, based on the sparsity properties of the subgraphs of core and periphery nodes. We present a class of sparse graphs with such properties, and provide methods to simulate from this class, and to perform posterior inference. We demonstrate that our model can detect core-periphery structure in simulated and real-world networks.
\end{abstract}



\end{frontmatter}

\section{Introduction} \label{introduction}
Network data arises in a number of areas, including social, biological and transport networks. Modelling of this type data has become an increasingly important field in recent times, partly due to the greater availability of the data, and also due to the fact that we increasingly want to model complex systems with many interacting components. A central topic within this field is the design of random graph models. Various models have been proposed, building on the early work of \citep{erdos1959random}. A key goal with these models is to capture important properties of real-world graphs. In particular, a lot of effort has gone into the investigation of meso-scale structures~\citep{rombach2017core}. These are intermediate scale structures such as communities of nodes within the overall network. There has been a large amount of work devoted to models that can capture community structure, such as the popular stochastic block-model~\citep{holland1983stochastic,snijders1997estimation,nowicki2001estimation}.

Here, we focus on a different type of meso-scale structure, known as a core-periphery structure, which we will define formally in Section \ref{sec:definitions}. The intuitive idea is of a network that consists of two classes of nodes --- \textit{core} nodes that are densely connected to each other, and \textit{periphery} nodes that are more loosely connected to core nodes, and sparsely connected between themselves. A number of algorithmic approaches have been proposed for the detection of core-periphery structure~\citep{della2013profiling,cucuringu2016detection,de2019detecting}, see~\citep{rombach2017core} for a review. Our focus here is on model-based approaches. A popular model-based approach is to consider a two-block stochastic block-model, where one block corresponds to the core, and the other one to the periphery~\cite{zhang2015identification}. However, such models are known to generate dense graphs\footnote{A graph is said to be dense if the number of edges scales quadratically with the number of nodes. Otherwise, it is said to be sparse.}, a property considered unrealistic for many real-world networks~\citep{Orbanz2015}.

In this work, we are particularly interested in models for sparse graphs. We use the statistical network modelling framework introduced by \citep{Caron2017}, based on representing the network as an exchangeable random measure, as this framework allows us to capture both sparse and dense networks. 
The contributions of our work are as follows. Firstly, we provide a precise definition of what it means for a graph to have a core-periphery structure, based on the sparsity properties of the subgraphs of core and periphery nodes.
Secondly, building on earlier work from \citep{todeschini2016exchangeable}, we present a class of sparse network models with such properties, and provide methods to simulate from this class, and to perform posterior inference.
Finally, we demonstrate that our approach can detect meaningful core-periphery structure in two real-world airport and trade networks, while providing a good fit to global structural properties of the networks.

The network of flight connections between airports is a typical example of a core-periphery network. The core nodes correspond to central hubs that flights are routed through, according to the so-called Spoke-hub distribution~\citep{holme2005core}, while other airports are more sparsely connected between each other. Other examples include the World Wide Web~\citep{csermely2013structure} and social \citep{white1976social}, biological~\citep{luo2009core}, transport~\citep{holme2005core}, citation~\citep{borgatti2000models}, trade~\citep{nemeth1985international} or financial~\citep{fricke2015core} networks. The importance of being able to identify core-periphery structure can be seen by considering the properties of these networks. In cases such as transport networks, this identification allows for the detection of hubs which may be the most important locations for additional development. In protein-protein interaction networks, identifying a core helps to determine which proteins are the most important for the development of the organism. In internet networks, core nodes could be the most important places to defend from cyber-attacks. The advantage of our core-periphery model is that the framework that we work in allows us to capture the sparsity that many of these networks have \citep{barabasi2016network}.

Sparse graph models with (overlapping) block structure have already been proposed within the framework of \citep{Caron2017} --- \citep{Herlau2015,todeschini2016exchangeable}. However, these models cannot be applied directly to model networks with a core-periphery structure, as they make the assumption that a single parameter tunes the overall sparsity properties of the graph, with the same structural sparsity properties across different blocks. This is an undesirable property for core-periphery networks, where the subgraph of core nodes is expected to have different, denser structural properties than the rest of the network. Our work builds on the framework of (multivariate) completely random measures (CRMs) \citep{kingman1967completely,kingman1993poisson} that have been widely used in the Bayesian nonparametric literature \citep{regazzini2003distributional,lijoi2010models,griffin2017compound} to construct graphs with heterogeneous sparsity properties.

The rest of the paper is organised as follows. In Section \ref{sparse-model} we define what a core-periphery structure means in our framework, and give the general construction of this type of network as well as the particular model that we employ. We also show how our framework can accommodate both core-periphery and community structure. In Section \ref{sec:properties} we present some important theoretical results about the new model, such as the sparsity properties that define the core-periphery structure. In Section~\ref{sec:discussion} we provide a discussion of related models.  
In Section \ref{inference}, we look at performing posterior inference using this model, and give the details of the Markov Chain Monte Carlo (MCMC) sampler used to do this. In Section \ref{experiments} we test our model on a variety of simulated and real data sets. We also compare it against a range of contemporary alternatives, and show that it provides an improvement in certain settings.

\paragraph{Notations. } We follow the asymptotic notations of \citep{janson2011probability}. Let $(X_\alpha)_{\alpha\geq 0}$ and $(Y_\alpha)_{\alpha\geq 0}$ be two stochastic processes defined on the same probability space with $X_\alpha,Y_\alpha\rightarrow\infty$ almost surely (a.s.) as $\alpha \to \infty$. We have
    $X_{\alpha} = O(Y_{\alpha})$ a.s. $\iff$ $\limsup_{\alpha \to \infty}\frac{X_{\alpha}}{Y_{\alpha}} < \infty$ a.s.;
    $X_{\alpha} = o(Y_{\alpha})$ a.s. $\iff$ $\limsup_{\alpha \to \infty}\frac{X_{\alpha}}{Y_{\alpha}} =0$ a.s.;
    $X_{\alpha} = \Theta(Y_{\alpha})$ a.s. $\iff$ $X_{\alpha} = O(Y_{\alpha}) \text{ and } Y_{\alpha} = O(X_{\alpha})$ a.s.

\section{Statistical Network Models with Core-Periphery Structure} \label{sparse-model}
\subsection{Definitions}
\label{sec:definitions}

In this section, we formally define what it means for graphs to be sparse and to have a core-periphery structure.

Let $G=(G_\alpha)_{\alpha\geq 0}$ be a family of growing undirected random graphs with no isolated vertices, where $\alpha\geq 0$ is interpreted as a size parameter. $G_\alpha=(V_\alpha,E_\alpha)$ where $V_\alpha$ and $E_\alpha$ are the set of vertices and edges respectively. Denote respectively $N_\alpha=|V_\alpha|$ and $N_\alpha^{(e)}=|E_\alpha|$ for the number of nodes and edges in $G_\alpha$, and assume $N_\alpha,N_\alpha^{(e)}\rightarrow\infty$ almost surely as $\alpha \to \infty$.

We first give the definition of sparsity for the family $G=(G_\alpha)_{\alpha\geq 0}$.

\begin{definition}[Sparse graph]\label{def:sparsity} We say that a graph family $G$ is \textit{dense} if the number of edges scales quadratically with the number of nodes
\begin{align}
N^{(e)}_{\alpha}= \Theta(N_{\alpha}^2)\label{eq:defdense}
\end{align}
almost surely as $\alpha \to \infty$. Conversely, it is \textit{sparse} if the number of edges scales subquadratically with the number of nodes
\begin{align}
N^{(e)}_{\alpha}= o(N_{\alpha}^2)\label{eq:defsparse}
\end{align}
almost surely as $\alpha \to \infty$.
\end{definition}

Let $(V_{\alpha,c})_{\alpha\geq 0}$ be a growing family of core nodes, where $V_{\alpha,c}\subseteq V_\alpha$ for all $\alpha$. Let $N_{\alpha,c}=|V_{\alpha,c}|$ be the number of core nodes, and $N_{\alpha,c-c}^{(e)}$ the number of edges between nodes in the core. Assume both $N_{\alpha,c},N_{\alpha,c-c}^{(e)}\rightarrow\infty$ almost surely as $\alpha \to \infty$.
\begin{definition}[Core-periphery structure]\label{def:coreperiphery} We say that a graph family $G=(G_\alpha)_{\alpha\geq 0}$ is sparse with core-periphery structure if the graph is sparse with a dense core subgraph, that is
\begin{equation}\label{eq:core-periph}
N^{(e)}_{\alpha,c-c}= \Theta(N_{\alpha,c}^2) \quad
N^{(e)}_{\alpha} = o(N_{\alpha}^2).
\end{equation}
\end{definition}

A consequence of \eqref{eq:core-periph} is that $N_{\alpha,c}= o(N_\alpha)$, since $N^{(e)}_{\alpha,c-c} \leq N^{(e)}_{\alpha}$.
In other words, the core corresponds to a small dense subgraph of a sparse graph, with sparse connections to the other part of the graph, which is called the periphery.

\subsection{A model for networks with core-periphery structure} \label{sec:particularmodel}
Having defined what we mean by core-periphery structure, we start by giving a generic construction of a core-periphery model in the case where the graph is otherwise unstructured; an extension to graphs which also exhibit some community structure is presented in Section \ref{community_and_core_periphery}. Following \citep{Caron2017}, we represent a graph by the point process on the plane
\begin{align}\label{eq:graphpointprocess}
Z=\sum_{i,j}z_{ij}\delta_{(\theta_i,\theta_j)}
\end{align}
where $z_{ij}=z_{ji}=1$ if $i$ and $j$ are connected, and $0$ otherwise. Here, each node $i$ is located at some point $\theta_i \in \mathbb{R}_+=[0,\infty)$. A finite graph $G_\alpha$ of size $\alpha>0$ is obtained by considering the restriction of $Z$ to $[0,\alpha]^2$, see \citep{Caron2017}.  As in \cite{todeschini2016exchangeable}, we consider that the probability of a connection between nodes $i$ and $j$ is given by the link function
\begin{align}\label{eq:community_model}
\Pr(z_{i,j}=1 | (w_{l1},w_{l2})_{l=1,2, \ldots}) = \begin{cases}
1-e^{-2(w_{i1}w_{j1}+w_{i2}w_{j2})} &i\neq j\\
1-e^{-w_{i1}^2-w_{i2}^2} &i= j
\end{cases}
\end{align}
where $w_{i1}\geq 0$ is the core parameter and $w_{i2}>0$ is an overall sociability parameter. The model parameters $(w_{i1},w_{i2},\theta_i)_{i=1,2,\ldots}$ are the points of a Poisson point process on $\mathbb R_+^3$ with mean measure $\rho(w_1,w_2)d\theta$ where $\rho$ is a $\sigma$-finite measure on $\mathbb R_+^2$, concentrated on $\mathbb R_+^2 \backslash \{0,0\} $, which satisfies
$\int_{\mathbb R_+^2} \min(1,w_1+w_2)\rho(dw_1,dw_2)<\infty.$ As shown in \cite{todeschini2016exchangeable}, the resulting graph is sparse if
\begin{equation}
\int_{[0,\infty)^2}\rho(dw_1,dw_2)=\infty,
\end{equation}
and dense otherwise. \cite{todeschini2016exchangeable} considered a specific class of models for $\rho$, based on compound completely random measures (CRMs)~\cite{griffin2017compound}, where $\rho$ is concentrated on $(0,\infty)^2$; that is, for all nodes $i$ we have $w_{i1}>0$ and $w_{i2}>0$. As a consequence, the sparsity pattern is homogeneous across the graph. We consider a different and more flexible construction here where $w_{i1}\geq 0$ and $w_{i2}>0$ . As will be shown in Theorem \ref{thm:cp_asymptotics1}, the core-periphery property can be enforced by making the following assumptions on the mean measure $\rho$:
\begin{align}
&\int_{(0,\infty)\times \{0\}}\rho(dw_1,dw_2)=0,\label{eq:cond1}\\
&\int_{\{0\}\times(0,\infty)}\rho(dw_1,dw_2)>0,\label{eq:condperiphery}\\
&0<\int_{(0,\infty)^2}\rho(dw_1,dw_2)<\infty.\label{eq:condcore}
\end{align}
We identify the set $\mathcal{C}=\{i \mid w_{i1}>0\}$ as the core nodes, and the remaining ones $\mathcal{P}=\{i \mid w_{i1}=0,w_{i2}>0\}$ as the periphery nodes. Assumption \eqref{eq:cond1} ensures that all nodes have a strictly positive sociability parameter. The strict positivity assumptions in Equations \eqref{eq:condperiphery} and \eqref{eq:condcore} ensure  that the size of the core or periphery is not empty with probability 1. The boundedness assumption in Equation~\eqref{eq:condcore} ensures that the subgraph of core nodes is dense, as will be shown in Theorem \ref{thm:cp_asymptotics1}. Note that the overall graph may be sparse or dense, depending whether the integral in Equation \eqref{eq:condperiphery} is finite or not. 

We now propose a way to construct such a function $\rho$:
\begin{align}\label{eq:levymeasurerho}
\rho(dw_1,dw_2)=\int_0^\infty \left ( (1-e^{-w_0})\delta_{w_0}(dw_1)+e^{-w_0}\delta_0(dw_1) \right )w_0^{-1}F\left (\frac{dw_2}{w_0}\right )\rho_0(dw_0)
\end{align}
where $\rho_0$ is a L\'evy measure on $\mathbb R_+$. In other words, with probability $1-e^{-w_0}$, $w_{i,1}=w_0>0$ and $i$ is a core node,  while with probability $e^{-w_0}$, $w_{i,1}=0$ and $i$ belongs to the periphery.

In this paper, we set $\rho_0$ to be the  mean measure of the jump part of a generalized gamma process (GGP)
\begin{align}\label{rho_0_ggp}
\rho_0(dw_0)=\frac{1}{\Gamma(1-\sigma)}w_0^{-1-\sigma}\exp(-w_0\tau)dw_0
\end{align}
where $(\sigma,\tau)$ satisfy
$
\sigma \in (0,1),\tau \geq 0 $ or $\sigma \in (-\infty,0], \tau >0.$
The GGP has been used extensively due to its flexibility, the interpretability of its parameters and its conjugacy properties.
The model then  admits the following equivalent representation: let  $(w_{i0},\theta_i)_{i\geq 1}$ be the points of a Poisson point process with mean measure $\rho_0(dw_0)d\theta$ and set
\begin{equation} \label{beta:parametrisation}
\begin{split}
w_{ik}&=\beta_{ik} w_{i0} , \quad k=1,2\\
\beta_{i1}\mid w_{i0} &\sim \text{Ber}(1-e^{-w_{i0}}), \quad \beta_{i2} \overset{iid}{\sim} F
\end{split}
\end{equation}
where the scores $\beta_{i1}\in\{0,1\}$ and $\beta_{i2}>0$ are mutually independent.
Note crucially that, although the formulation \eqref{beta:parametrisation} resembles the formulation of  the class of compound CRMs introduced by \cite{griffin2017compound}; in our construction, the scores $\beta_{i1}$ are not identically distributed, and depend on the base parameter $w_{i0}$. This key difference enables the model to have different sparsity properties, as shown in Section~\ref{sec:properties}.

Of particular  interest, for computational reasons, is the case where $F$ is set to be the gamma distribution
\begin{align} \label{GammaF}
F(dx)=x^{a-1}e^{-bx}\frac{b^a}{\Gamma(a)}dx, \quad a,b>0.
\end{align}

We simulate a network from the above model with parameters $\alpha = 200, \sigma= 0.2, \tau = 1, b=0.5, a=0.2$. For comparison, we also simulate a network from the Sparse Network with Overlapping Communities (SNetOC) model of \citep{todeschini2016exchangeable} with $K=2$ communities and  parameters $\alpha = 100, \sigma= 0.2, \tau = 2.5, b=(0.2,0.05), a=(0.4,0.1)$. The parameters are chosen so that the graphs have similar sizes and overall degree distributions.
\begin{figure}[h!t]
\centering
\subfloat[SNetOC graph with 2 communities \label{fig:SNetOC_network}]{\includegraphics[width=0.4\textwidth]{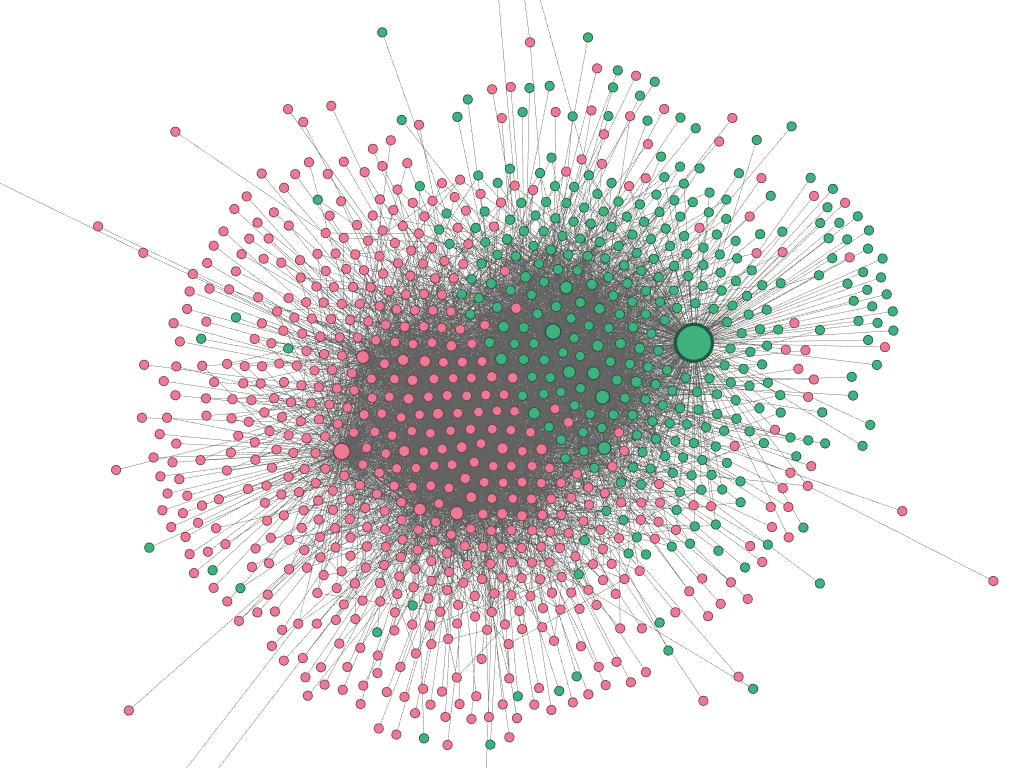}}
\subfloat[Core-periphery graph \label{fig:CP_network}]{\includegraphics[width=0.4\textwidth]{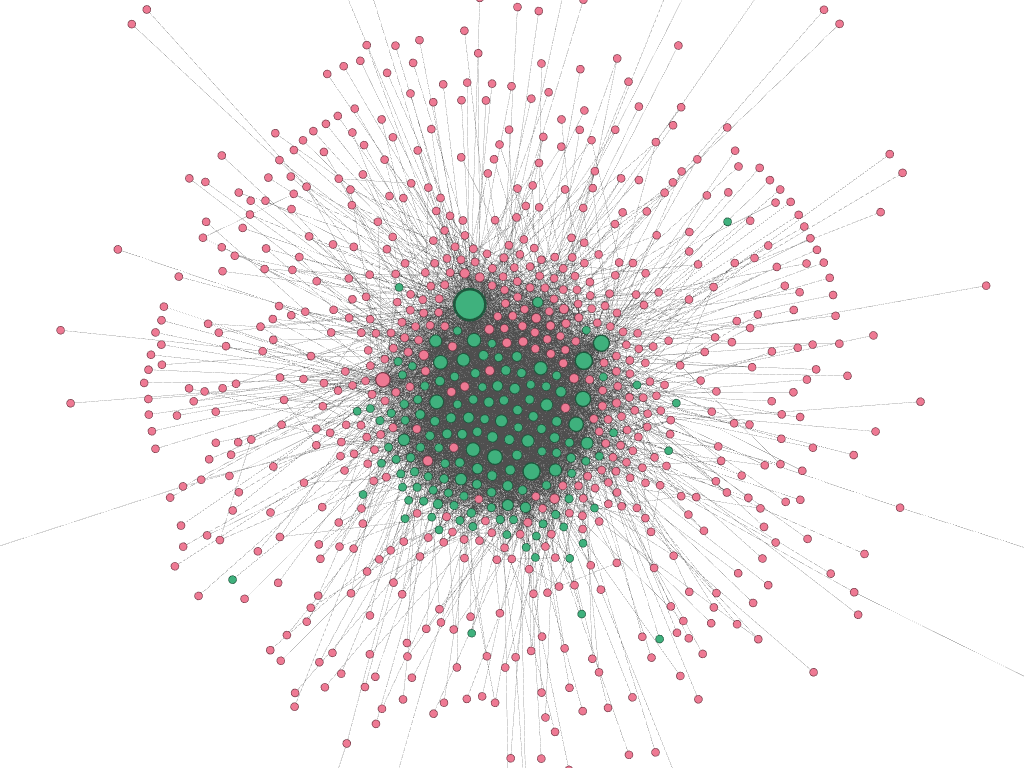}}\\
\subfloat[SNetOC adjacency matrix \label{fig:SNetOC_adjacency}]{\includegraphics[width=0.4\textwidth]{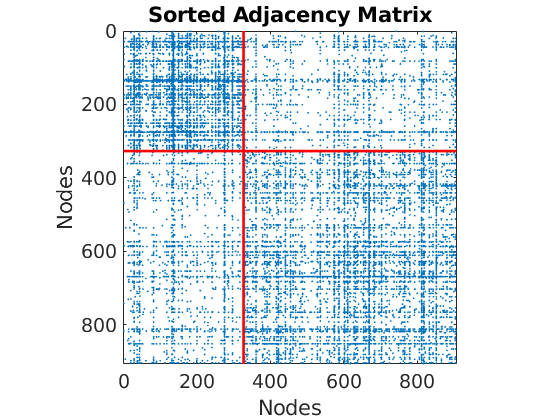}}
\subfloat[Core-periphery adjacency matrix \label{fig:CP_adjacency}]{\includegraphics[width=0.4\textwidth]{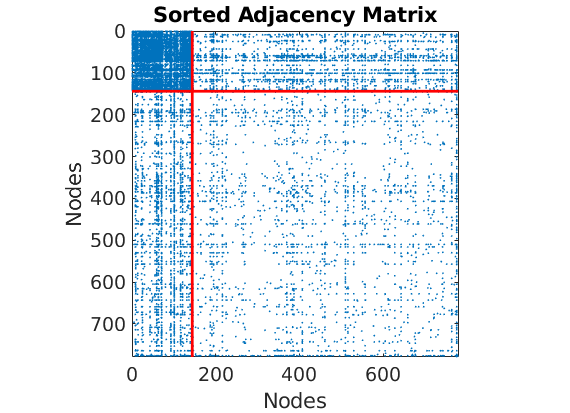}}\\
\subfloat[SNetOC subgraph degree distributions \label{fig:SNetOC_subgraphdegree}]{\includegraphics[width=0.35\textwidth]{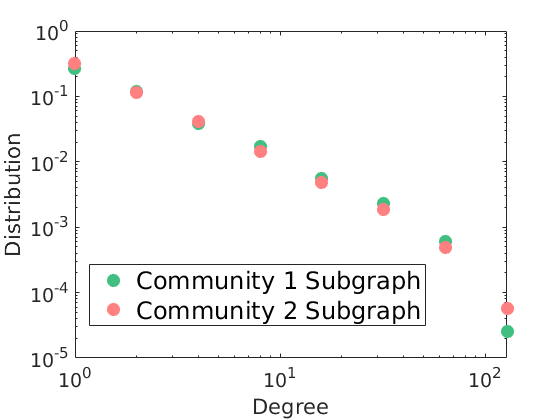}}~~
~\subfloat[Core-periphery subgraph degree distributions \label{fig:CP_subgraphdegree}]{\includegraphics[width=0.35\textwidth]{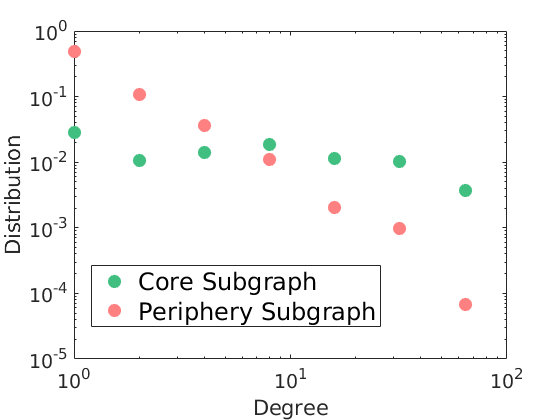}}~~\\
\caption{Comparing a core-periphery graph simulated from out model to a graph with two communities from the SNetOC model of \citep{todeschini2016exchangeable}. Plots of $(a)$  the graph with two communities, with nodes from each community in green and red respectively, and $(b)$ the core-periphery graph with core nodes in green and periphery nodes in red. In each case the size of the nodes is proportional to their mean sociability. Adjacency matrices for $(c)$ the graph with two communities and $(d)$ the core-periphery graph. Degree distributions of $(e)$ the subgraphs of nodes within communities 1 and 2 for the two community graph, and $(f)$ the subgraphs of core and periphery nodes for the core-periphery graph.}
\label{fig:gephi_plot}
\end{figure}

\begin{figure}[t!]
\centering
\includegraphics[width=0.5\textwidth]{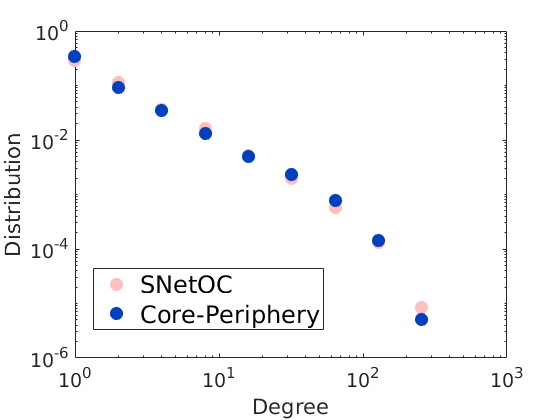}
\caption{Overall degree distribution of the overlapping community and core-periphery graphs.}
\label{fig:overall_degree_comp}
\end{figure}

Figures~\ref{fig:gephi_plot}\subref{fig:SNetOC_network} and \ref{fig:gephi_plot}\subref{fig:CP_network} show the networks sampled from both models. Figures~\ref{fig:gephi_plot}\subref{fig:SNetOC_adjacency} and \ref{fig:gephi_plot}\subref{fig:CP_adjacency} show the associated adjacency matrices, where nodes are ordered according to the community where they have the largest affiliation for the SNetOC model, and by core ($w_{i1}>0$) or periphery $w_{i1}=0$ affiliation for our model.
In Figures~\ref{fig:gephi_plot}\subref{fig:SNetOC_subgraphdegree} and \ref{fig:gephi_plot}\subref{fig:CP_subgraphdegree} we represent the empirical degree distributions of nodes within each community for SNetOC, and for the core and periphery subgraphs for our model. Although the overall degree distributions for both models are very close (see Figure~\ref{fig:overall_degree_comp}), the degree distributions of the subgraphs are very different: the degree distributions within each subgraph have a similar power-law behaviour for SNetOC, whereas for the core-periphery model, it exhibits a Poisson-like behaviour for the core, and a power-law behaviour for the periphery.

The model introduced in this section can be easily extended to allow for both overlapping communities and a core-periphery structure by adding further sociability parameters. This is explained in the following section.

\subsection{Model with core-periphery and overlapping community structure}
\label{community_and_core_periphery}
The conditional probability of connection between 2 nodes is given by:
\begin{equation}\label{eq:core_community_model}
\begin{split}
\Pr(z_{i,j}=1 | (w_{l1},w_{l2},\ldots,w_{lK})_{l=1,2, \ldots}) =
	 \begin{cases}
 		1-e^{-2\sum_{k=1}^K w_{ik}w_{jk}} &i\neq j\\
 		1-e^{-\sum_{k=1}^K w_{ik}^2} &i= j
	\end{cases}
\end{split}
\end{equation}

and assume that
\begin{equation}\label{eq:core_community_model2}
\begin{split}
w_{ik} & =\beta_{ik}w_{i0}\qquad k=1,\ldots,K \\
\beta_{i1}\mid w_{i0}& \sim \text{Ber}(1-e^{-w_{i0}}), \quad
(\beta_{i2},\ldots,\beta_{iK}) \overset{iid}{\sim} F
\end{split}
\end{equation}
and for computational reason we consider independent Gamma distributions,
\begin{equation*}
F(d\beta_2,\ldots,d\beta_K)=\prod_{k=2}^K\beta_k^{a_k-1}e^{-b_k\beta_k}\frac{b_k^{a_k}}{\Gamma(a_k)}d\beta_k
\end{equation*}
as in the overlapping communities model of \citep{todeschini2016exchangeable}. The parameters $w_{ik}$, for $k=2,\ldots,K$ are interpreted as the degree of affiliation of node $i$ to community $k$. In this case the L\'evy measure $\rho$ is given by
\begin{align}\label{eq:levymeasurerho_communities}
\begin{split}
\rho(dw_1,dw_2,\ldots,dw_K)=\int_{0}^\infty \bigg[&\left ( (1-e^{-w_0})\delta_{w_0}(dw_1)+e^{-w_0}\delta_0(dw_1) \right ) \\
&\times w_0^{-(K-1)}F\left (\frac{dw_2}{w_0},\ldots,\frac{dw_K}{w_0}\right) \bigg]\rho_0(dw_0)
\end{split}
\end{align}

We finally give a brief descriptions of how the different parameters in the model control the structure of the network.
\begin{itemize}
\item $\mathbf{w_{i1}}$ --- We interpret $w_{i1}$ as a \textit{coreness} parameter, with $w_{i1}>0$ indicating that node $i$ is in the core.
\item $\mathbf{w_{i2},\ldots,w_{iK}}$ --- When $K=2$, we can interpret $w_{i2}$ as an overall sociability parameter. Otherwise $w_{i2},\ldots,w_{iK}$ indicate the affiliation to each of the respective communities, and we can interpret them as sociabilities within each of these communities.
\item $\boldsymbol{\sigma}$ --- As we will see from Theorem \ref{thm:cp_asymptotics2}, $\sigma$ controls the overall sparsity of the network, with a higher value leading to sparser networks. It also controls the size of the core, with a larger value of $\sigma$ leading to a smaller relative core size.
\item $\boldsymbol{\tau}$ --- As we can see from the theoretical results, $\tau$ does not affect the asymptotic rates for the densities of the different regions. Its main effect is to induce an exponential tilting of large degrees in the degree distribution, as we see in Figure \ref{base_degree_graphs} in Appendix \ref{app:simulations}.
\item $\mathbf{a_2,\ldots,a_K}, \mathbf{b_2,\ldots,b_K}$ --- The hyperparameters of the community sociability parameters control the distribution $F$ defined in Section \ref{community_and_core_periphery}, which is the product of $\Gamma(a_i,b_i)$ distributions. Furthermore increasing the $a_i$ decreases the relative size of the core by changing the sizes of the $w_{ik}$ relative to $w_{i1}$, while increasing the $b_i$ increases the relative size.
\item $\boldsymbol{\alpha}$ ---  This is a size-parameter, tuning the number of nodes and edges in the network.
\end{itemize}

In the following section we show theoretically and through simulations that these models recover the core-periphery structure defined in Definition \ref{def:coreperiphery}.

\section{Core-Periphery and Sparsity Properties}
\label{sec:properties}
\label{sparsity_asymptotics}

In this section we study the asymptotic behaviour of the number of nodes $N_\alpha$, the number of nodes in the core  $N_{\alpha,c}$ and in the periphery $N_{\alpha,p}$, together with the number of edges $N^{(e)}_{\alpha}$, the number of edges between core nodes $N^{(e)}_{\alpha, c-c}$, between periphery nodes $N^{(e)}_{\alpha, p-p}$ and between core and periphery nodes $N^{(e)}_{\alpha, c-p}$, which are the key quantities to understand the core periphery structure.  They are defined as
\begin{equation*} 
\begin{split}
N_{\alpha} &=\sum_i\mathds{1}_{\theta_i \leq \alpha}\mathds{1}_{(\sum_j z_{ij}\mathds{1}_{\theta_j \leq \alpha})\geq 1},\\
N^{(e)}_{\alpha}&=  \sum_{i \leq j}z_{ij}\mathds{1}_{\theta_i \leq \alpha}\mathds{1}_{\theta_j \leq \alpha} \\
N_{\alpha,c} &=\sum_i\mathds{1}_{\theta_i \leq \alpha}\mathds{1}_{w_{i1}>0}  \mathds{1}_{(\sum_j z_{ij}\mathds{1}_{\theta_j \leq \alpha})\geq 1},\\
N^{(e)}_{\alpha, c-c}&=  \sum_{i \leq j}z_{ij}\mathds{1}_{w_{i1}>0}  \mathds{1}_{w_{j1}>0} \mathds{1}_{\theta_i \leq \alpha}\mathds{1}_{\theta_j \leq \alpha}
\end{split}
\end{equation*}
and similarly for the other quantities. In Theorem \ref{thm:cp_asymptotics1} we study the generic core - periphery model as defined  by Equations~\eqref{eq:graphpointprocess} and \eqref{eq:community_model}, where the L\'evy measure $\rho$ satisfies Assumptions \eqref{eq:cond1}, \eqref{eq:condperiphery} and \eqref{eq:condcore}.

\begin{theorem}\label{thm:cp_asymptotics1}
Consider the graph family defined by Equations~\eqref{eq:graphpointprocess} and \eqref{eq:community_model}, where the L\'evy measure $\rho$ satisfies Assumptions \eqref{eq:cond1}, \eqref{eq:condperiphery} and \eqref{eq:condcore}. Then the graph has a dense core subgraph
\begin{align}
N^{(e)}_{\alpha,c-c}&= \Theta(N_{\alpha,c}^2)\text{ almost surely as }\alpha\rightarrow\infty.
\end{align}
If
$$
\int_{\{0\}\times(0,\infty)}\rho(dw_1,dw_2)=\infty
$$
then the graph is sparse overall and \eqref{eq:defsparse} holds. Otherwise, it is dense, and \eqref{eq:defdense} holds.
\end{theorem}
We now characterize more precisely the sparsity properties for the particular model described by Equations \eqref{eq:levymeasurerho}, \eqref{rho_0_ggp} together with \eqref{GammaF}, in Section~\ref{sec:particularmodel}.

\begin{theorem}\label{thm:cp_asymptotics2}
Consider the graph family defined by Equations~\eqref{eq:graphpointprocess} and \eqref{eq:community_model}, where the L\'evy measure $\rho$  takes the form of Equation~\eqref{eq:levymeasurerho}, with a generalized gamma process base measure $\rho_0$, and $F$ a gamma distribution. Assume further that $\sigma \in (0,1)$ and $\tau >0$, so that
\begin{align}
\int_0^{\infty}\rho_0(dw_0)=\infty, \qquad \int_0^{\infty}w_0\rho_0(dw_0)<\infty.
\end{align}
Then, almost surely as $\alpha$ tends to infinity,
\begin{equation}\label{density_rates}
\begin{aligned}
N_{\alpha}^{(e)}&=O \left(N_{\alpha}^{\frac{2}{1+\sigma}}\right), \quad
N_{\alpha,c-c}^{(e)} =\Theta \left(N_{\alpha,c}^2\right)\\
N_{\alpha,p-p}^{(e)}&=O \left(N_{\alpha,p}^{\frac{2}{1+\sigma}}\right), \quad
N_{\alpha,c-p}^{(e)}=O \left(\left(\sqrt{N_{\alpha,c}N_{\alpha,p}}\right)^{\frac{2}{1+\sigma/2}}\right).
\end{aligned}
\end{equation}
\end{theorem}

A natural consequence of the definition of core-periphery structure that we use is that the relative size of the core tends to zero, since the overall graph must be sparse. In corollary \ref{thm:core_size} we confirm this, noting that $\sigma \in (0,1)$.
\begin{corollary}\label{thm:core_size}
In the same setting as Theorem \ref{thm:cp_asymptotics2}, we have that
\begin{align}\label{eq:relative_core}
N_{\alpha,c}=O\left(N_{\alpha}^{\frac{1}{1+\sigma}}\right)
\end{align}
and furthermore
\begin{align}\label{core_prop}
\frac{N_{\alpha,c}}{N_{\alpha}} = O \left(\alpha^{-\sigma}\right)\qquad \text{almost surely }as\ \alpha \to \infty
\end{align}
\end{corollary}

We know from Equation (\ref{density_rates}) that when $\sigma \in (0,1)$, $N_{\alpha}^{(e)}=O \left(N_{\alpha}^{\beta}\right)$ in each region, where $\sigma$ is the parameter of the base L\'evy measure. We then have that
\begin{align*}
\beta=\begin{cases}
\frac{2}{1+\sigma} &\text{overall}\\
2 &\text{in the core region} \\
\frac{2}{1+\sigma} &\text{in the periphery region}\\
\frac{2}{1+\sigma/2} &\text{in the core-periphery region}
\end{cases}
\end{align*}
When $\sigma <0$, $\beta=2$ in each region.

\begin{figure}[h]
\centering
\subfloat[Overall graph \label{fig:edgesvsnodes}]{\includegraphics[width=0.35\textwidth]{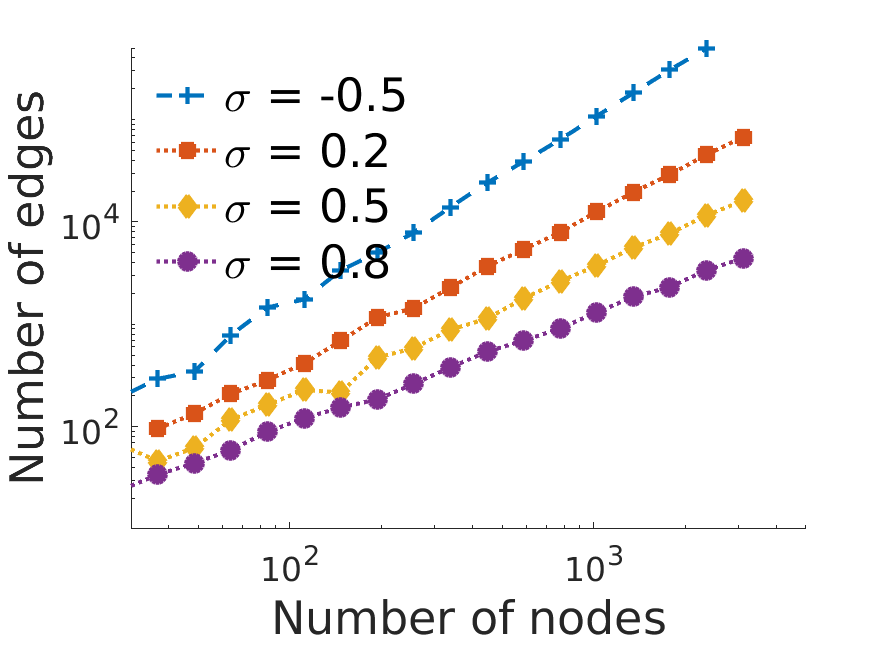}}
\subfloat[Core region \label{fig:ccedgesvsnodes}]{\includegraphics[width=0.35\textwidth]{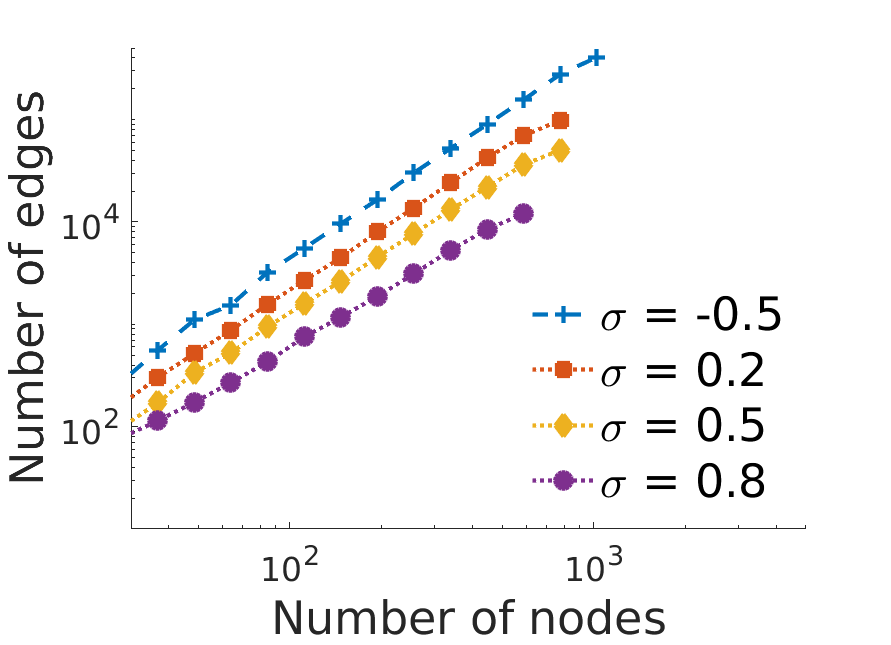}}\\
\subfloat[Core-periphery region\label{fig:cpedgesvsnodes}]{\includegraphics[width=0.35\textwidth]{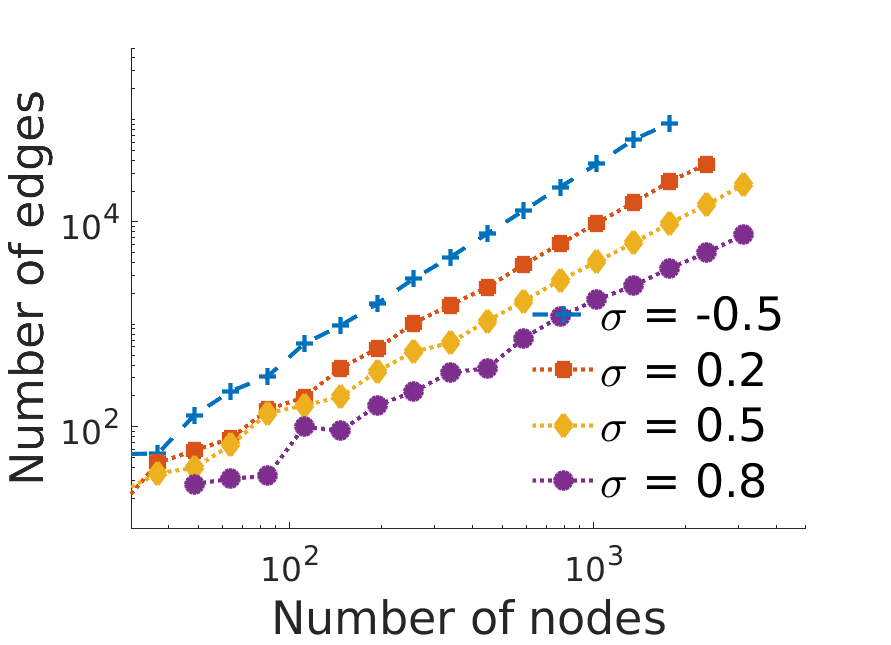}}
\subfloat[Periphery region\label{fig:ppedgesvsnodes}]{\includegraphics[width=0.35\textwidth]{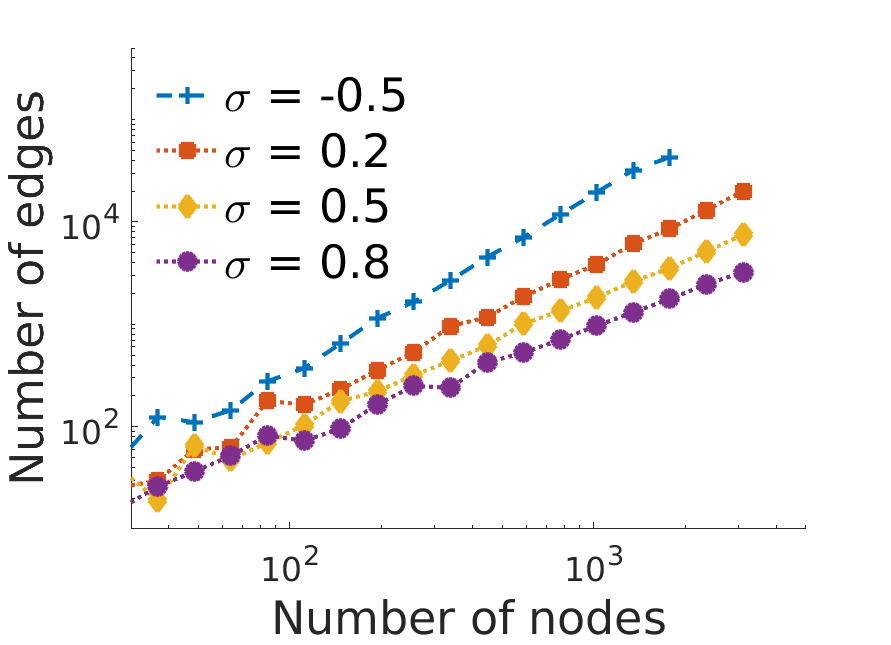}}
\caption{Relationship between nodes and edges for varying $\sigma$ for $(a)$ the overall graph, $(b)$ the core, $(c)$ the core-periphery region and $(d)$ the periphery. In each case we use $K=2, \tau = 1, a= 0.2, b = \frac{1}{K}$ and vary $\alpha$ to generate different sized graphs.}
\label{asymptotic_graphs}
\end{figure}

In Figure \ref{asymptotic_graphs}, we see a graphical representation of these results. From Figure \ref{asymptotic_graphs}\subref{fig:edgesvsnodes} we see that the number of edges grows quadratically with the number of nodes when $\sigma<0$ (and thus the graph is dense) and otherwise grows with power-law exponent $\frac{2}{1+\sigma}$. We see similar behaviour for the periphery and core-periphery regions, but in Figure \ref{asymptotic_graphs}\subref{fig:ccedgesvsnodes} we see that the core region is dense for any value of $\sigma$. In Appendix \ref{app:simulations} we present some more empirical results on the effects of varying the model parameters on the degree distribution, sparsity properties and core proportion.

In the next theorem, we present the results in the case of both core-periphery and community structure:
\begin{theorem}\label{thm:cp_asymptotics_community}
Consider the graph family defined by Equations~\eqref{eq:graphpointprocess} and \eqref{eq:core_community_model}. Assume that the mean measure $\rho$ satisfies the assumptions
\begin{align}
\int_{(0,\infty)\times \{0\}^{K-1}} \rho(dw_1,\ldots,dw_K)=0,\label{eq:cond1_community}\\
\int_{\{0\}\times (0,\infty)^{K-1} }  \rho(dw_1,\ldots,dw_K)>0,\label{eq:condperiphery_community}\\
0<\int_{(0,\infty)^{K}}\rho(dw_1,\ldots, dw_{K})<\infty.\label{eq:condcore_community}
\end{align}
Then the same results as in Theorem \ref{thm:cp_asymptotics1} hold.
 Furthermore, if the L\'evy measure $\rho$  takes the form of Equation~\eqref{eq:levymeasurerho_communities}, with a generalized gamma process base measure $\rho_0$, and $F$ a product of independent gamma distributions, and $\sigma \in (0,1)$ and $\tau >0$, so that
\begin{align}
\int_0^{\infty}\rho_0(dw_0)=\infty, \qquad \int_0^{\infty}w_0\rho_0(dw_0)<\infty.
\end{align}
Then the results of Theorem \ref{thm:cp_asymptotics2} and Corollary \ref{thm:core_size} also hold.
\end{theorem}

\section{Discussion}
\label{sec:discussion}

A standard and natural model-based approach for detecting a core-periphery structure, described in \cite{zhang2015identification}, is to consider a two-groups stochastic blockmodel where
$$
\Pr(z_{ij}=1\mid \Pi, c_i,c_j)=\Pi_{c_i,c_j}
$$
where $c_i=1$ if node $i$ is in the core and $c_i=2$ otherwise, $$\Pi=\left(
             \begin{array}{cc}
               \Pi_{11} & \Pi_{12} \\
               \Pi_{12} & \Pi_{22} \\
             \end{array}
           \right)
$$
with $\Pi_{11}$, $\Pi_{11}$ and $\Pi_{22}$ respectively the core-core, core-periphery and periphery-periphery probabilities of connection, where typically  $\Pi_{12},\Pi_{22}\ll\Pi_{11}$. Such models however cannot produce sparse graphs (in the sense of Definition~\ref{def:sparsity}) with power-law degree distribution; see e.g. \cite{Caron2017}. To obtain sparse graph sequences, one can have the matrix $\Pi$ to depend on $n$, e.g. $\Pi_n=\frac{\Pi_1}{n}$ for some initial matrix $\Pi_1$~\cite{zhang2015identification}. However in this case, the graph family is not projective any more for different network sizes. The approach considered in this paper allows to have both a sparse and projective graph family.

\citep{Williamson2018} recently developed a class of sparse graphs with locally dense subgraphs. The construction is based on Poisson random measures as in our case, but the objective and properties of both models are rather different. The graphs of \citep{Williamson2018} have a growing number of dense subgraphs, where each subgraph has a bounded number of nodes, and no subgraph is identified as a periphery. In contrast, our model has a single dense core, whose size is unbounded, and a sparse periphery.

\section{Posterior Inference} \label{inference}
We design an MCMC algorithm to perform posterior inference with the core-periphery model, based on that of \citep{todeschini2016exchangeable}. Although we have seen from Section \ref{sec:particularmodel} that our model does not fall into the compound CRM framework used there, we can adapt the algorithm to our setting. In Section \ref{sec:sampler} we give the basic structure of the sampler, with more details provided in Appendix \ref{app:mcmc_alg}.

\subsection{Characterization of conditionals and MCMC sampler}\label{sec:sampler}

In this Section, we consider the model of Section \ref{community_and_core_periphery}.
As in \cite{todeschini2016exchangeable}, we assume that we have observed a set of connections $(z_{ij})_{1 \leq i,j,\leq N_{\alpha}}$ where $N_{\alpha}$ is the number of nodes with at least one connection. We want to infer the parameters $(w_{i1},\ldots,w_{iK})_{i=1,\ldots,N_{\alpha}}$. We also want to estimate the sums of the parameters for the nodes with no connection $(w_{\ast 1},\ldots,w_{\ast K})$, the hyperparameters $\phi$ of the mean intensity $\rho$ and the parameter $\alpha$ which is also assumed to be unknown. Thus the aim is to sample from the posterior
\begin{align}
p((w_{1k},\ldots,w_{N_{\alpha}k},w_{\ast k})_{k=1,\ldots,K},\phi,\alpha \mid(z_{ij})_{1 \leq i,j,\leq N_{\alpha}})
\end{align}
As in \citep{todeschini2016exchangeable}, we introduce the latent count variables $\tilde{n}_{ijk}$ with
\begin{equation} \label{eq:augmentation}
\begin{aligned}
(\tilde{n}_{ij1},\ldots,\tilde{n}_{ijK})\mid w,z &\sim \begin{cases}
\delta_{(0,\ldots,0)} &z_{ij}=0\\
\mbox{tPoisson}(2w_{i1}w_{j1},\ldots,2w_{iK}w_{jK}) &z_{ij}=1,i \neq j
\end{cases} \\
\left(\frac{\tilde{n}_{ij1}}{2},\ldots,\frac{\tilde{n}_{ijK}}{2}\right)\mid w,z &\sim \mbox{tPoisson}(2w_{i1}^2,\ldots,2w_{iK}^2) \qquad z_{ij}=1,i = j
\end{aligned}
\end{equation}
where $\mbox{tPoisson}(\lambda_1,\ldots,\lambda_K)$ is the multivariate Poisson distribution truncated at $(0,\ldots,0)$. We note that in the overlapping communities model, all the parameters $\lambda_k$ would be strictly positive. In our case there is a non-zero probability that $\lambda_1=0$. We allow for this by defining a $\mbox{Poisson}(0)$ random variable to be identically $0$.

Then similarly to \citep{Caron2017},  using the data augmentation scheme we can define an algorithm which uses Metropolis-Hastings (MH) and Hamiltonian Monte Carlo (HMC) updates with a Gibbs sampler to perform posterior inference. At each iteration of the Gibbs sampler we update using Algorithm \ref{alg:MCMC_sampler}.

\begin{algorithm}
  \caption{MCMC sampler for posterior inference}\label{alg:MCMC_sampler}
  At each iteration:
  \begin{algorithmic}[1]
  \State Update $(w_{i0}, \beta_{i,1},\ldots,\beta_{iK})_{i=1,\ldots,N_{\alpha}}$  given the rest at the same time using HMC.
  \State Update the hyperparameters $(\phi,\alpha)$ and the total masses $(w_{\ast 1},\ldots,w_{\ast K})$ given the rest using MH.
  \State Update the latent variables $\tilde{n}_{ijk}$ given the rest using (\ref{eq:augmentation}).
  \end{algorithmic}
\end{algorithm}
In the first step of the algorithm, we use the conditional distribution given the latent variable counts as defined in \citep{todeschini2016exchangeable}:
\begin{equation}\label{ccrm_posterior_characterization}
\begin{aligned}
p((&w_{10},\ldots,w_{N_{\alpha}0}),(\beta_{1k},\ldots,\beta_{N_{\alpha}k},w_{\ast k})_{k=1,\ldots,K}\mid(n_{ijk})_{1 \leq i,j,\leq N_{\alpha},k=1,\ldots,K},\phi,\alpha)\\
\propto & \left[\prod_{i=1}^{N_{\alpha}}  w_{i0}^{m_{i}}\right]\left[\prod_{i=1}^{N_{\alpha}} \prod_{k=1}^K \beta_{ik}^{m_{ik}}\right] e^{-\sum_{k=1}^K(w_{\ast k}+\sum_{i=1}^{N_{\alpha}}w_{i0}\beta_{ik})^2}\\
&\times \left[\prod_{i=1}^{N_{\alpha}}f(\beta_{i2},\ldots,\beta_{iK};\phi)\right]\left[\prod_{i=1}^{N_{\alpha}} (1-e^{-w_{i0}})^{\beta_{i1}}e^{-w_{i0}(1-\beta_{i1})} \right]\\
&~~~~\times \left[\prod_{i=1}^{N_{\alpha}}\rho_0(w_{i0};\phi)\right]\alpha^{N_{\alpha}}g_{\ast \alpha}(w_{\ast 1},\ldots,w_{\ast K};\phi)
\end{aligned}
\end{equation}
where $m_i=\sum_{k=1}^Km_{ik}$, $m_{ik} = \sum_{j=1}^{N_\alpha} \tilde{n}_{ijk}$ and $f$ is the density function corresponding to the distribution $F$ (in this case the product of gamma densities). The key difference in our case compared to \citep{todeschini2016exchangeable}, as we see in Section \ref{sec:particularmodel}, is that the $\beta_{i1}$ are not identically distributed, and depend on $w_{i0}$. This gives us the separate term involving the $\beta_{i1}$ in \eqref{ccrm_posterior_characterization}. However, we see that we can still follow the same algorithm, with some modifications, and we give the details of the steps in Appendix \ref{app:mcmc_alg}. The pdf $g_{\ast \alpha}(w_{\ast 1},\ldots,w_{\ast K};\phi)$ has no analytic expression, and we use an approximation to update the total masses $(w_{\ast 1},\ldots,w_{\ast K})$ in Step 2 of Algorithm \ref{alg:MCMC_sampler}.

\section{Experiments} \label{experiments}
\subsection{Simulated data}
\label{sec:experiments_simu}
In order to test our posterior inference algorithm, we first generate synthetic data from the model described in Section \ref{sparse-model}, with the construction given in \ref{sec:particularmodel}. We generate a graph with $K=2$, i.e. with a core-periphery structure but no community structure. We use the same parameters $\alpha = 200$, $\sigma = 0.2$, $\tau = 1$, $b = \frac{1}{K}$, $a=0.2$ as used by \citep{todeschini2016exchangeable}. In our case, the sampled graph has $778$ nodes and 5984 edges.

We fit our model to the simulated network, placing a vague $\Gamma(0.01,0.01)$ priors on the unknown parameters $\alpha$, $1-\sigma$, $\tau$, $a_k$ and $b_k$. We run 3 parallel MCMC chains with an initialization run of $10000$  steps and full chain lengths of  $500000$. We then discard the first $375000$ samples as burn in and thin the remaining $125000$ to give a sample size of $500$. Trace plots and convergence diagnostics are given in Section \ref{sec:app:simucore} of Appendix \ref{app:mcmc_diagnostic}.

\begin{figure}[h]
\centering
\subfloat[Mean sociability credible intervals for high degree nodes  \label{fig:high_credible}]{\includegraphics[width=0.35\textwidth]{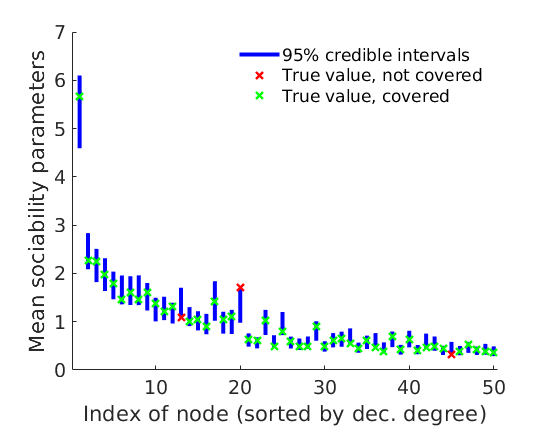}}
\hspace{0.01\textwidth}
\subfloat[Mean sociability credible intervals for low degree nodes \label{fig:low_credible}]{\includegraphics[width=0.35\textwidth]{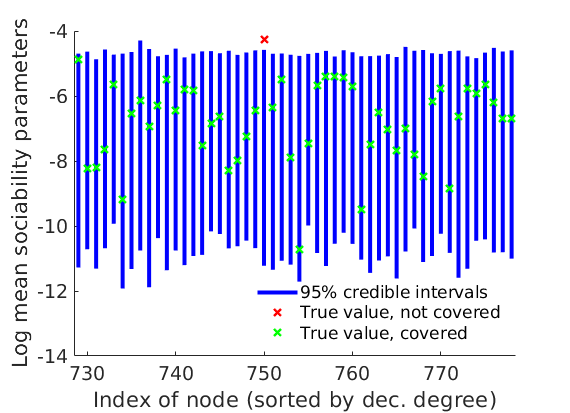}}\\
\subfloat[Posterior predictive degree distribution \label{fig:post_pred_deg}]{\includegraphics[width=0.35\textwidth]{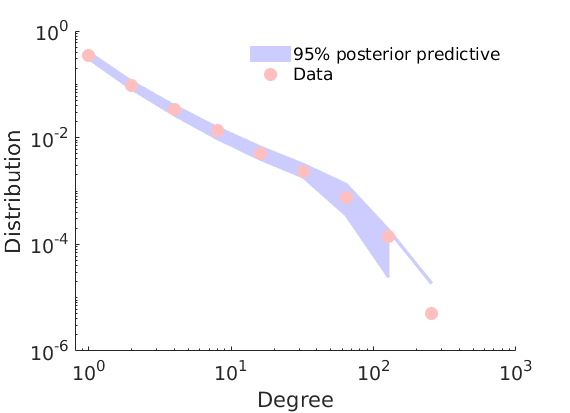}}
\hspace{0.01\textwidth}
\subfloat[Core sociability ($w_{i1}$)  credible intervals for high degree nodes \label{fig:core_credible}]{\includegraphics[width=0.35\textwidth]{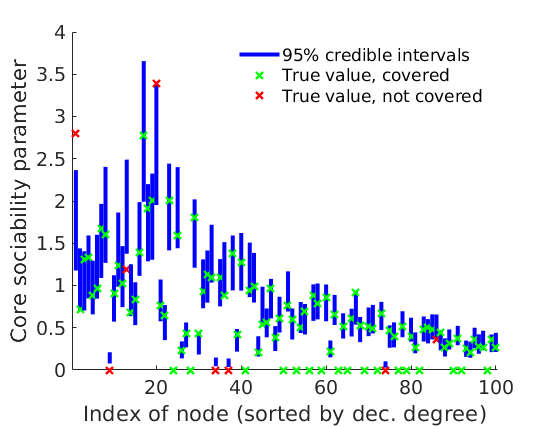}}

\caption{Fitting our model to data simulated from a graph generated with $K=2, \alpha = 200, \sigma= 0.2, \tau = 1, b=\frac{1}{K}, a=0.2$}\label{fig:credible}
\end{figure}

Our model accurately recovers the  mean sociability parameters (in this case, the mean $\bar{w}_i=\frac{1}{2}(w_{i1}+w_{i2})$ of the core and overall sociability parameters) of both high and low degree nodes, providing reasonable credible intervals in each case as shown in Figures \ref{fig:credible}\subref{fig:high_credible} and \ref{fig:credible}\subref{fig:low_credible}. By generating $5000$ graphs from the posterior predictive distribution, we also see that the model fits the empirical power-law distribution of the generated graph, as shown in \ref{fig:credible}\subref{fig:post_pred_deg}. In Figure \ref{fig:credible}\subref{fig:core_credible} we see that some high degree periphery nodes in particular being mis-classified into the core. However, we expect that these high degree periphery nodes will be some of the hardest to correctly classify.

Moreover, we see that we are able to very accurately recover the classification into core and periphery. In our generated graph, 137 out of 144 nodes are classified correctly into the core, with 631 out of 634 are classified correctly into the periphery. Importantly, the core is not simply comprised of the nodes with the highest degrees. In  Figure \ref{fig:credible}\subref{fig:core_credible} we see the credible intervals for the core sociability parameters, and we see that there are several high degree nodes in the periphery which are identified correctly. This means that our algorithm is not simply classifying the nodes with the highest degrees into the core.

\begin{figure}[h]
\centering
\subfloat[Adjacency Matrix  \label{fig:adj_4}]{\includegraphics[width=0.35\textwidth]{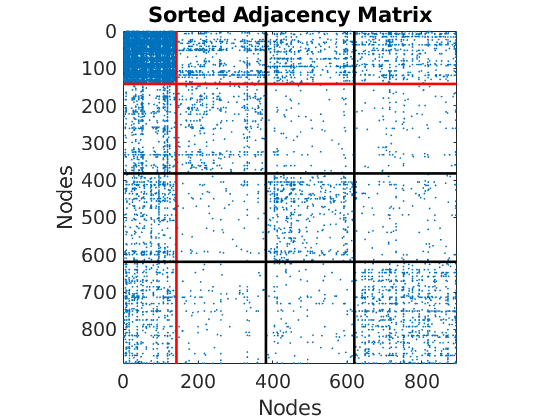}}
\hspace{0.01\textwidth}
\subfloat[Posterior predictive degree distribution \label{fig:post_pred_4}]{\includegraphics[width=0.35\textwidth]{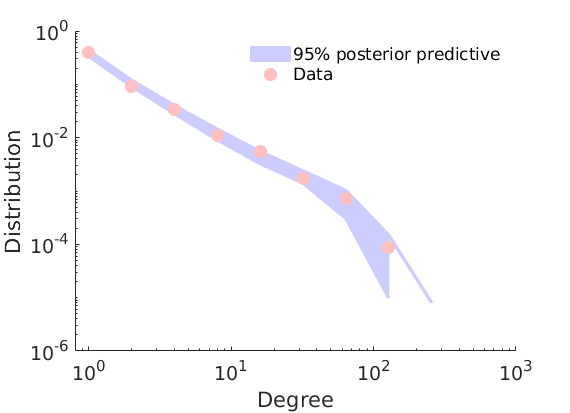}}\\
\subfloat[Mean sociability credible intervals for high degree nodes \label{fig:high_credible_4}]{\includegraphics[width=0.35\textwidth]{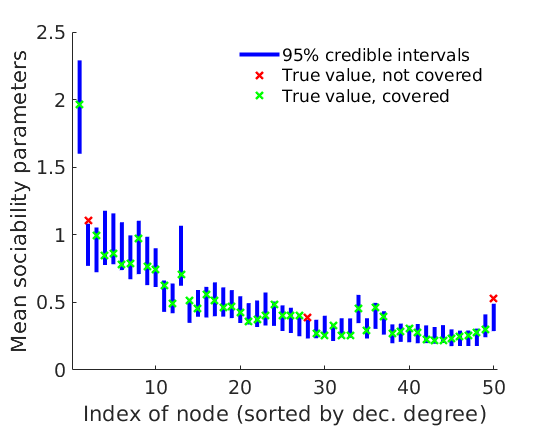}}
\hspace{0.01\textwidth}
\subfloat[Mean sociability credible intervals for low degree nodes \label{fig:low_credible_4}]{\includegraphics[width=0.35\textwidth]{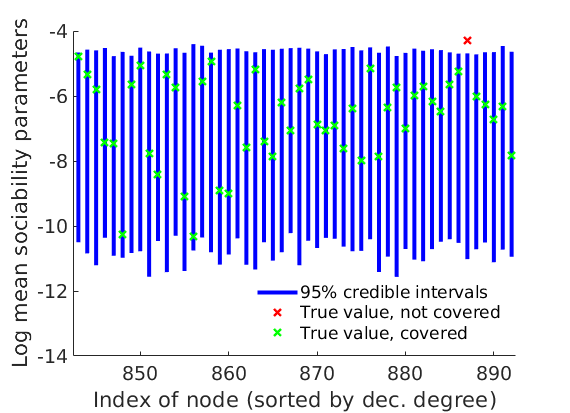}}
\caption{Fitting our model to simulated data with core-periphery and community structure, generated with $K=4, \alpha = 200, \sigma= 0.2, \tau = 1, b_i=b=\frac{4}{K}, a_i=a=0.2$.}\label{fig:crediblecorecommunity}
\end{figure}

We also test our model on a graph with core-periphery and community structure. We take $K=4$ in order to generate a graph with a core-periphery structure and 3 overlapping communities. In Figure~\ref{fig:crediblecorecommunity}\subref{fig:adj_4} we see the adjacency matrix for the simulated network, sorted into core and periphery (indicated by the red lines) and then by largest community sociability (indicated by the black lines). We see a clear global core-periphery structure and community structure. As before, we see from Figures~\ref{fig:crediblecorecommunity}\subref{fig:post_pred_4} and~\ref{fig:crediblecorecommunity}\subref{fig:low_credible_4} that we are able to recover the degree distribution, as well as the mean sociabilities of the high and low degree nodes. 
Additional trace plots, convergence diagnostics and results on core detection are reported in Section~\ref{sec:app:simucorecommunity} of Appendix \ref{app:mcmc_diagnostic}.

\subsection{Real world networks}
\label{sec:experiments_real}

 There are many classical examples of networks with core-periphery structure. Our model is designed to detect this structure in sparse networks with a power-law degree distribution. We apply our method on two real world networks, one that has a power-law distribution and one that does not. We will see that our method performs well in the power-law setting but also produces reasonable and interpretable results in the non-power-law case. The power-law network we consider is the \textbf{USairport} network of airports with at least one connection to a US airport in 2010\footnote{\url{http://www.transtats.bts.gov/DL_SelectFields.asp?Table_ID=292}}. This network was previously considered by \citep{todeschini2016exchangeable}, and we compare the methods to see the benefits of using a core-periphery model in this case. The other network we consider is the \textbf{Trade} network of historical international trade\footnote{\url{https://web.stanford.edu/~jacksonm/Data.html}} for which the core-periphery structure has been previously studied \citep{della2013profiling,fagiolo2010evolution}. In Table \ref{table:real_networks}, we give the size of these networks, the value of $K$ we use, the estimated relative size of the core and the estimated value of $\sigma$. Following the work of \citep{todeschini2016exchangeable}, we take $K=4$ for the \textbf{USairport} data set, whilst for the \textbf{Trade} network we take $K=2$. In each case, we assume vague $\Gamma(0.01,0.01)$ priors on the unknown parameters $\alpha$, $1-\sigma$, $\tau$, $a_k$ and $b_k$.

 \begin{table}[ht]
    \centering
    \begin{tabular}{|r||rrrrr|}
      \hline
     Name & Nb Nodes& Nb Edges& $K$ & Est Core Prop & $\widehat\sigma$ \\
      \hline
   \textbf{USairport}& 1574 & 17215 & 4 & 0.13 & 0.22  \\
  \textbf{Trade}& 158 & 1897 & 2 & 0.39 & -0.77 \\
   \hline
    \end{tabular}
    \caption{Real World Networks and Summary Statistics}
    \label{table:real_networks}
    \end{table}

\subsubsection{US Airport network}\label{sec:us_airport}
The first real world network we consider is the \textbf{USairport} network of airports with at least one connection to a US airport in 2010. Airport networks such as this have been seen to have a core-periphery structure \cite{holme2005core,della2013profiling}. Furthermore, one of the communities identified by \citep{todeschini2016exchangeable} are the \textit{Hub} airports, highly connected airports with no preferred location. It seems plausible that the network could more accurately be modelled by a core-periphery model, whilst retaining the three other communities. Thus, we take $K=4$ and fit our model to the network.

\begin{figure}[h]
\centering
\subfloat{\includegraphics[width=0.4\textwidth]{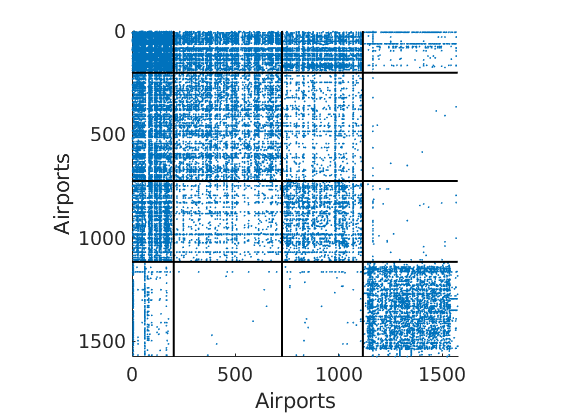}}
\subfloat{\includegraphics[width=0.4\textwidth]{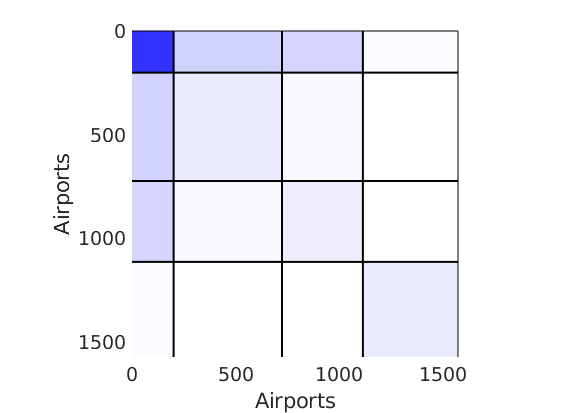}}
\caption{Adjacency matrix for the US Airport Network, ordered into core and periphery, and then by highest community weight. In $(a)$ we give the standard adjacency matrix, $(b)$ shows the densities in each of the regions}
\label{fig:airports_adj}
\end{figure}

\begin{figure}[h]
\centering
\subfloat[Airports with highest core weights \label{fig:airports_top_core}]{\includegraphics[width=0.4\textwidth]{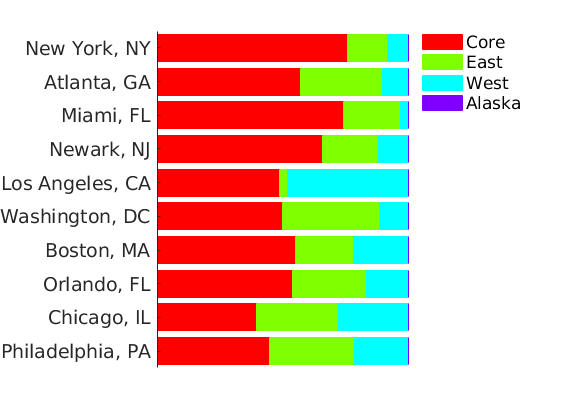}}~~
\subfloat[Periphery airports with highest degrees \label{fig:airports_top_periphery}]{\includegraphics[width=0.4\textwidth]{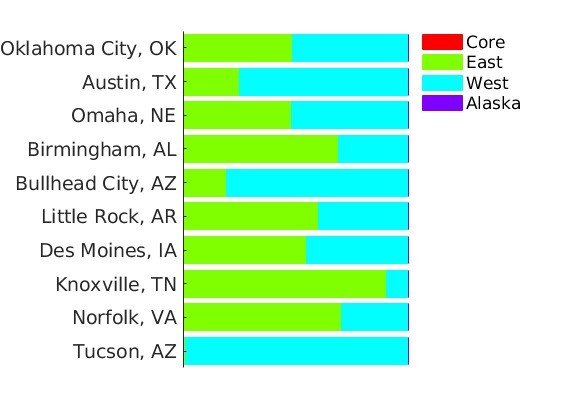}}
\caption{Relative values of the weights for ``top'' core and periphery nodes }
\label{fig:airports_post_ranks}
\end{figure}

We run 3 parallel MCMC chains with an initialization run of $10000$  iterations followed by $10^7$ iterations. We then discard the first $5000000$ samples as burn in and thin the remaining $5000000$ to give a sample size of $500$. Trace plots and convergence diagnostics are reported in Section~\ref{sec:app:airports} of Appendix \ref{app:mcmc_diagnostic}. Note that despite the large number of iterations, it seems that the sampler has not fully converged yet. Nonetheless, we observed that increasing the number of iterations does not the change significantly the values of the core-periphery parameters of interest and the overall interpretation of the network. 
In Figure \ref{fig:airports_adj} we see the adjacency matrix formed by ordering the nodes firstly by core and periphery, and then by their highest community sociability. We recover similar communities to those found in the previous work. Furthermore, the \textit{core} comprises similar nodes to those previously placed in the \textit{Hub} community by \citep{todeschini2016exchangeable}. However, as mentioned by \citep{rombach2017core}, this hub community can be better explained by a core-periphery structure, and we also find that to be the case here. In Figure~\ref{fig:airports_post_ranks}\subref{fig:airports_top_core} we see the relative values of the weights for the core airports with the highest core weights. As we expect, these are all major international hubs. Conversely, in Figure~\ref{fig:airports_post_ranks}\subref{fig:airports_top_periphery} we see the same thing for the airports in the periphery with the highest degree. We see that these are generally large regional airports, with many connections within the \textit{East} or \textit{West} communities. The final community corresponds to Alaskan airports, and we can see from Figure~\ref{fig:airports_adj} that airports in this community generally do not have many connections to the core.

If we investigate the classification into core and periphery in more detail, we see several interesting features. Firstly, when looking at the airports placed in the core, we can compare these to the list of hub airports for various airline companies\footnote{\url{https://en.wikipedia.org/wiki/List_of_hub_airports}}. We see that $38$ out of $47$ of these airline hub airports are placed in the core, with some interesting exceptions. Three of the airline hub airports not placed in the core are only hubs for the parcel delivery services Fedex and UPS, whilst one is only a hub for a small charter airline. The most interesting exceptions are Chicago Midway and Dallas Love Field airports, both of which are focus cities for major airline company Southwest Airlines but are not placed in the core. A possible explanation for this is that Southwest Airlines does not operate using a traditional spoke-hub model, instead utilising a point-to-point system.

Secondly, we see that our model does not simply classify by degree, the airports that we see in Figure~\ref{fig:airports_post_ranks}\subref{fig:airports_top_periphery} have degrees higher than many core nodes. If we look at the core nodes with low degrees, we can see a clear pattern here. Most of these airports are major international hubs such as in London Heathrow, Frankfurt International, Zurich and Tokyo Narita. Another airport that appears here is Honolulu International. In each case, it is not surprising that these airports are placed in the core, as most of  the connections are long-distance flights to the major US hub airports.

\begin{figure}[h]
\centering
\subfloat[Core-periphery posterior  degree \label{fig:airports_cp_post_10m}]{\includegraphics[width=0.35\textwidth]{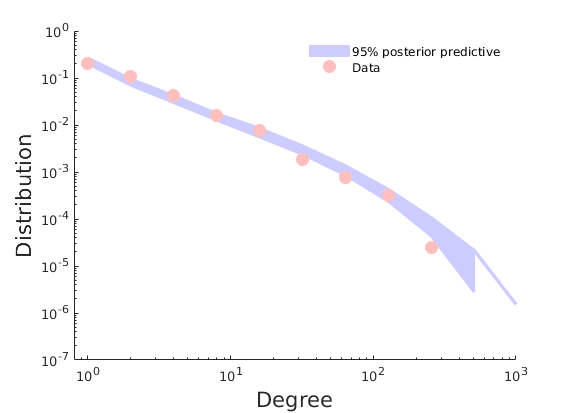}}~~
\subfloat[SNetOC posterior  degree\label{fig:airports_gamma_post_10m}]{\includegraphics[width=0.35\textwidth]{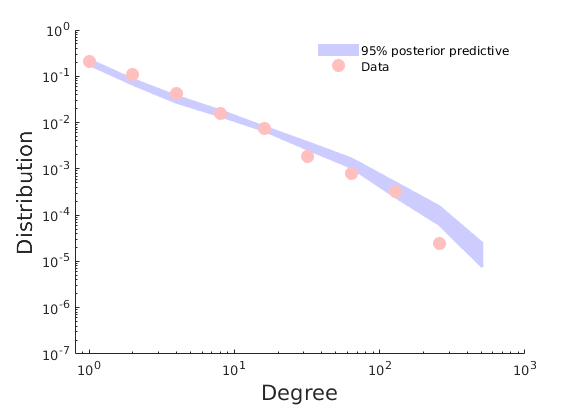}}

\caption{Posterior Predictive Degree Distributions for the US Airport Network}
\label{fig:airports_post_ranks_10m}
\end{figure}

One advantage of our model over that of \citep{todeschini2016exchangeable} in this setting is that ours gives a discrete classification into core and periphery, whilst also modelling community affiliations. Another advantage can be seen when examining the fit of the two models to the degree distribution. In Figures~\ref{fig:airports_post_ranks_10m}\subref{fig:airports_cp_post_10m} and~\ref{fig:airports_post_ranks_10m}\subref{fig:airports_gamma_post_10m}, we see the $95\%$ posterior predictive intervals for the degree distribution for both models. We see that while the core-periphery model slightly overestimates the number of nodes with high degrees, the communities model is significantly overestimating the number of these nodes.

In Table \ref{table:airports_gof} we see some statistics to measure the distance between the posterior predictive degree distributions from our model and the true degree distribution, as well as the corresponding difference for the model of \citep{todeschini2016exchangeable}. These illustrate the benefit of using our core-periphery model. We first report a standard statistic, the Kolgomorov-Smirnov (KS) statistic. However, it is well known that the KS statistic has poor sensitivity to deviations from the true distribution that occur in the tails \cite{mason1983modified}. So, we also give here the reweighted KS statistic suggested by \citep{clauset2009power}, which weights the tails more strongly:
\begin{align}
    D = \max_{x \geq x_{min}} \frac{|S(x) - P(x)|}{\sqrt{P(x)(1-P(x))}}
\end{align}
where $S(x)$ is the CDF of observed degrees, $P(x)$ is the CDF of degrees of graphs sampled from the posterior predictive distribution and $x_{min}$ is the minimum value among the observed and sampled degrees. For the reweighted statistic we also see that our model is performing significantly better.

However, we may also want to know precisely what deviations are occurring, and in which tails. \citep{mason1983modified} devise another version of the KS test, which uses the R\'enyi statistics $(L_1,U_1)$ and $(L_2,U_2)$ to test for light and heavy tailed alternatives respectively. Here our interest lies on the heavy tail aspect of degree distributions, so  we only consider the R\'enyi statistics $L_2$ and $U_2$. We do not need the full estimator of \citep{mason1983modified} as we are not performing a goodness of fit test, but simply using these statistics as a measure of distance for comparison purposes. In Table \ref{table:airports_gof} we confirm that the problem of overestimating the high degree nodes is worse for the communities model.

\begin{table}[htb]
\centering
\begin{tabular}{ |p{2cm}||p{2cm}|p{2cm}|p{1.9cm}|p{1.9cm}|  }
 \hline
\multicolumn{5}{|c|}{Distance Measures} \\
 \hline
\multirow{2}{4em}{Method} & Reweighted KS& Unweighted KS & \multicolumn{2}{c|}{R\'enyi Statistics} \\
\cline{4-5}
 &  $D$&  $K$&  $L_{2}$&  $U_{2}$\\
 \hline
 Core-periphery& $\mathbf{0.186 \pm 0.048}$  &   $\mathbf{3.695 \pm 1.217}$ & $0.029 \pm 0.002$    &$\mathbf{0.758 \pm 0.222}$  \\ 
 Communities &$0.251 \pm 0.081 $   &$5.360 \pm 1.461 $ &   $\mathbf{0.027 \pm 0.001}$   & $1.019 \pm 0.237$      \\
  \hline
\end{tabular}
\caption{Distance Measures for the US Airport Network}
\label{table:airports_gof}
\end{table}

Investigating the overestimation of high degree nodes more carefully, we see that for our model and that of \citep{todeschini2016exchangeable}, the posterior distribution on $b_4$, the parameter entering the Gamma distribution  $F$ corresponding to the Alaskan community (4th community) concentrates on values very close to 0 (see Figure \ref{fig:mcmc_trace_airports} in Appendix \ref{app:mcmc_diagnostic}). This increases the posterior variance of the sociabilities $w_{i4}$, and leads to some graphs with very high degree nodes being generated when simulating from the posterior predictive distribution. A possible reason for this is that the Alaskan airports exhibit a different type of behaviour, not well explained by our model. A small number of nodes (the Alaskan airports) are strongly connected to each other but not to the other communities, or the core nodes. Furthermore, hardly any of the Alaskan airports are in the core. Conversely, the \textit{East} and \textit{West} communities contain large numbers of core nodes, and also have more connections between communities.

In order to investigate this possible source of model misspecification further, we repeat our analysis of this network, taking out all of the Alaskan airports, and any airport which only has connections to Alaskan airports. In Appendix \ref{app:us_airport_no_alaska} we present the results. We see that the overestimation problem is no longer present, but overall the fit to the degree distribution is no better than for the full model. Furthermore, if we examine the nodes that are placed into the core and periphery, there is not very much difference.

\subsubsection{World Trade network}
The next network we consider is the dyadic trade network between countries\footnote{\url{https://web.stanford.edu/~jacksonm/Data.html}}. The original data details the flow of trade between pairs of countries from 1870 to 2009. In our case, we consider the single simple, undirected network found by aggregating the data over all the years. However, as shown by \citep{della2013profiling}, when the trade data set is considered as an unweighted network, with a link between countries if there was any flow of trade between them, then the core-periphery structure is quite weak. There is a strong core-periphery structure if the network is weighted by volume of trade is considered. As our method currently cannot deal with weighted networks, we instead use a cutoff method, forming a binary network by only considering trade links over a certain volume.

In order to perform posterior inference, we run 3 MCMC chains with an initialization run of $10000$ steps and full chain lengths of $200000$. Trace plots and convergence diagnostics are reported in Section~\ref{sec:app:WT} of Appendix \ref{app:mcmc_diagnostic}, suggesting the convergence of the MCMC sampler.

\begin{figure}[h]
\centering
\subfloat{\includegraphics[width=0.35\textwidth]{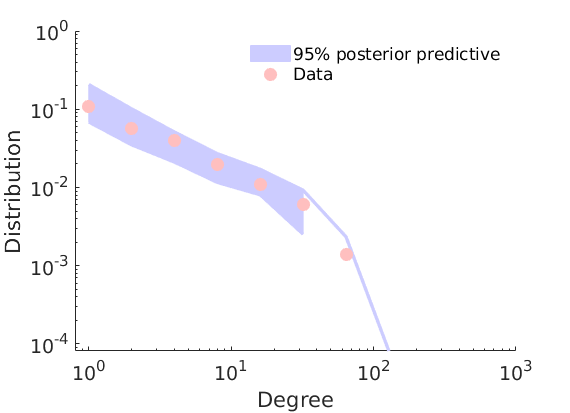}}
\subfloat{\includegraphics[width=0.35\textwidth]{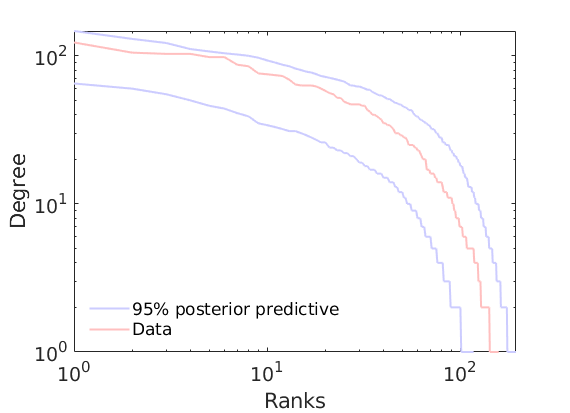}}
\caption{Posterior Predictive Degree Distribution and Ranked Degrees for the World Trade Network}
\label{fig:trade_post_ranks}
\end{figure}

As we see in Table \ref{table:real_networks}, $\sigma$ is estimated to be negative in this case. This means that this network does not fit into our definition of a sparse graph as defined in Definition \ref{def:coreperiphery}. Nevertheless, in Figure \ref{fig:trade_post_ranks} we plot the posterior predictive distribution and the posterior predictive intervals for the ranked degrees\footnote{This is formed by generating graphs from the posterior predictive distribution, and calculating the sequence of the ordered degrees of the nodes.} and we see that we can estimate the degree distribution fairly well, albeit with a large posterior predictive interval. Furthermore, we will see that we can obtain an interpretable classification of countries into core and periphery. Therefore, our model is still producing useful results despite the network not technically fitting into our framework.

\begin{figure}[h]
\centering
\subfloat{\includegraphics[width=0.4\textwidth]{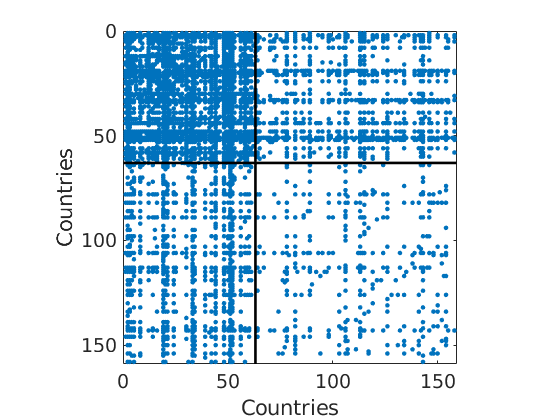}}
\subfloat{\includegraphics[width=0.4\textwidth]{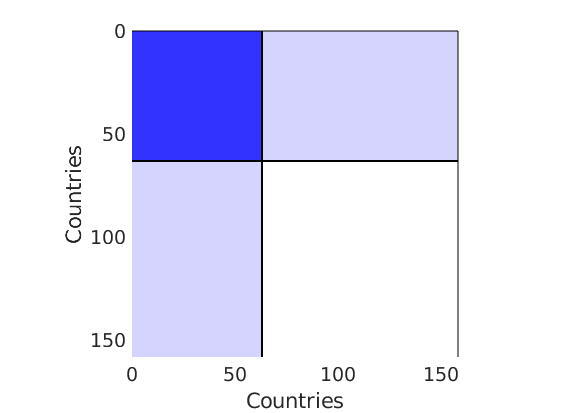}}
\caption{Adjacency matrix for the World Trade network, ordered into core and periphery. In $(a)$ we give the standard adjacency matrix, $(b)$ shows the densities in each of the regions}
\label{fig:trade_adj}
\end{figure}

As with the US airport network, we calculate both the unweighted and reweighted KS statistics in Table \ref{table:trade_gof}. We again compare against the model of \citep{todeschini2016exchangeable}, with two communities. Here we see that our model does not give as good a fit, however the large standard deviations show that there is no significant difference between the two. However, the advantage of our model is that it gives a discrete classification between core an periphery, which we see clearly from the adjacency matrix in Figure \ref{fig:trade_adj}.

\begin{table}[ht]
\centering
\begin{tabular}{ |p{2cm}||p{2cm}|p{2cm}|  }
 \hline
\multicolumn{3}{|c|}{Distance Measures} \\
 \hline
\multirow{2}{4em}{Method} & Reweighted KS& Unweighted KS \\
 &  $D$&  $K$\\
 \hline
 Core-periphery& $0.505 \pm 0.288$  &   $1.407 \pm 0.488$  \\ 
 SNetOC &$\mathbf{0.454 \pm 0.270 }$  &$\mathbf{1.308 \pm 0.521}$ \\
  \hline
\end{tabular}
\caption{Distance Measures for the World Trade Network}
\label{table:trade_gof}
\end{table}

\begin{figure}[h]
\centering
\subfloat[Mean Sociabilities by Country \label{fig:trade_mean_map}]{\includegraphics[width=0.6\textwidth, trim={0, 5cm, 0 2cm},clip]{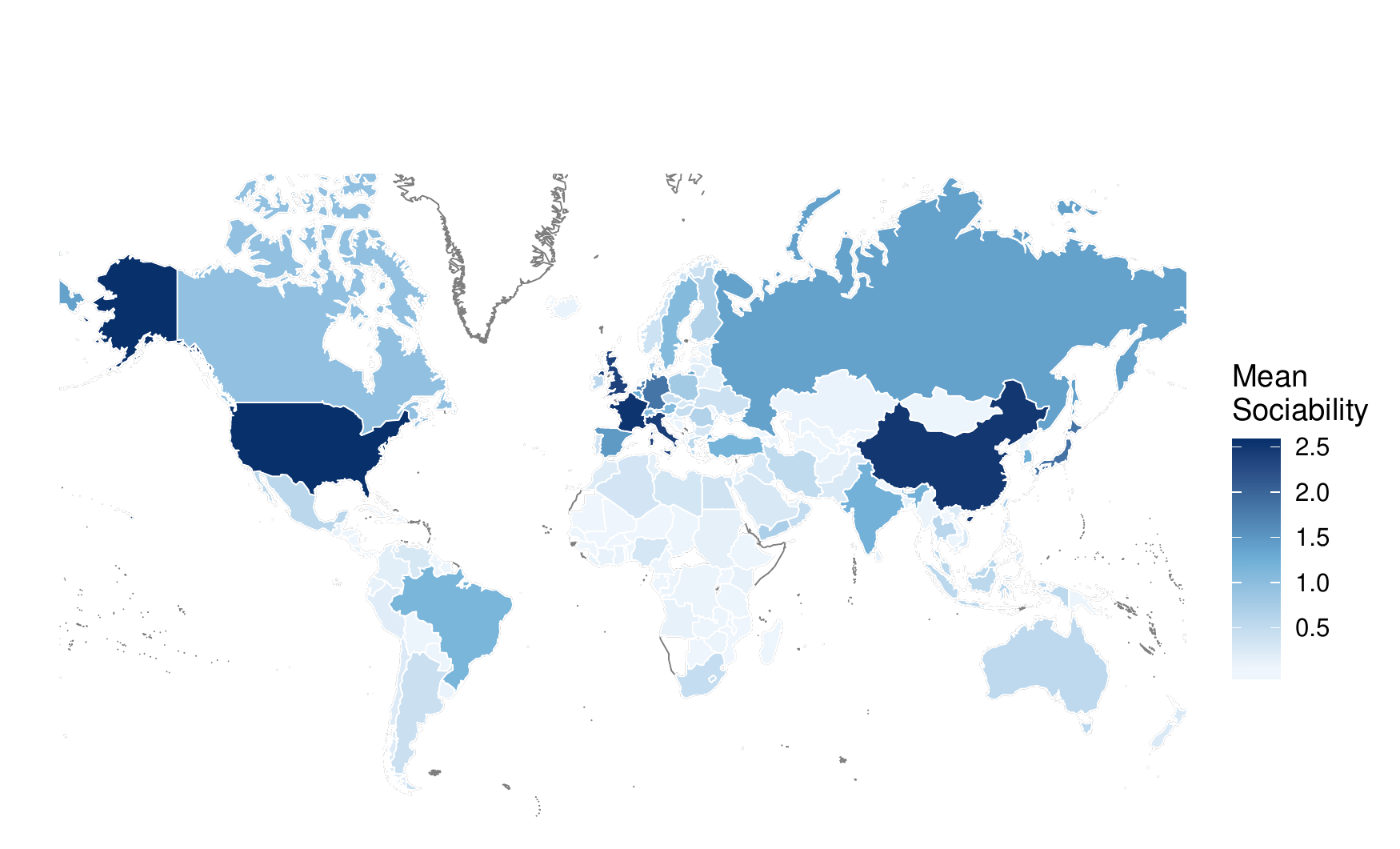}} \\
\subfloat[Core Sociabilities by Country \label{fig:trade_core_map}]{\includegraphics[width=0.6\textwidth, trim={0, 5cm, 0 2cm},clip]{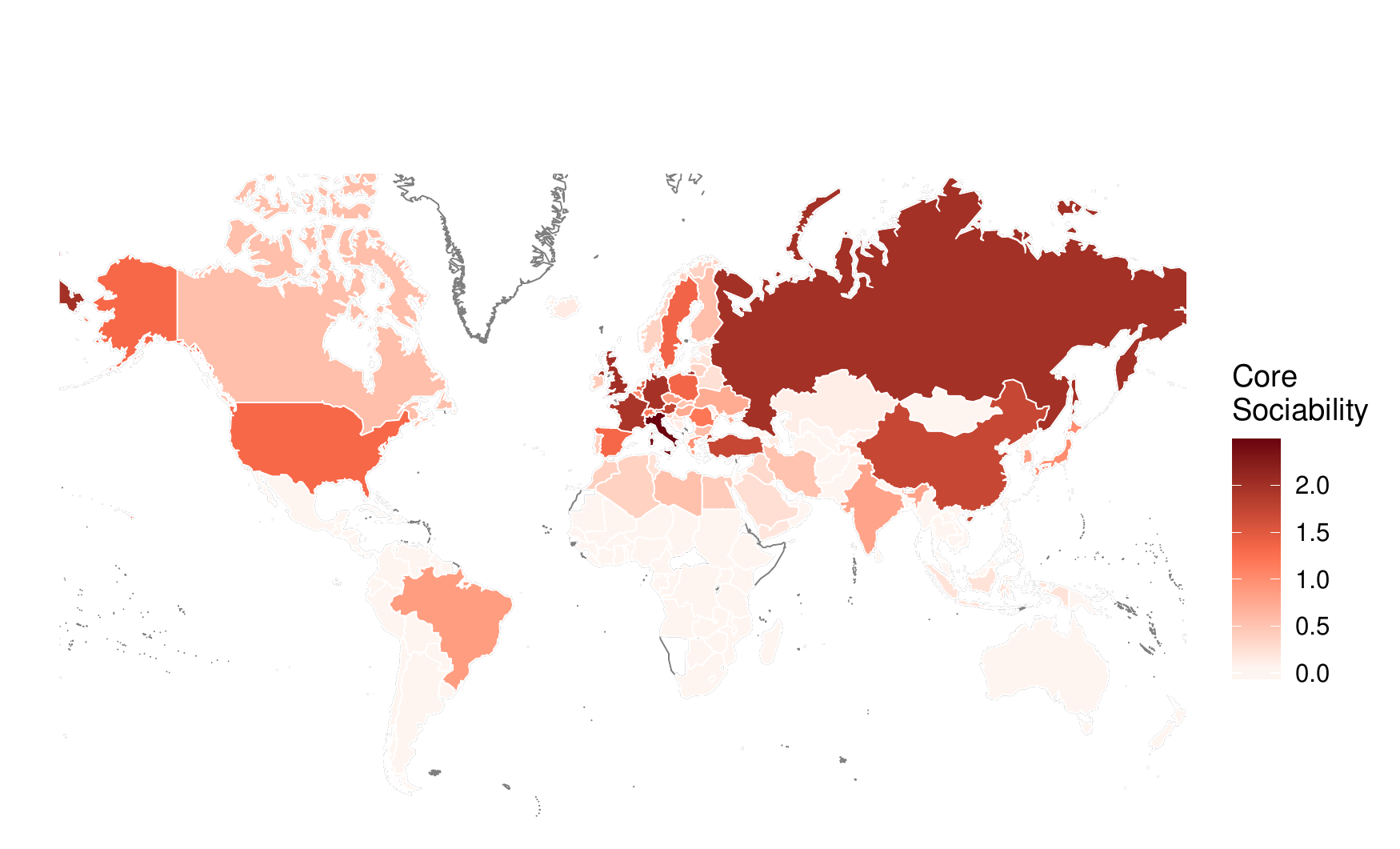}}
\caption{Core Periphery Structure of the World Trade network. $(a)$ gives the mean sociabilities for each country, $(b)$ gives the core sociabilities.}
\label{fig:trade_map}
\end{figure}

In Figure \ref{fig:trade_map} we see the world map, coloured by the value of mean sociability parameter in Figure~\ref{fig:trade_map}\subref{fig:trade_mean_map}, and by the value of the core sociability parameter in Figure \ref{fig:trade_map}\subref{fig:trade_core_map}. We see that the core consists largely of the large, developed countries that we expect, as well as some smaller European countries. This fits with other results that have been obtained in the literature \citep{della2013profiling}. Comparing the two plots, some of the more interesting results are countries that have a relatively high core sociability compared to their mean sociability. These include countries such as Russia, and indicate countries that trade predominantly with other countries in the core. Conversely, we see other countries such as the US and China which have relatively lower core sociabilities. These are countries which trade more internationally, with countries in the core and periphery.

\subsection{Comparisons}
Finally, we want to compare our method against other standard methods of core-periphery detection. In order to do this we generate simulated core-periphery networks, and run each algorithm on them. We can then calculate the Area Under the Curve (AUC) of the  Receiver Operating Characteristic (ROC) curve \citep{hanley1982meaning} in order to determine the accuracy of the classification of each algorithm. In this case, we simply compare the binary core-periphery classification of each method.  The methods that we compare against are: the algorithm of \citep{borgatti2000models}, the MINRES algorithm of \citep{boyd2010computing} and \citep{lip2011fast}, the Stochastic Block-Model (SBM)  of \citep{zhang2015identification}, the three different methods of \citep{cucuringu2016detection}, the aggregate core score method of \citep{rombach2017core} and the method of \citep{de2019detecting}. In each case, we implement the methods using the software found in the \verb|cpalgorithm| Python package\footnote{\url{https://core-periphery-detection-in-networks.readthedocs.io/en/latest/}}. While the method of \citep{rombach2017core} gives continuous coreness parameters, we convert these to binary classifications in the same way that they do in their simulation study.

As we have previously noted, our focus here is on model based approaches. The only other model based approach amongst these is that of \citep{zhang2015identification}. Our approach already provides several advantages over the rest of the models considered, such as the ability to model the degree distribution of the networks being studied. However, here we focus on the  classification accuracy and how it compares to some classic and more contemporary alternatives. Specifically, we are interested in comparing the accuracy in the setting for which our model is designed: the modelling of sparse graphs with a core-periphery structure and power-law degree distribution.

\subsubsection{Comparison on in-model simulated data}
The first comparison we do is using our model to generate power-law, core-periphery networks. Of course, we expect our model to perform very well here, and indeed we see this as we compare the AUC for each model. We vary the strength of the core-periphery structure by varying $b$ in our model, which varies the relative sociabilities of the nodes in the core and periphery. We run $20$ simulations for each value of $b$, and adjusting the value of $\alpha$ to keep the number of nodes roughly equal in each case. We then measure the average classification accuracy in each case. We can see from Table \ref{table:our_comparison} that as $b$ increases, the relative size of the core increases, as does the ease of classifying core and periphery. This second point we can see from the fact that the accuracy of the methods we compare against generally increases as $b$ is increased. However, we see that our method, which we call \textbf{Sparse CP}, achieves the highest accuracy in each case.
\begin{table}[ht]
    \centering
    \begin{tabular}{|c||ccccc|}
      \hline
     \multirow{2}{4em}{Method} & \multicolumn{5}{|c|}{$b$} \\
     \cline{2-6}
      & 0.2 & 0.5 & 1 & 2 & 5 \\
      \hline
      Average Number of Nodes & 796 & 786 & 785 & 790 & 778 \\
    Average Relative Core Size & 0.12 & 0.19 & 0.26 & 0.37 & 0.57  \\
    \hline
    \textbf{Sparse CP} & \textbf{0.97} &  \textbf{0.98} &  \textbf{0.98} &  \textbf{0.98} &  \textbf{0.98}  \\
    Borgatti-Everett & 0.57 & 0.60 & 0.62 & 0.63 & 0.63  \\
    MINRES & 0.72 & 0.78 & 0.78 & 0.82 & 0.52  \\
    SBM & 0.70 & 0.84 & 0.88 & 0.90 & 0.90  \\
    LowRankCore & 0.77 & 0.79 & 0.80 & 0.75 & 0.66  \\
    LapCore & 0.58 & 0.63 & 0.62 & 0.60 & 0.50  \\
    LapSgnCore & 0.56 & 0.50 & 0.46 & 0.44 & 0.43  \\
    Rombach & 0.69 & 0.84 & 0.90 & 0.94 &  0.95  \\
    Surprise & 0.76 & 0.82 & 0.86 & 0.88 & 0.87  \\
   \hline
    \end{tabular}
    \caption{AUC for different methods and different values of $b$}
    \label{table:our_comparison}
    \end{table}

\begin{figure}[h]
\centering
\includegraphics[width=0.5\textwidth]{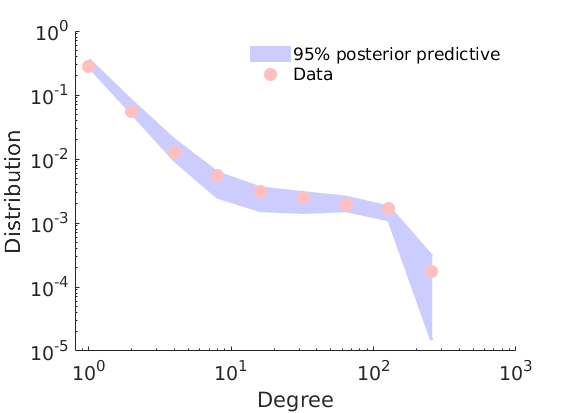}
\caption{Posterior predictive degree distribution for the graph generated with $K=2, \alpha = 200, \sigma= 0.2, \tau = 1, b=5, a=0.2$.}
\label{fig:b_5_post_pred}
\end{figure}

The case $b=5$ is an extreme case, with the core comprising almost $60\%$ of the network, and in this case some of the other methods perform badly, with some methods placing all the nodes in the core on some of the simulations. Recalling our definition of core-periphery networks, we see from Figure \ref{fig:b_5_post_pred} that the degree distribution is very far from a power-law, due to the large number of core nodes. Nevertheless, it is encouraging to see that in this case we are still able to accurately recover the degree distribution.

There are a few reasons why we might expect to see this difference in classification performance. The only other model-based approach, based on the SBM~\citep{zhang2015identification}, cannot account for power-law degree distributions and tend to classify nodes according to their degree. Other non-model based approaches have similar issues.

In an attempt to allow a fairer comparison, we also test our method against the alternatives mentioned above on another simulated network, again with a power-law degree distribution, but in this case not generated using our model.

\subsubsection{Comparison on out-of-model simulated data}
\begin{figure}[h]
\centering
\includegraphics[width=0.6\textwidth]{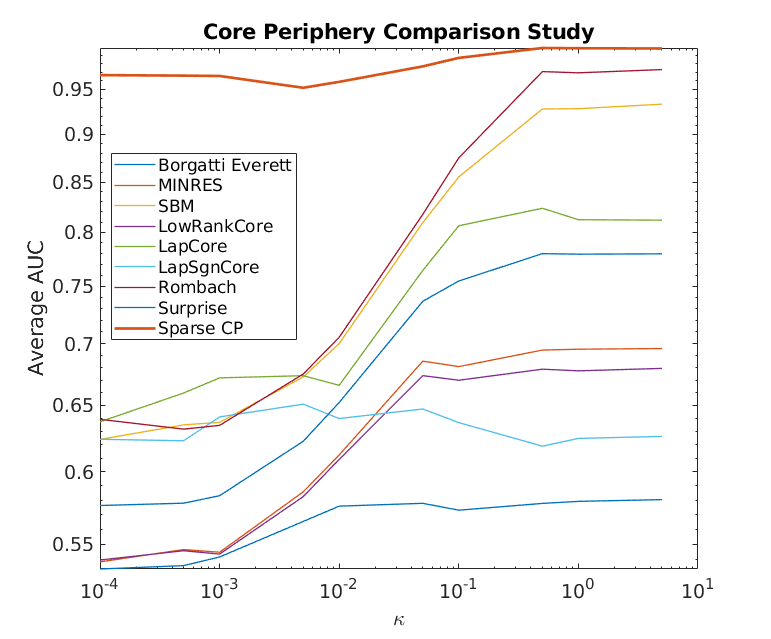}
\caption{AUC for different methods and different values of $\kappa$}
\label{fig:config_comparison2}
\end{figure}

The second comparison we do is using the simulated core-periphery networks of \citep{holme2005core}. Our algorithm is based on theirs, but differs slightly to allow us to tune how well the core and periphery can be separated by degree. We construct networks as follows:

\begin{enumerate}
    \item Generate degrees $m_1 < m_2< \ldots < m_N$ from a power-law distribution. These are the desired degrees and can be thought of as stubs, as in the configuration model \citep{newman2010networks}.
    \item Place node $i$ in the core with probability $q_i$, where $q_i$ is given by
    \begin{align*}
        q_i = \frac{1}{1+\exp \left(-2\kappa(m_i-m_{min})\right)}
    \end{align*}
    where $\kappa$ and $m_{min}$ are parameters that we use to tune the model. Call the set of core nodes $\mathcal{C_H}$ and the set of periphery nodes $\mathcal{P_H}$.
    \item Go through each of the nodes $i \in \mathcal{C_H}$ in increasing order of degree (node $i$ has degree $m_i$) and for each node $i$ attach its stubs to those of nodes $j$ with $m_j\geq m_i$ as long as the degree of $j$ is less than $m_j$.
    \item Attach the remaining stubs randomly and make them into edges if they do not form loops or multiple edges.
\end{enumerate}
The form of $q_i$ is a standard approximation to a Heaviside step function. When $\kappa$ is large, the function approximates a step function with a jump at $m_{min}$. In this case, the core is comprised of only the nodes with degrees $m_i\geq m_{min}$. However, for smaller $\kappa$  we have low degree nodes entering the core and high degree nodes entering the periphery. In Figure \ref{fig:config_comparison2} we compare our method to the alternatives for varying $\kappa$ As in our previous simulation study, we calculate the area under the ROC curve, averaged over 20 realisations of networks for each value of $\kappa$.

We see that when $\kappa$ is large, and the core and periphery are essentially divided by degree, our method performs very well, but so do those of \citep{zhang2015identification} and \citep{rombach2017core}. As we decrease $\kappa$, the other methods fail to perform as well, while ours retains a high accuracy. Therefore, we see that our method also outperforms the alternatives when applied to simulated networks that are not generated using our model. Specifically, we do better especially when the core and periphery nodes are not split simply by degree.

We must also note here that some of the methods we compare against here are very fast, taking only seconds to produce a classification into core and periphery. However, our sampler generally takes $\sim5$ minutes to run on a standard desktop computer. Furthermore, as discussed earlier in our case the aim is not just to perform classification, but to estimate the parameters of a generative model.

\section{Conclusions}
In this work, we provide a precise definition of a sparse network with a core-periphery structure, based on the sparsity properties of the subgraphs of core and periphery nodes. Building on earlier work from \citep{Caron2017, todeschini2016exchangeable}, we then present a class of sparse graphs with such properties. We obtain theoretical results on the sparsity properties of our model. Specifically, we see that our model generates a core region which is dense, and a periphery region which is sparse. Theoretical results on the relative size of the core are also obtained.

We provide methods to simulate from this class of graphs, and to perform posterior inference with this class of models. We demonstrate that our approach can detect interpretable core-periphery structure in two real-world airport and trade networks, while providing a good fit to global structural properties of the networks. When restricting ourselves to simply looking at core-periphery classification accuracy, we see that it compares favourably against various alternatives when tested in the power-law setting.  

A property of our model is that the relative size of the core tends to zero as the size of the graph increases. In some applications, we may instead want to have a network which is overall dense, but with a sparse periphery region, where the relative size of the core is bounded about zero. Furthermore, whilst our model can currently accommodate the existence of multiple communities as well as a core-periphery structure, it currently cannot be used to model networks with multiple core-periphery pairs. Work has been done in the literature on the detection of multiple such pairs \citep{kojaku2017finding} and it could be valuable to extend our model to this setting.

\bibliography{sample}
\newpage
\appendix

\section{Proofs}
\subsection{Asymptotic notation}
We first describe some further asymptotic notation used in the proof. As before we follow the notation of \citep{janson2011probability}, where if $X=(X_\alpha)_{\alpha\geq 0}$ and $Y=(Y_\alpha)_{\alpha\geq 0}$ are two stochastic processes defined on the same probability space with $X_\alpha,Y_\alpha\rightarrow\infty$ a.s. as $\alpha \to \infty$, we have
\begin{align*}
    X_{\alpha} = O(Y_{\alpha}) \text{ a.s.} &\iff \limsup_{\alpha \to \infty}\frac{X_{\alpha}}{Y_{\alpha}} < \infty \text{ a.s.}\\
    X_{\alpha} = o(Y_{\alpha}) \text{ a.s.} &\iff \limsup_{\alpha \to \infty}\frac{X_{\alpha}}{Y_{\alpha}} =0 \text{ a.s.}\\
    X_{\alpha} = \Omega(Y_{\alpha}) \text{ a.s.} &\iff  Y_{\alpha} = O(X_{\alpha}) \text{ a.s.}\\
    X_{\alpha} = \omega(Y_{\alpha}) \text{ a.s.} &\iff Y_{\alpha} = o(X_{\alpha}) \text{ a.s.}\\
    X_{\alpha} = \Theta(Y_{\alpha}) \text{ a.s.} &\iff  X_{\alpha} \asymp Y_{\alpha} \text{ a.s.} \iff X_{\alpha} = O(Y_{\alpha}) \text{ and } Y_{\alpha} = O(X_{\alpha}) \text{ a.s.}
\end{align*}

\subsection{Proof of Theorems \ref{thm:cp_asymptotics1}, \ref{thm:cp_asymptotics2} and corollary \ref{thm:core_size}} \label{proof:cp_asumptotics}

Let
\begin{align*}
\rho_c(dw_1,dw_2)=\rho(dw_1,dw_2)1_{w_1>0}\\
\rho_p(dw_1,dw_2)=\rho(dw_1,dw_2)1_{w_1=0}
\end{align*}

A straightforward adaptation of Theorem 1 in \cite{Caron2017} and Proposition 1 in \cite{naulet2017estimator} yields
\begin{align*}
N_{\alpha}^{(e)}&\sim \frac{\alpha^2}{2}\int \left (1-e^{-2w_1 w'_1-2w_2 w'_2}\right )\rho(dw_1,dw_2)\rho(dw'_1,dw'_2)\\
N_{\alpha,c-c}^{(e)}&\sim \frac{\alpha^2}{2}\int \left (1-e^{-2w_1 w'_1-2w_2 w'_2}\right )\rho_c(dw_1,dw_2)\rho_c(dw'_1,dw'_2)\\
N_{\alpha,p-p}^{(e)}&\sim \frac{\alpha^2}{2}\int \left (1-e^{-2w_1 w'_1-2w_2 w'_2}\right )\rho_p(dw_1,dw_2)\rho_p(dw'_1,dw'_2)\\
N_{\alpha,c-p}^{(e)}&\sim \alpha^2\int \left (1-e^{-2w_1 w'_1-2w_2 w'_2}\right )\rho_c(dw_1,dw_2)\rho_p(dw'_1,dw'_2)
\end{align*}
almost surely as $\alpha$ tends to infinity. It follows that
$$
N_{\alpha}^{(e)}\asymp N_{\alpha,c-c}^{(e)} \asymp N_{\alpha,p-p}^{(e)}\asymp N_{\alpha,c-p}^{(e)}\asymp \alpha^2\text{ almost surely as }\alpha\rightarrow\infty.
$$

Let $Q_{\alpha,c}=\sum_{i} 1_{w_{i1}>0} 1_{\theta_i\leq \alpha}$. Note that $Q_{\alpha,c}$ is a homogeneous Poisson process on $\mathbb R_+$ with intensity $\int_{(0,\infty)^2} \rho(dw_1,dw_2)<\infty$ (by assumption \eqref{eq:condcore}), hence the law of large numbers implies $Q_{\alpha,c}\sim \alpha \int_{(0,\infty)^2} \rho(dw_1,dw_2)$. As
$$
\sqrt{N_\alpha^{(e)}}\leq N_{\alpha,c}\leq Q_{\alpha,c}
$$
it follows that, almost surely as $\alpha\rightarrow\infty$,
$$
N_{\alpha,c}\asymp \alpha.
$$
Similarly, if $\int_{\{0\}\times(0,\infty)} \rho(dw_1,dw_2)<\infty$, then $N_{\alpha,p}\asymp \alpha$ and therefore $N_\alpha\asymp \alpha$.
Otherwise, define
$$
\widetilde N_{\alpha,p}=\sum_{i} 1_{w_{i1}=0}1_{\theta_i\leq\alpha} 1_{\sum_j 1_{w_{i1}=0}z_{ij}1_{\theta_j\leq \alpha}>0  }
$$
the number of periphery nodes with at least one connection to a periphery node. Note that
$$
N_{\alpha,p}\geq \widetilde N_{\alpha,p}.
$$

Two nodes $i$ and $j$ in the periphery connect with probability
\begin{align}\begin{cases}
1-e^{-2 w_{i2}w_{j2}} &i\neq j\\
1-e^{-w_{i2}^2} &i= j
\end{cases}
\end{align}
where $(w_{i2})_{i\geq 1,w_{i1}=0}$ are the points of a Poisson point process with mean $\rho_p(dw_2)\allowbreak =\int_{w_1\in [0,\infty)}\rho_p(dw_1,dw_2)$. This is therefore the same model as the model of \citep{Caron2017}. We therefore have, using Theorem 4 in  \cite{Caron2017}
\begin{align}
\widetilde N_{\alpha,p}=\omega(\alpha)\text{ if }\int\rho_p(dw_2)=\infty
\end{align}
and therefore $N_{\alpha,p}=\omega(\alpha)$ and $N_{\alpha}=\omega(\alpha)$. Noting that 
\begin{align*}
    \int\rho_p(dw_2)= \int_{\{0\}\times(0,\infty)}\rho(dw_1,dw_2)
\end{align*}
finishes the proof of Theorem \eqref{thm:cp_asymptotics1}.

We now consider the particular case of the L\'evy measure \eqref{eq:levymeasurerho}. In the next section, we give the more general proof of Theorem \ref{thm:cp_asymptotics_community} for generic $K$. Taking $K=2$ in that case gives us the proof of Theorem \ref{thm:cp_asymptotics2}. Here, we give a more direct proof of the asymptotics for the periphery nodes and overall graph. 

Using a slight abuse of notation, let $\rho_p(w_2)$ and $\rho_0(w_0)$ denote the intensity functions of the measures $\rho_p(dw_2)$ and $\rho_0(dw_0)$. Let $f$ denote the pdf of a gamma random variable with parameters $a$ and $b$. We have
\begin{align}
\rho_p(w_2)&=\int_0^\infty e^{-w_0}w_0^{-1}f(dw_2/w_0)\rho_0(w_0)dw_0.
\end{align}
As $e^{-w_0}\rho_0(w_0)\sim w_0^{-1-\sigma}\Gamma(1-\sigma)^{-1}$ as $w_0$ tends to 0 when $\sigma\in(0,1)$ we have, using \citep[Corollary 6]{Ayed2019a} (see also \citep[Theorem 4.1.6 page 201]{Bingham1989}),
$$
\rho_p(w_2)\sim w_2^{-1-\sigma}\Gamma(1-\sigma)^{-1}\mathbb E[\beta_2^\sigma]
$$
where $\beta_2$ has cdf $F$. Finally, using \citep[Proposition 1.5.8. p. 26]{Bingham1989},  we obtain
$$
\int_x^\infty \rho_p(w_2)dw_2\sim x^{-\sigma}\sigma^{-1}\Gamma(1-\sigma)^{-1}\mathbb E[\beta_2^\sigma]
$$
and it follows from \citep[Theorem 4]{Caron2017} that
$$
\widetilde N_{\alpha,p}=\Omega(\alpha^{1+\sigma})
$$
Since $N_{\alpha} \geq N_{\alpha,p}\geq \widetilde N_{\alpha,p}$ we also have that
$$
N_{\alpha,p}=\Omega(\alpha^{1+\sigma}), \qquad N_{\alpha}=\Omega(\alpha^{1+\sigma})
$$
which gives the desired results for the periphery nodes and the overall graph. The corresponding results for the core nodes, which tell us that $N_{\alpha,c} \asymp \alpha$, can be found by taking $K=2$ in the proof of Theorem \ref{thm:cp_asymptotics_community}. This then completes the proof of Theorem \ref{thm:cp_asymptotics2}.

The results of Corollary \ref{thm:core_size}, both follow directly from $N_{\alpha,c} \asymp \alpha$
and the fact that $N_{\alpha}=\Omega\left(\alpha^{1+\sigma}\right)$ for the overall graph.

\subsection{Proof of Theorem \ref{thm:cp_asymptotics_community}}

In this case, we let
\begin{align*}
\rho_c(dw_1,dw_2,\ldots,dw_K)=\rho(dw_1,dw_2,\ldots,dw_K)1_{w_1>0}\\
\rho_p(dw_1,dw_2,\ldots,dw_K)=\rho(dw_1,dw_2,\ldots,dw_K)1_{w_1=0}
\end{align*}

As before, a straightforward adaptation of Theorem 1 in \cite{Caron2017} and Proposition 1 in \cite{naulet2017estimator} yields
$$
N_{\alpha}^{(e)}\asymp N_{\alpha,c-c}^{(e)} \asymp N_{\alpha,p-p}^{(e)}\asymp N_{\alpha,c-p}^{(e)}\asymp \alpha^2\text{ almost surely as }\alpha\rightarrow\infty.
$$

If we define $Q_{\alpha,c}=\sum_{i} 1_{w_{i1}>0} 1_{\theta_i\leq \alpha}$ then $Q_{\alpha,c}$ is a homogeneous Poisson process on $\mathbb R_+$ with intensity $\int_{(0,\infty)^K} \rho(dw_1,dw_2,\ldots,dw_K)<\infty$ (by assumption \eqref{eq:condcore_community}). By the the law of large numbers, $Q_{\alpha,c}\sim \alpha \int_{(0,\infty)^K} \rho(dw_1,dw_2,\ldots,dw_K)$. As
$$
\sqrt{N_\alpha^{(e)}}\leq N_{\alpha,c}\leq Q_{\alpha,c}
$$
it follows that, almost surely as $\alpha\rightarrow\infty$,
$$
N_{\alpha,c}\asymp \alpha.
$$
Similarly, if $\int_{\{0\}\times(0,\infty)^K} \rho(dw_1,dw_2,\ldots,dw_K)<\infty$, then $N_{\alpha,p}\asymp \alpha$ and therefore $N_\alpha\asymp \alpha$.
Otherwise, define
$$
\widetilde N_{\alpha,p}=\sum_{i} 1_{w_{i1}=0}1_{\theta_i\leq\alpha} 1_{\sum_j 1_{w_{i1}=0}z_{ij}1_{\theta_j\leq \alpha}>0  }
$$
the number of periphery nodes with at least one connection to a periphery node. Note that
$$
N_{\alpha,p}\geq \widetilde N_{\alpha,p}.
$$

Two nodes $i$ and $j$ in the periphery connect with probability
\begin{align}\begin{cases}
1-e^{-2 \sum_{k=2}^K w_{ik}w_{jk}} &i\neq j\\
1-e^{-\sum_{k=2}^K w_{ik}^2} &i= j
\end{cases}
\end{align}
where $(w_{i2},\ldots,w_{iK})_{i\geq 1,w_{i1}=0}$ are the points of a Poisson point process with mean $\rho_p(dw_2,\ldots,dw_{K})=\int_{w_1\in [0,\infty)}\rho_p(dw_1,dw_2,\ldots,dw_{K})$. This is therefore the same model as the model of \citep{todeschini2016exchangeable}. We therefore have, using Proposition 4 in  \cite{todeschini2016exchangeable}
\begin{align}
\widetilde N_{\alpha,p}=\omega(\alpha)\text{ if }\int\rho_p(dw_2,\ldots,dw_{K})=\infty
\end{align}
and therefore $N_{\alpha,p}=\omega(\alpha)$ and $N_{\alpha}=\omega(\alpha)$. Noting that $\int\rho_p(dw_2,\ldots,dw_{K})=\int_{\{0\}\times(0,\infty)^{K-1}}\rho(dw_1,dw_2,\ldots,dw_{K})$ finishes the equivalent result of Theorem \eqref{thm:cp_asymptotics1} in Theorem \eqref{thm:cp_asymptotics_community}.

We now consider the particular case of the L\'evy measure \eqref{eq:levymeasurerho_communities}. We have
\begin{align}
\rho_p(dw_2,\ldots,dw_{K})&=\int_0^\infty e^{-w_0}w_0^{-(K-1)}F\left (\frac{dw_2}{w_0},\ldots,\frac{dw_K}{w_0}\right )\rho_0(dw_0).
\end{align}
where $F$ is a product of independent gamma distributions, and $\rho_0$ is the jump part of a GGP with parameters $\sigma$ and $\tau$, as before. We recognise this as the particular compound CRM model of \citep{todeschini2016exchangeable}, except that we now have a base measure
\begin{align*}
    \rho_{0,p}(dw_0)=e^{-w_0}\rho_0(dw_0)
\end{align*}
Then $\int\rho_0(dw_0)=\infty \implies \int\rho_{0,p}(dw_0)=\infty$ for our particular choice of $\rho_0$. Hence, the results of Proposition 5 of \cite{todeschini2016exchangeable} tell us that
$$
\widetilde N_{\alpha,p}=\Omega(\alpha^{1+\sigma})
$$
Since $N_{\alpha} \geq N_{\alpha,p}\geq \widetilde N_{\alpha,p}$ we also have that
$$
N_{\alpha,p}=\Omega(\alpha^{1+\sigma}), \qquad  N_{\alpha}=\Omega(\alpha^{1+\sigma})
$$
Returning to the core nodes, we know that $N_{\alpha,c}\asymp \alpha$ if
$$
\int_{(0,\infty)^K} \rho(dw_1,dw_2,\ldots,dw_K)<\infty
$$
But $\int_{(0,\infty)^K} \rho(dw_1,dw_2,\ldots,dw_K) = \int\rho_c(dw_2,\ldots,dw_{K})$, where 
\begin{align*}
    \rho_c(dw_2,\ldots,dw_{K})=\int_{w_1\in [0,\infty)}\rho_c(dw_1,dw_2,\ldots,dw_{K}).
\end{align*}
For the particular case of the L\'evy measure \eqref{eq:levymeasurerho_communities}. We have
\begin{align}
\rho_c(dw_2,\ldots,dw_{K})&=\int_0^\infty \left(1-e^{-w_0}\right)w_0^{-(K-1)}F\left (\frac{dw_2}{w_0},\ldots,\frac{dw_K}{w_0}\right )\rho_0(dw_0).
\end{align}
As before, this is the same as the particular compound CRM model of \citep{todeschini2016exchangeable}, except that we now have a base measure
\begin{align*}
    \rho_{0,c}(dw_0)=\left(1-e^{-w_0}\right)\rho_0(dw_0)
\end{align*}
In this case $\int\rho_{0,c}(dw_0)$ is the Laplace exponent $\psi(1)$ of the base L\'evy measure $\rho_0$, which is finite. Hence, the results of Proposition 5 of \cite{todeschini2016exchangeable} tell us that
$\int_{(0,\infty)^K} \rho(dw_1,dw_2,\ldots,dw_K)<\infty$ as desired, and thus $N_{\alpha,c}\asymp \alpha$. This completes the proof of Theorem \ref{thm:cp_asymptotics_community}, and taking $K=2$ also gives us the proof of Theorem \ref{thm:cp_asymptotics2}.

As before, the equivalent results of Corollary \ref{thm:core_size}, follow directly from $N_{\alpha,c} \asymp \alpha$
and the fact that $N_{\alpha}=\Omega\left(\alpha^{1+\sigma}\right)$ for the overall graph.

\section{More Simulation Results}\label{app:simulations}

We show here empirical results on the effects that changing various parameters of the model have on the degree distributions, sparsity properties and core proportion. We restrict ourselves to the $K=2$ case here, for ease of visualisation.

\subsection{Degree distributions}
\begin{figure}
\centering
\subfloat[All nodes \label{fig:sigmadegree}]{\includegraphics[width=0.33\textwidth]{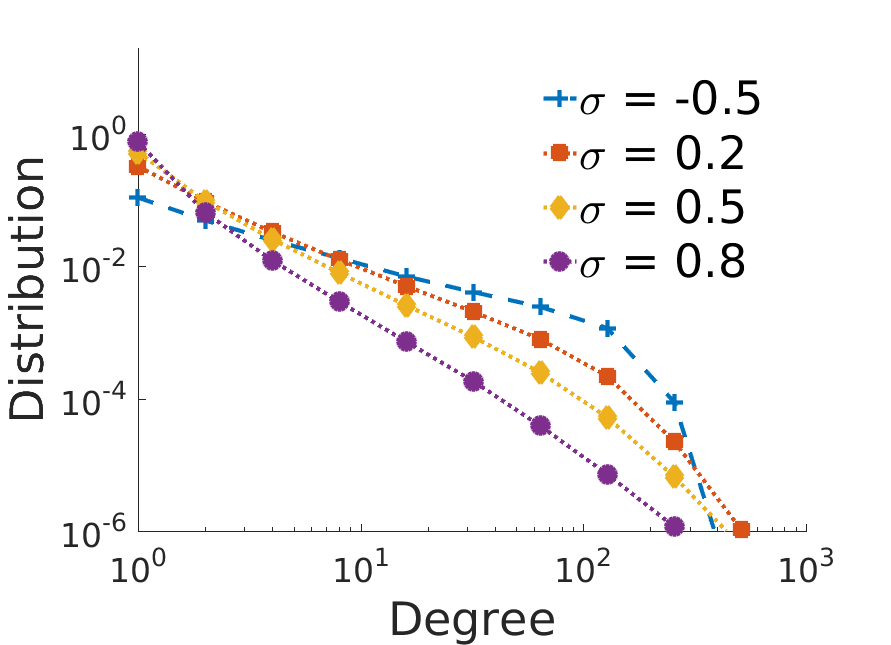}}
\subfloat[Core nodes \label{fig:sigmadegreecore}]{\includegraphics[width=0.33\textwidth]{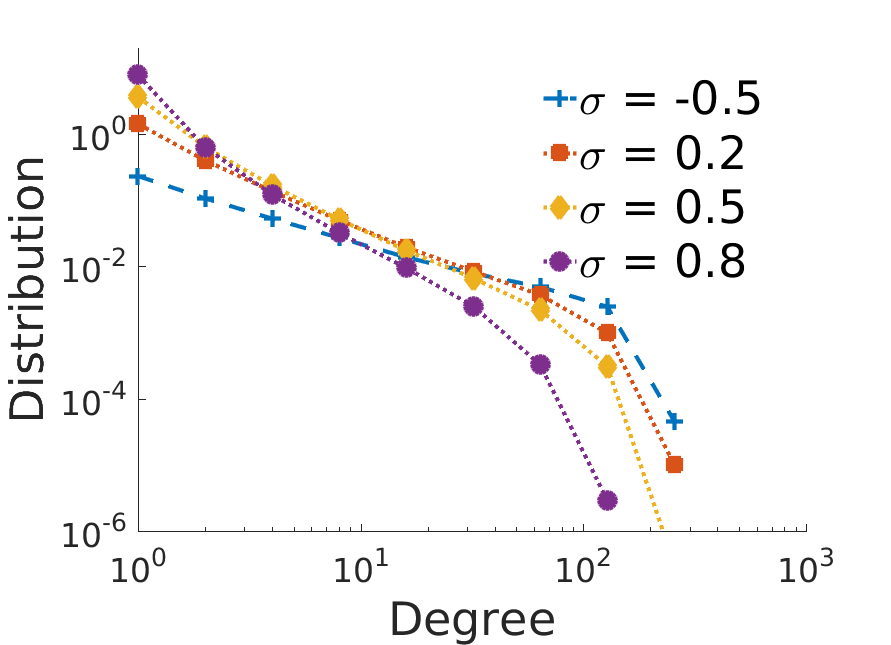}}
\subfloat[Periphery nodes\label{fig:sigmadegreeperiph}]{\includegraphics[width=0.33\textwidth]{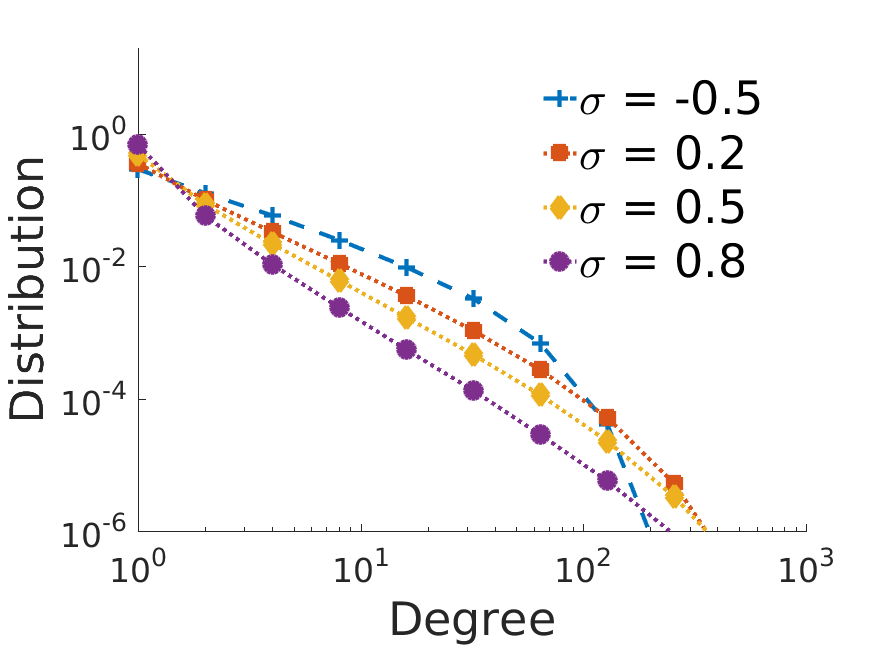}}\\
\subfloat[All nodes\label{fig:tau_degree}]{\includegraphics[width=0.33\textwidth]{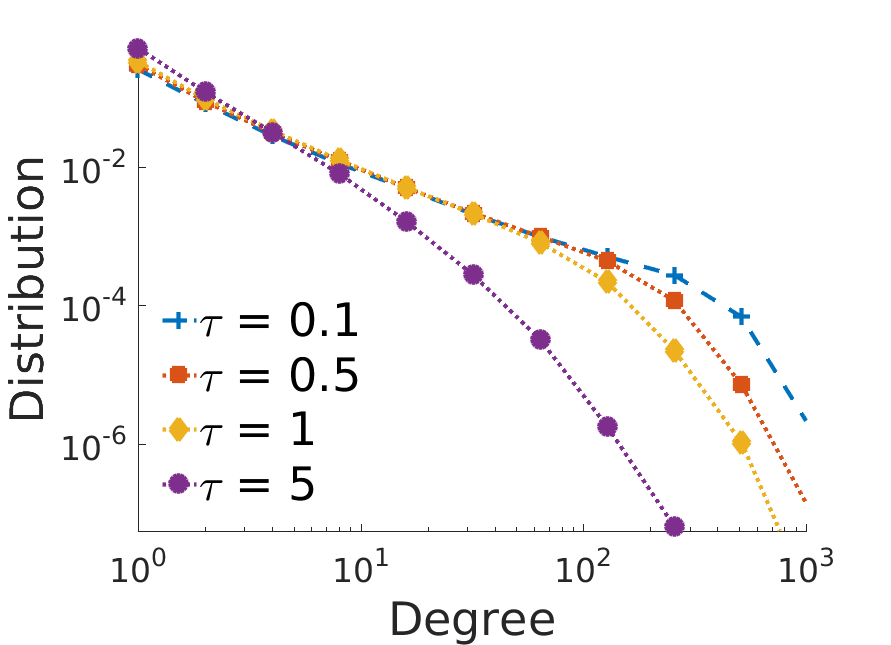}}
\subfloat[Core nodes\label{fig:tau_degreecore}]{\includegraphics[width=0.33\textwidth]{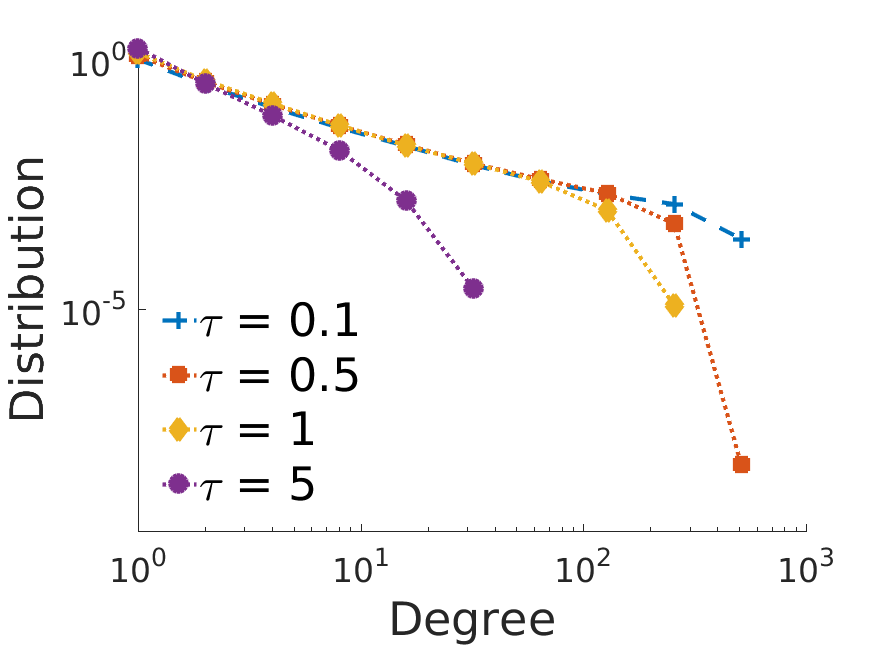}}
\subfloat[Periphery nodes\label{fig:tau_degreeperiph}]{\includegraphics[width=0.33\textwidth]{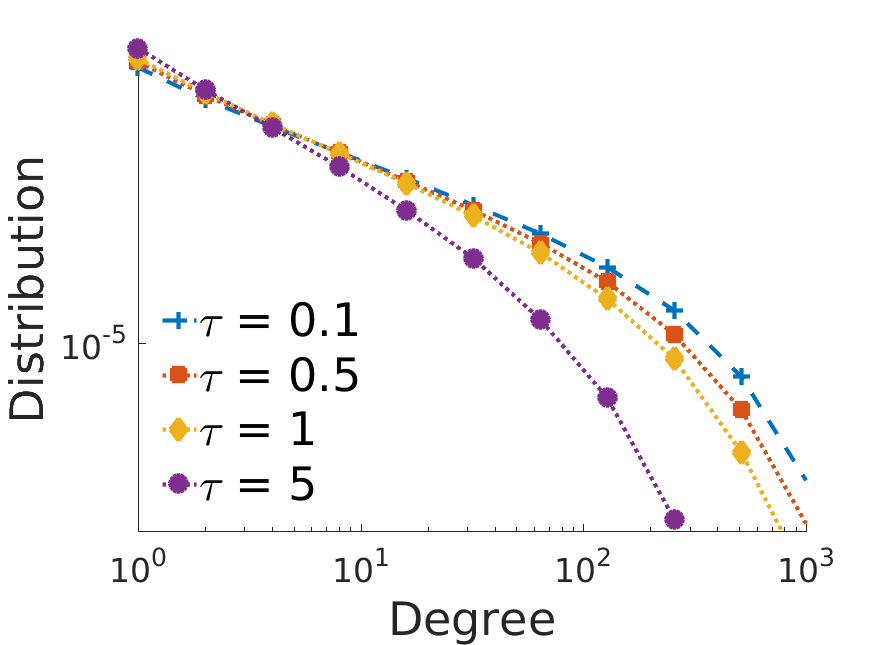}}
\caption{Degree distributions of core and periphery nodes for varying values of $\sigma$ and $\tau$}
\label{base_degree_graphs}
\end{figure}

We first look at the degree distributions for the overall graph, and for the core and periphery nodes separately. The first thing we notice is that, as we expect, the core and periphery nodes have very different degree distributions. In Figure \ref{base_degree_graphs} we see the results for varying $\sigma$ and $\tau$. Here we see that increasing $\sigma$ leads to lower degree nodes in the overall graph, as well as both core and periphery regions. We expect this, since we know that a larger value of $\sigma$ means that the graph is more sparse. The distribution begins to look closer to a pure power-law for large $\sigma$. We see further that increasing $\tau$ has little affect on the degree distribution in the core (for $\tau=5$ the core size is very small, leading to less interpretable results). We also see fewer nodes with a high degree, and again behaviour more closely resembling a power-law.

\begin{figure}
\centering
\subfloat[All nodes \label{fig:a_degree}]{\includegraphics[width=0.33\textwidth]{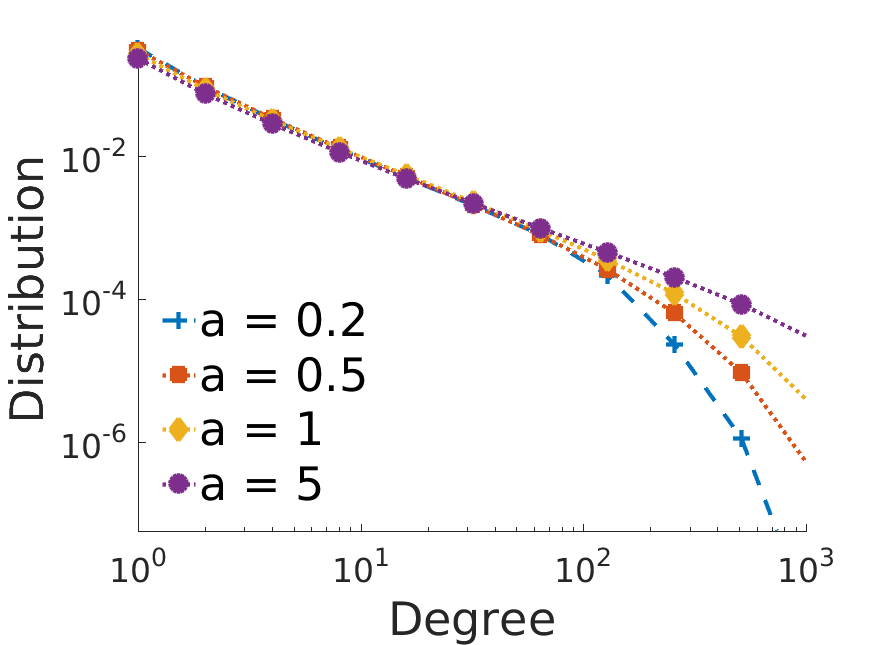}}
\subfloat[Core nodes \label{fig:a_degreecore}]{\includegraphics[width=0.33\textwidth]{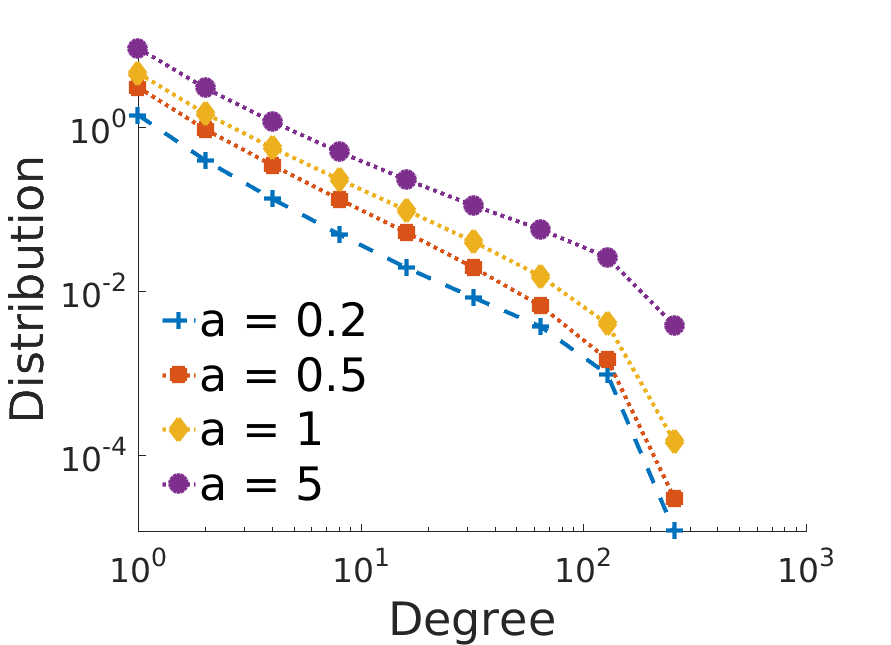}}
\subfloat[Periphery nodes\label{fig:a_degreeperiph}]{\includegraphics[width=0.33\textwidth]{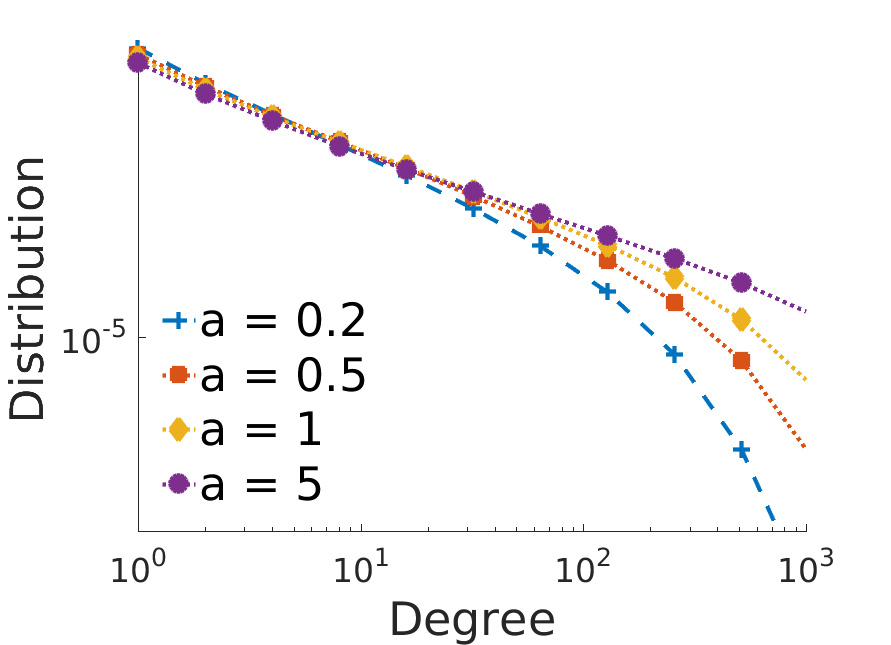}}\\
\subfloat[All nodes\label{fig:b_degree}]{\includegraphics[width=0.33\textwidth]{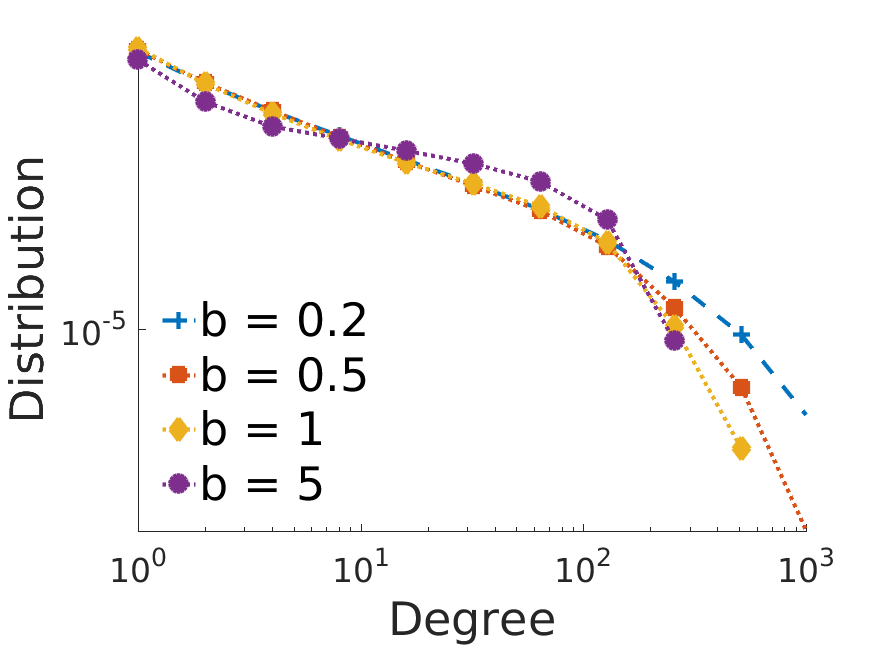}}
\subfloat[Core nodes\label{fig:b_degreecore}]{\includegraphics[width=0.33\textwidth]{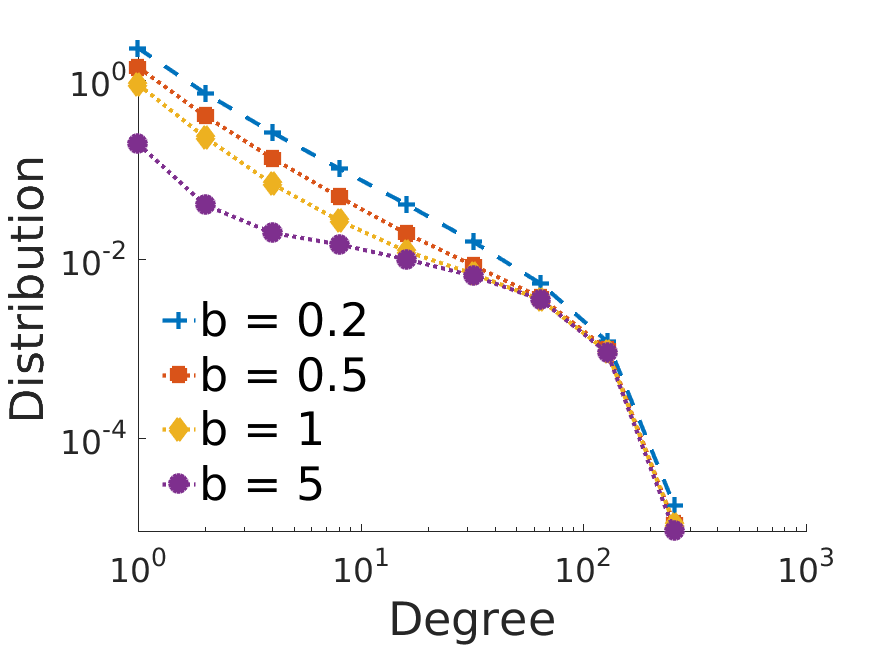}}
\subfloat[Periphery nodes\label{fig:b_degreeperiph}]{\includegraphics[width=0.33\textwidth]{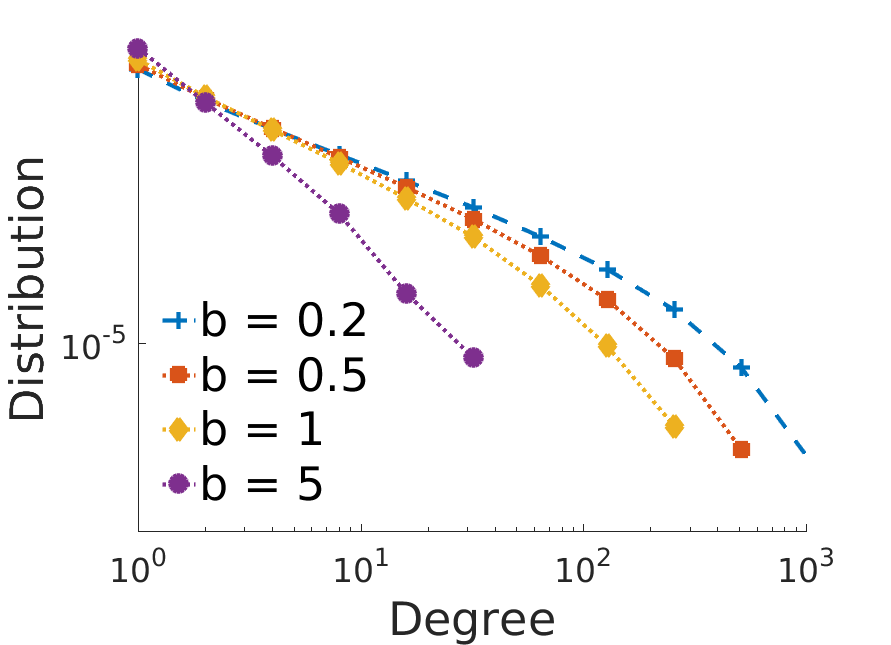}}
\caption{Degree distributions of core and periphery nodes for varying values of $a$ and $b$}
\label{gamma_degree_graphs}
\end{figure}

Figure \ref{gamma_degree_graphs} shows the results for varying $a$ and $b$. When increasing $a$ we see that the shape of the degree distribution for the core does not change, although the number of nodes in the core increases. Overall and in the periphery we little difference in the degree distribution apart from nodes with high degree. For larger values of $a$ there are more of these. The degree distributions for $b$ are similar, except that increasing $b$ decreases the number of nodes with a high degree. 

\subsection{Sparsity}
\begin{figure}
\centering
\subfloat[Overall graph \label{fig:a_edgesvsnodes}]{\includegraphics[width=0.35\textwidth]{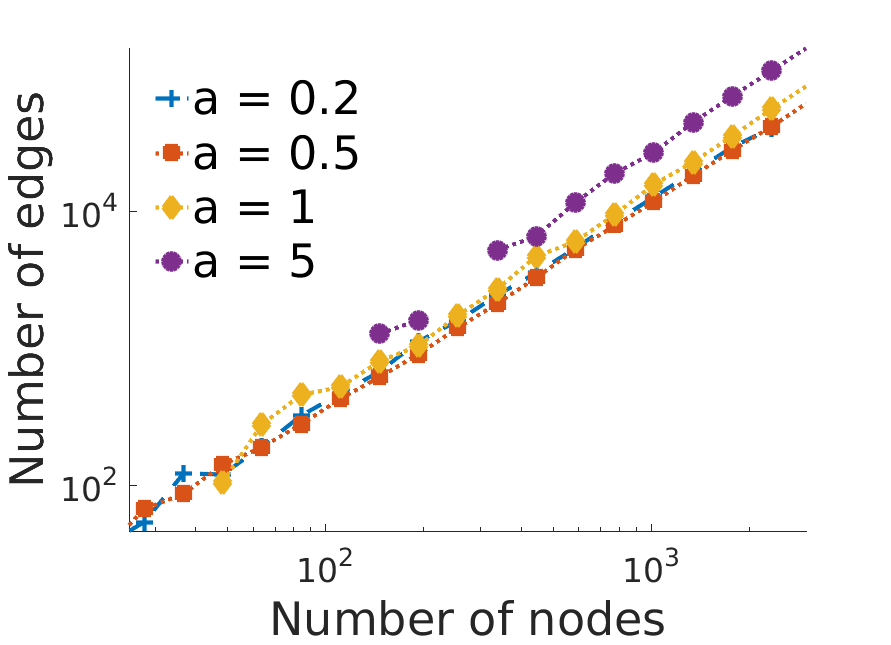}}
\subfloat[Core region \label{fig:a_ccedgesvsnodes}]{\includegraphics[width=0.35\textwidth]{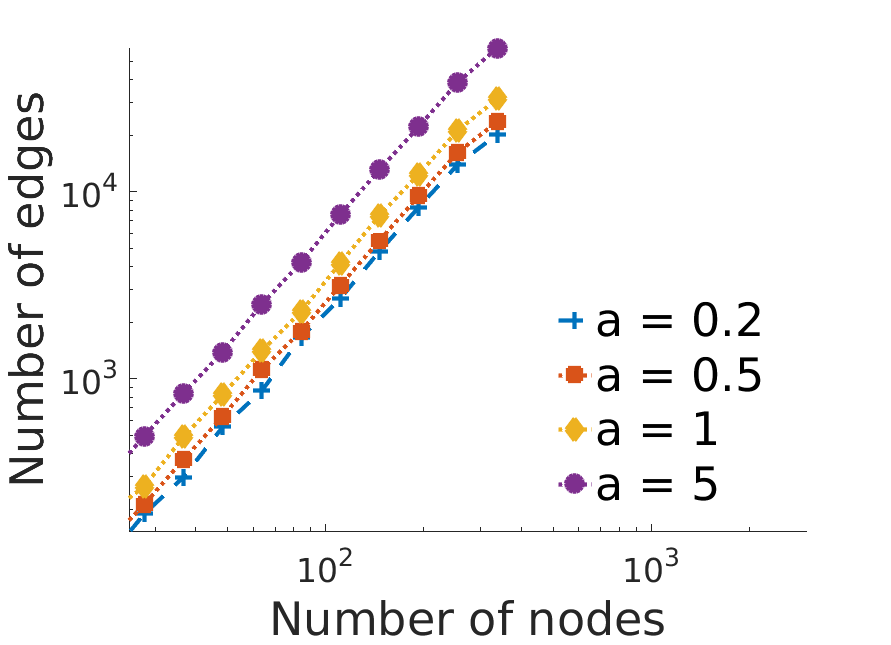}}\\
\subfloat[Core-periphery region\label{fig:a_cpedgesvsnodes}]{\includegraphics[width=0.35\textwidth]{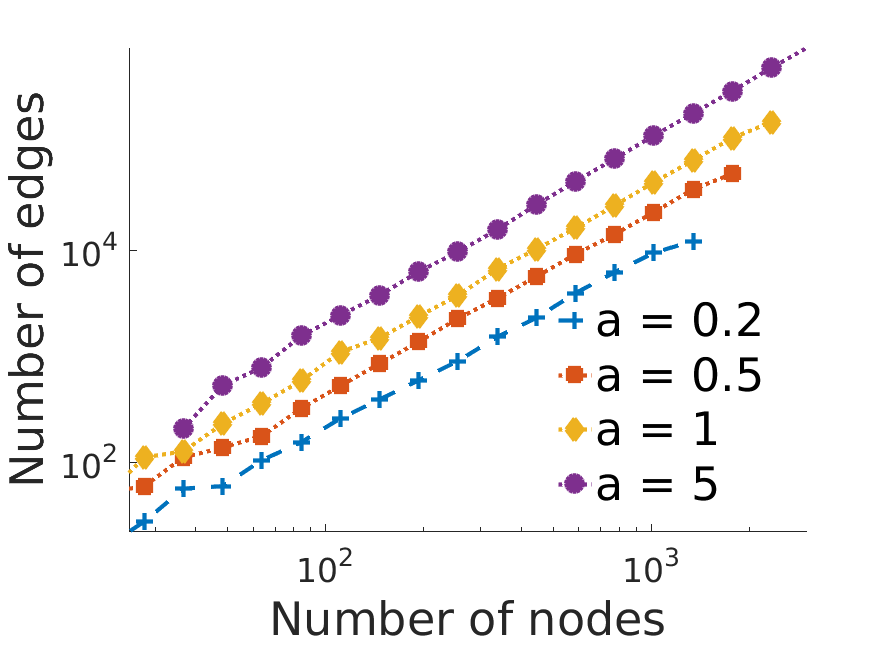}}
\subfloat[Periphery region\label{fig:a_ppedgesvsnodes}]{\includegraphics[width=0.35\textwidth]{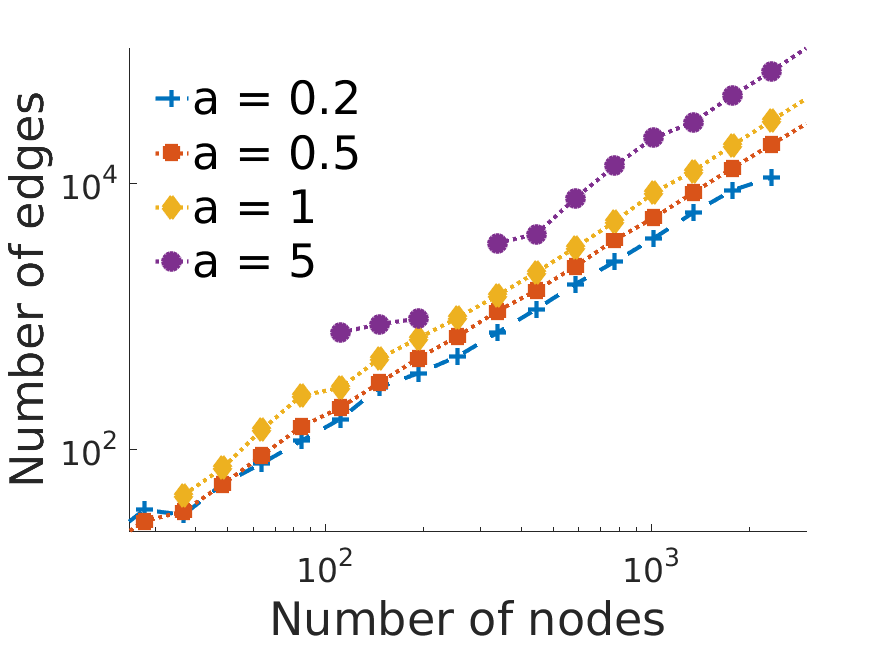}}
\caption{Relationship between nodes and edges for varying $a$}
\label{a_asymptotic_graphs}
\end{figure}

\begin{figure}
\centering
\subfloat[Overall graph \label{fig:b_edgesvsnodes}]{\includegraphics[width=0.35\textwidth]{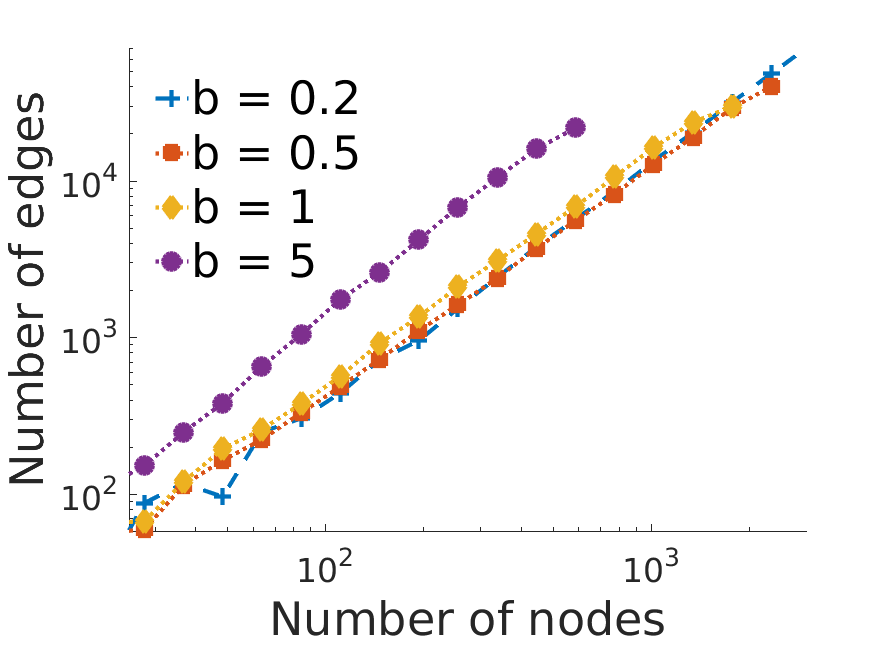}}
\subfloat[Core region \label{fig:b_ccedgesvsnodes}]{\includegraphics[width=0.35\textwidth]{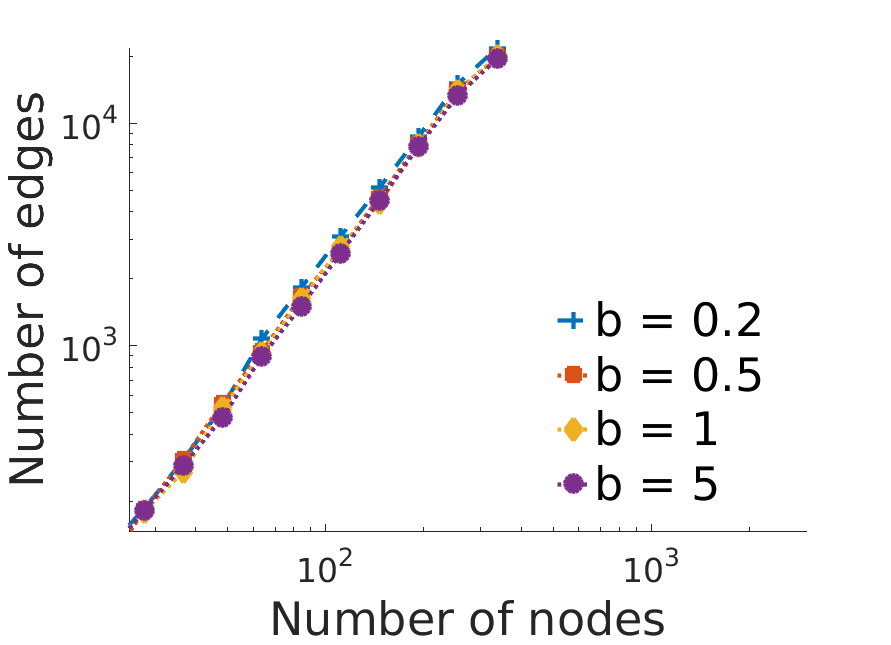}}\\
\subfloat[Core-periphery region\label{fig:b_cpedgesvsnodes}]{\includegraphics[width=0.35\textwidth]{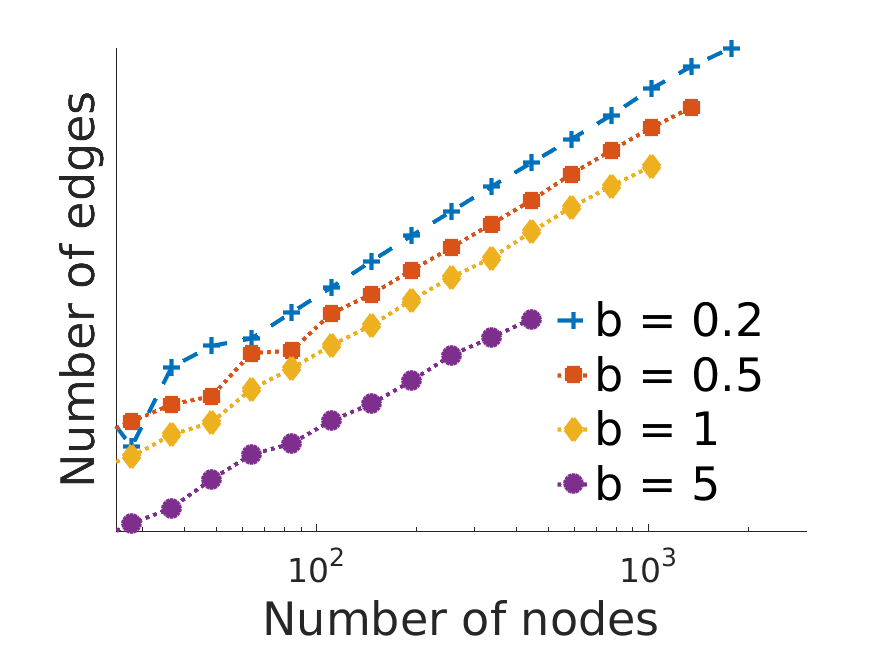}}
\subfloat[Periphery region\label{fig:b_ppedgesvsnodes}]{\includegraphics[width=0.35\textwidth]{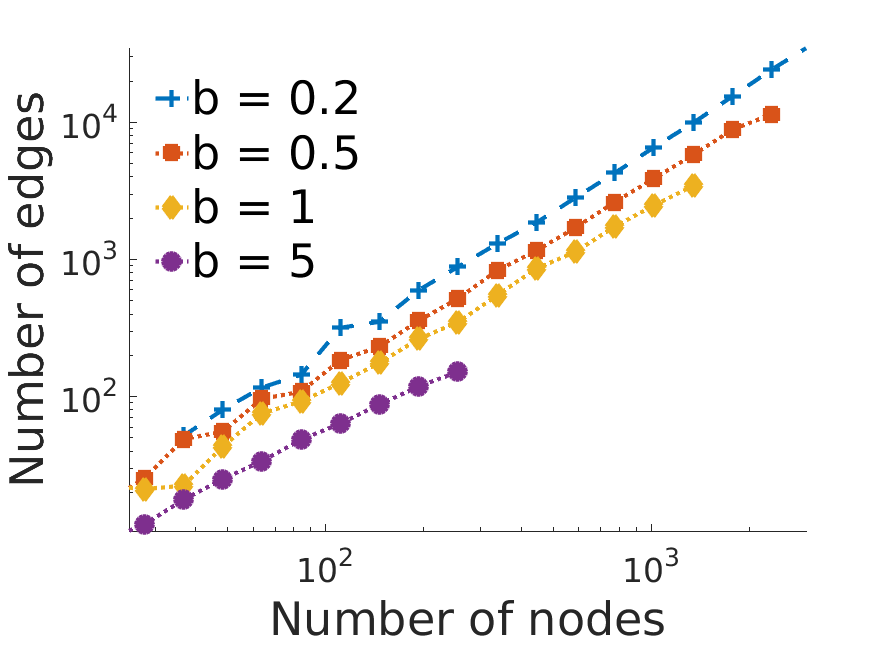}}
\caption{Relationship between nodes and edges for varying $b$}
\label{b_asymptotic_graphs}
\end{figure}

We have seen how the sparsity properties in the different regions are controlled by $\sigma$. From Figures \ref{a_asymptotic_graphs} and \ref{b_asymptotic_graphs} we see that, as expected, changing the parameters $a$ and $b$ does not affect this, since the gradients of the lines in each case are the same. However, we do see that for fixed sized graphs, increasing $a$ increases the density in the overall graph, and in the different regions. Conversely, increasing $b$ decreases the density in the core-periphery and periphery regions. The density in the core region is also decreased, but this effect appears to be far less significant.

\subsection{Core proportion}
\begin{figure}
\centering
\subfloat[$a$ \label{fig:a_corenodesvsnodes}]{\includegraphics[width=0.4\textwidth]{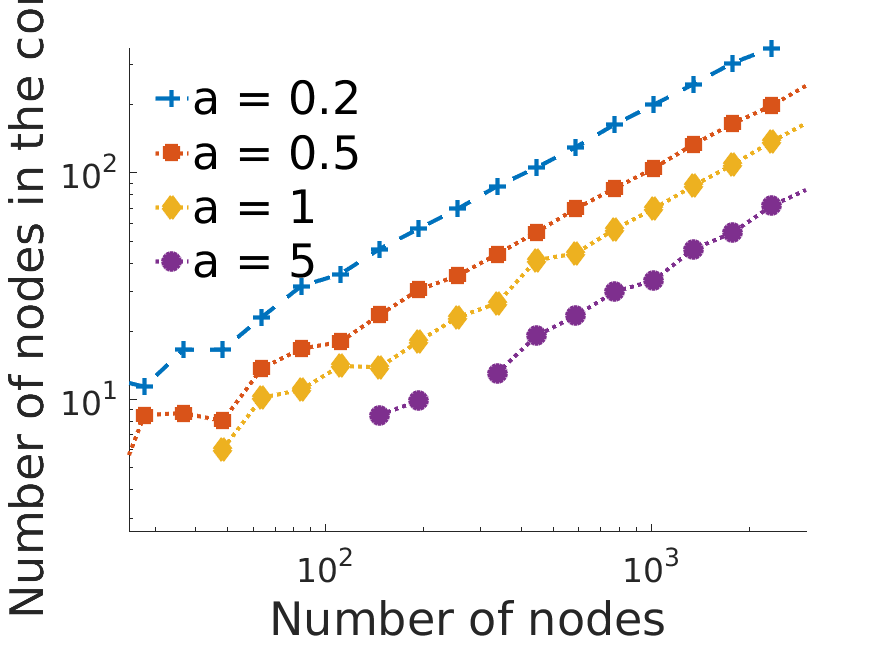}}
\subfloat[$b$ \label{fig:b_corenodesvsnodes}]{\includegraphics[width=0.4\textwidth]{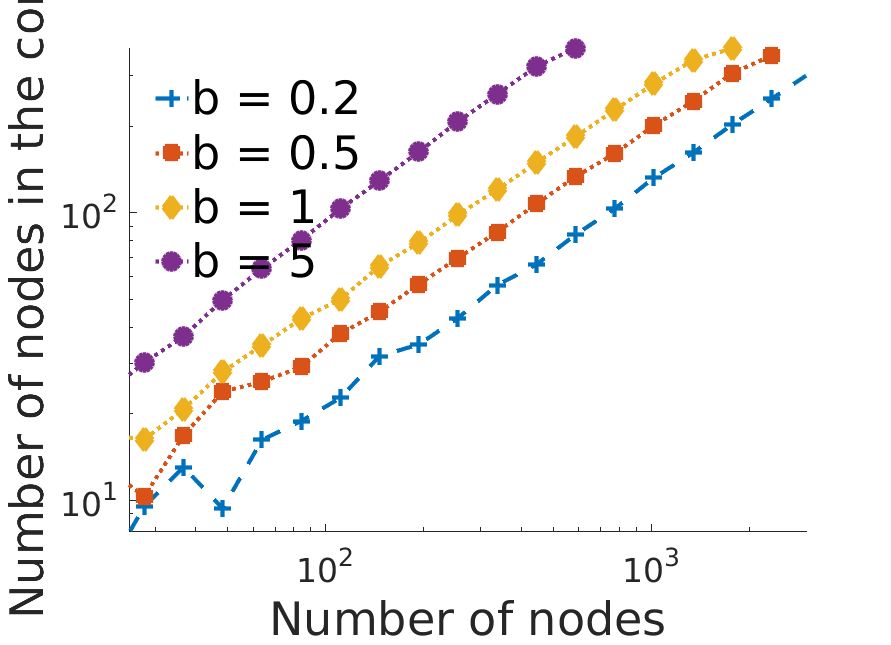}}
\caption{Relative core size for varying $a$ and $b$}
\label{a_b_asymptotic_graphs}
\end{figure}

We are finally interested in how the proportion of nodes in the core is affected  by changing $a$ and $b$. From Figure \ref{a_b_asymptotic_graphs} we see that, as expected, the asymptotic rates are not affected by changing these hyperparameters. However, for fixed size graphs, increasing $a$ decreases the relative size of the core region, whilst the opposite is true for $b$.

\section{MCMC Algorithm Details}\label{app:mcmc_alg}
We give here the details of the various steps of Algorithm \ref{alg:MCMC_sampler}, in particular where they differ from the Algorithm of \citep{todeschini2016exchangeable}.
\subsection{Step 1}
In general, if $\rho$ can be evaluated pointwise, a MH update can be used for this step, however this will scale poorly with the number of nodes. In the compound CRM case, if $f$ and $\rho_0$ are differentiable then a HMC update can be used. For the overlapping communities model where $f$ is the product of independent gamma distributions and $\rho_0$ is the mean measure of a generalized gamma process this differentiability condition does hold. Thus, a  HMC sampler is used to update $(w_{i0},\beta_{i1},\ldots,\beta_{iK})_{i=1,\ldots,N_{\alpha}}$ all at once, with the potential energy function $U$ defined as
\begin{align}
U(w_0,\beta)=\log\left(p((\log(w_{i0}))_{i=1,\ldots,N_{\alpha}},(\log(\beta_{ik}))_{i=1,\ldots,N_{\alpha};k=1,\ldots,K}\mid rest)\right)
\end{align}
$U(w_0,\beta)$ can be calculated from (\ref{ccrm_posterior_characterization}) using a simple change of variables. Furthermore its derivatives have simple analytic forms, and so the HMC step can be computed exactly.

In our case this strategy will not work, because $\beta_1$ is a binary variable, and so both the log transformation and the HMC update of the $(\beta_{i1})_{i=1,\ldots,N_{\alpha}}$ cannot be used. We propose an alternative solution, using a Gibbs step to update $\beta_1$ separately from the others as follows
\begin{enumerate}
\item Update $(w_{i0},\beta_{i2},\ldots,\beta_{iK})_{i=1,\ldots,N_{\alpha}}$ from \newline $p((w_{i0},\beta_{i2},\ldots,\beta_{iK})_{i=1,\ldots,N_{\alpha}}\mid (\beta_{i1})_{i=1,\ldots,N_{\alpha}}, \text{rest})$
\item Update $(\beta_{i1})_{i=1,\ldots,N_{\alpha}}$ from \newline $p((\beta_{i1})_{i=1,\ldots,N_{\alpha}}\mid (w_{i0},\beta_{i2},\ldots,\beta_{iK})_{i=1,\ldots,N_{\alpha}}, \text{rest})$
\end{enumerate}
This method also allows us to deal with the fact that the $\beta_{i1}$ are not identically distributed, and depend on the $w_{i0}$. The two conditional distributions can be found from (\ref{ccrm_posterior_characterization}) as follows. Ignoring the terms not involving $(w_{0},\beta_{2},\ldots,\beta_{K})$, the first conditional is proportional to
\begin{equation}
\begin{aligned}
& \left[\prod_{i=1}^{N_{\alpha}}  w_{i0}^{m_{i}}\right]\left[\prod_{i=1}^{N_{\alpha}} \prod_{k=2}^K \beta_{ik}^{m_{ik}}\right] e^{-\sum_{k=1}^K(w_{\ast k}+\sum_{i=1}^{N_{\alpha}}w_{i0}\beta_{ik})^2}\left[\prod_{i=1}^{N_{\alpha}}\rho_0(w_{i0};\phi)\right]\\
&\times \left[\prod_{i=1}^{N_{\alpha}}\prod_{k=2}^K \beta_{ik}^{a_k-1}e^{-b_k\beta_{ik}}\frac{b_k^{a_k}}{\Gamma(a_k)}\right]\left[\prod_{i=1}^{N_{\alpha}} (1-e^{-w_{i0}})^{\beta_{i1}}e^{-w_{i0}(1-\beta_{i1})} \right]
\end{aligned}
\end{equation}
We can simulate from this using HMC as before, except that now $\beta_1$ is considered to be constant, and we have to incorporate an extra term involving the $w_{i0}$ coming from the distribution of $\beta_{i0} \mid w_{i0}$. The details of the exact HMC sampler are therefore very similar to the overlapping communities model, and we omit them here.

The second conditional distribution is proportional to
\begin{align}
\left[\prod_{i=1}^{N_{\alpha}} \beta_{i1}^{m_{i1}}\right] e^{-(w_{\ast 1}+\sum_{i=1}^{N_{\alpha}}w_{i0}\beta_{i1})^2}\left[\prod_{i=1}^{N_{\alpha}} (1-e^{-w_{i0}})^{\beta_{i1}}e^{-w_{i0}(1-\beta_{i1})} \right]
\end{align}
In order to simulate from this distribution we first approximate the conditional by linearizing the exponent as follows
\begin{align*}
&\exp\left\{ -\left(w_{\ast 1}+\sum_{i=1}^{N_{\alpha}}w_{i0}\beta_{i1}\right)^2 \right\} \\
&\approx \exp\left\{ -\left(w_{\ast 1}+\sum_{i=1}^{N_{\alpha}}w_{i0}\beta^{\ast}_{i1}\right)\left(w_{\ast }+\sum_{i=1}^{N_{\alpha}}w_{i0}\beta_{i1}\right) \right\}
\end{align*}
where $\beta^{\ast}_{i1}$ is the sampled value of $\beta_{i1}$ from Step 1 of the previous iteration of the overall Gibbs sampler in Algorithm \ref{alg:MCMC_sampler}. We thus treat $\beta^{\ast}_{i1}$ as a constant here. Using this approximation, and defining
\begin{align*}
c_1=\left(w_{\ast 1}+\sum_{i=1}^{N_{\alpha}}w_{i0}\beta^{\ast}_{i1}\right)
\end{align*}
we have that the conditional distribution of $(\beta_{i1})_{i=1,\ldots,N_{\alpha}}$ given the rest is proportional to
\begin{align*}
&\exp\left\{ - c_1\sum_{i=1}^{N_{\alpha}}w_{i0}\beta_{i1} \right\}\left[\prod_{i=1}^{N_{\alpha}} \beta_{i1}^{m_{i1}} (1-e^{-w_{i0}})^{\beta_{i1}}e^{-w_{i0}(1-\beta_{i1})} \right]\\
&=\prod_{i=1}^{N_{\alpha}} \beta_{i1}^{m_{i1}} \left[e^{-c_1w_{i0}}(1-e^{-w_{i0}})\right]^{\beta_{i1}}e^{-w_{i0}(1-\beta_{i1})} \\
&=\prod_{i=1}^{N_{\alpha}} f_i(\beta_{i1})
\end{align*}
We see that we can simulate each of the $\beta_{i1}$ individually. The distribution $f_i(\beta_{i1})$ depends on $m_{i1}$, and there are two cases.
\begin{enumerate}
\item If $m_{i1}>0$ then, recalling that $\beta_{i1}$ is a binary variable, $Pr(\beta_{i1}=0)=f_i(0)=0$ and so $Pr(\beta_{i1}=1)=1$. This means that if the sum of the latent counts $\sum_j\tilde{n}_{ij1}$ from the previous iteration is not equal to $0$ then $\beta_{i1}$ will be updated to $1$. We also know from (\ref{eq:augmentation}) that if $\beta_{i1}=0$ then $\tilde{n}_{ij1}=0\quad \forall j$. However, this does not lead to a problem of losing irreducibility of the Markov chain, since the reverse is not true.
\item If $m_{i1}=0$ then we have that
\begin{align*}
f_i(\beta_{i1}) \propto \left[e^{-c_1w_{i0}}(1-e^{-w_{i0}})\right]^{\beta_{i1}}e^{-w_{i0}(1-\beta_{i1})}
\end{align*}
which we recognize as a Bernoulli distribution with parameter
\begin{align*}
p_1=\frac{1-e^{-w_{i0}}}{1-e^{-w_{i0}}+e^{-(1-c_1)w_{i0}}}
\end{align*}
Thus we can update the $\beta_{i1}$ given the rest by sampling
\begin{align*}
\beta_{i1} \sim \mbox{Bernoulli}(p_1)
\end{align*}
\end{enumerate}

Of course, we are conditioning on $m_{i1}$ when performing this step, so we will always know what case we are in. Thus, the above can be used to sample $\beta_{1}$ from the relevant conditional distribution.
\subsection{Step 2}
In this step we want to sample from $p((w_{\ast k})_{k=1,\ldots,K},\phi,\alpha \mid rest)$, which we know from (\ref{ccrm_posterior_characterization}) is proportional to
\begin{align*}
p(\phi)&p(\alpha)e^{-\sum_{k=1}^K(w_{\ast k}+\sum_{i=1}^{N_{\alpha}}w_{ik})^2}\left[\prod_{i=1}^{N_{\alpha}}\rho(w_{i1},\ldots,w_{iK};\phi)\right]\\ &\times \alpha^{N_{\alpha}}g_{\ast \alpha}(w_{\ast 1},\ldots,w_{\ast K};\phi)
\end{align*}
this distribution is not of a standard form, and so for the overlapping communities model a MH step is used, with a proposal
\begin{equation}
\begin{aligned}
q(\tilde{w}_{\ast 1:K},\tilde{\phi},\tilde{\alpha}\mid w_{\ast 1:K},\phi,\alpha)=&q(\tilde{w}_{\ast 1:K}\mid w_{\ast 1:K},\tilde{\phi},\tilde{\alpha}) \times q(\tilde{\phi}\mid \phi)\\
&\times q(\tilde{\alpha}\mid w_{\ast 1:K},\tilde{\phi},\alpha)
\end{aligned}
\end{equation}
where
\begin{align}
q(\tilde{\alpha}\mid w_{\ast 1:K},\tilde{\phi},\alpha)=Gamma(\tilde{\alpha};a_{\alpha}+N_{\alpha},b_{\alpha}+\psi(\lambda_1,\ldots,\lambda_K;\tilde{\phi}))
\end{align}
and $q(\tilde{w}_{\ast 1:K}\mid w_{\ast 1:K},\tilde{\phi},\tilde{\alpha})$ is an exponentially tilted version of $g_{\ast \alpha}$.
Here, $\lambda_k= w_{\ast k}+2\sum_{i=1}^{N_{\alpha}}w_{ik}$, and $\psi$ is the Laplace exponent. In the compound CRM case $\psi$ takes a simple form that only involved evaluating a one-dimensional integral. The calculation of $\psi$ in our case can be done similarly, the only difference being that we need the moment generating function of a Bernoulli distribution for the first component. The details are omitted here. Similarly, since the distribution of $\beta_{1}$ has no hyperparameter, the same $q(\tilde{\phi}\mid \phi)$ can be used in our case as the compound CRM case of the overlapping community model, ignoring the $k=1$ term.

The other challenging part of the step is to simulate $w_{\ast 1:K}$ from the distribution $q(\tilde{w}_{\ast 1:K}\mid w_{\ast 1:K},\tilde{\phi},\tilde{\alpha})$. Again we can do this as in the compound CRM case of \citep{todeschini2016exchangeable}, by setting $(w_{\ast 1},\ldots,w_{\ast K})=X_{\epsilon}+X^{\epsilon}$, with $X_{\epsilon}=\sum_{i|w_{i0}<\epsilon}w_{i0}(\beta_{i1},\ldots,\beta_{iK})$ and $X^{\epsilon}=\sum_{i|w_{i0}>\epsilon}w_{i0}(\beta_{i1},\ldots,\beta_{iK})$. Then, a realization of $X^{\epsilon}$ can be simulated exactly from a Poisson process with mean measure
\begin{align}
\alpha e^{-w_0\sum_{k=1}^K(\gamma_k+\lambda_k)\beta_k}f(\beta_1,\ldots,\beta_K)\rho_0(w_0)\mathds{1}_{w_0>\epsilon}
\end{align}
This is done using an adaptive thinning procedure as detailed in Appendix D of \citep{todeschini2016exchangeable}. We can use the same approach here, using the same adaptive bound by noting that $(1-e^{-w_0})\leq 1$. $X_{\epsilon}$ can be approximated as before using a truncated Gaussian random vector. All that changes are the exact forms of the mean and variance of the approximating random vector, and we omit the details here.
\subsection{Step 3}
This step can be done as in \citep{todeschini2016exchangeable} by introducing the latent variables $\widetilde{n}_{ijk}=n_{ijk}+n_{jik}$
where
\begin{align*}
\left(\widetilde{n}_{ij1},\ldots,\widetilde{n}_{ijK}\right)\vert z,w & \sim\begin{cases}
\delta_{(0)} & \mbox{if }z_{ij}=0\\
\mbox{tPoisson}\left(2w_{i1}w_{j1},\ldots,2w_{iK}w_{jK}\right) & \mbox{if }z_{ij}=1,\,i<j
\end{cases}
\end{align*}

\begin{align*}
\left(\frac{\widetilde{n}_{ii1}}{2},\ldots,\frac{\widetilde{n}_{iiK}}{2}\right)\vert z,w & \sim\mbox{tPoisson}\left(w_{i1}^{2},\ldots,w_{iK}^{2}\right)\mbox{ if }z_{ii}=1
\end{align*}

By convention we set $\widetilde{n}_{ijk}=\widetilde{n}_{jik}$ for
all $i>j$, and we have that $m_{ik}=\sum_{j}\widetilde{n}_{ijk}$. The pmf of a $\mbox{tPoisson}(\lambda_1,\ldots,\lambda_K)$ distribution is
\begin{align*}
\mbox{tPoisson}(x_1,\ldots,x_K; \lambda_1,\ldots,\lambda_K)= \frac{\prod_{k=1}^K \mbox{Poisson}(x_k;\lambda_k)}{1-\exp(-\sum_{k=1}^K x_k\lambda_k)}\mathds{1}_{\{\sum_{k=1}^K x_k >0\}}
\end{align*}
which can be sampled from by sampling $x=\sum_{k=1}^K x_k$ from a univariate zero-truncated Poisson distribution with rate $\sum_{k=1}^K\lambda_k$, and then sampling 
\begin{align*}
    (x_1,\ldots,x_K)\mid( \lambda_1,\ldots,\lambda_K),x \sim \mbox{Multinomial}\left(x,\left(\frac{\lambda_1}{\sum \lambda_k},\ldots,\frac{\lambda_K}{\sum \lambda_k}\right)\right). 
\end{align*}
The only difference in our case is that $\lambda_1$ may be $0$, in which case we set $x_1=0$.

In the update for $\beta_{1}$, we see that if $m_{i1}>0$ then $\beta_{i1}$ is set identically to $1$, because there is no posterior mass at $0$. In order to get better mixing, we can instead update the $\widetilde{n}_{ijk}$ via a Metropolis-Hastings step that proposes $m_{i1}=0$ more often. We do this as follows
\begin{enumerate}
    \item Choose a set $I=\{i_1,\ldots,i_N\},\ i_1<\ldots<i_N$, with $N=|I|$ of indices for which we will propose $m_{i1} = 0\ \forall i \in I$.
    \item Calculate the set $A$ of edges $(i,j)$ such that $\widetilde{n}_{ij1}=0\ \forall (i,j) \in A \iff m_{i1} = 0\ \forall i \in I$.
    \item For $(i,j) \not \in A$, update $\left(\widetilde{n}_{ij1},\ldots,\widetilde{n}_{ijK}\right)$ as normal, using the conditional distribution given $z,w$.
    \item For $(i,j) \in A$, propose an update for $\left(\widetilde{n}_{ij1},\ldots,\widetilde{n}_{ijK}\right)$ from the mixture distribution:
    \begin{align*}
        p_{lat}\left( \mathbbm{1}{(\widetilde{n}_{ij1}=0)}\times \mbox{tPoisson}\left(2w_{i2}w_{j2},\ldots,2w_{iK}w_{jK}\right)\right)\\
        +(1-p_{lat})\left(\mbox{tPoisson}\left(2w_{i1}w_{j1},\ldots,2w_{iK}w_{jK}\right) \right)
    \end{align*}
    for $i\neq j$, and similarly for $i=j$. This means that with probability $p_{lat}$ we set $\widetilde{n}_{ij1}$ to $0$ and simulate $\widetilde{n}_{ij2},\ldots, \widetilde{n}_{ijK}$ from a truncated Poisson distribution. With probability $1-p_{lat}$, $\widetilde{n}_{ij1},\ldots, \widetilde{n}_{ijK}$ are simulated from a truncated Poisson distribution as normal.
    \item Accept the proposal using the standard Metropolis Hastings acceptance rate given by
    \begin{align*}
        \alpha=\min \left(1, \frac{P(\widetilde{n})}{P(\widetilde{n}_{old})}\frac{Q(\widetilde{n}_{old})}{Q(\widetilde{n})} \right)
    \end{align*}
    where $\widetilde{n}=(\widetilde{n})_{1 \leq i,j\leq N_{\alpha},k=1,\ldots,K}$, $P$ is the true distribution of the $\widetilde{n}_{ijk}$ as given in (\ref{eq:augmentation}), $Q$ is the proposal distribution detailed above and $\widetilde{n}_{old}$ is the value of $\widetilde{n}$ from the previous iteration of the overall sampler.
\end{enumerate}
In practice, we find that this method allows for faster mixing of the algorithm.
\subsection{Local maxima problem}
When testing our model on simulated data sets with $K=2$ (i.e. with a core parameter $w_1$ and overall sociability parameter $w_2$) we find that when the core-periphery structure is particularly weak our estimated values of $w_1$ are close to the true values of $w_2$, and vice versa. Similarly, when taking $K>2$, we sometimes find that the estimated values of the core parameter are instead close to the values of one of the community parameters.

In order to prevent the chains getting stuck in these local maxima, we introduce a new initialization procedure for these cases, in order to ensure that our core parameter $w_1$ is estimating the core-periphery structure. The initialization method employed by \citep{todeschini2016exchangeable} performs a short run with $K=1$, using the parameter estimates obtained there as initial values for the full run. In our case, we instead perform a short initialization run using the full communities model, Having done this, we use the identified communities in order to initialize the parameter values for $w_k, k>1$. In the case that $K=2$, we find that when running the communities model with $K=2$, one of the features approximated the core parameter, while the other approximated the overall sociability parameter, and so the method works in this case as well. As in \citep{todeschini2016exchangeable}, this step in practice requires human interpretation for real data, in order to select which sociability parameter is approximating the core-periphery structure and thus should be used as initial values for $w_1$. For simulated data sets, the different distributions for $w_1$ and $w_2,\ldots,w_K$ allow us to do this step automatically, and also aid in the interpretation for real data.

\subsection{Approximation of the log-posterior density}
\label{subsec:app:jointposterior}
The posterior probability density function, up to a normalizing constant, takes the form
\begin{align}
&p\left( \left(w_{i0},\beta_{i1},\ldots,\beta_{iK}\right)_{i=1,\ldots,N_{\alpha}},\phi,\alpha\mid  (z_{ij})_{1\leq i,j\leq N_\alpha}\right)\nonumber\\
& \propto \left [\prod_{i=1}^{N_\alpha} \prod_{j=1}^{N_\alpha} \left (\frac{1-e^{-\sum_k w_{i0}\beta_{ik} w_{j0}\beta_{jk}}}{e^{-\sum_k w_{i0}\beta_{ik} w_{j0}\beta_{jk}}}\right )^{z_{ij}}\right ] e^{-\sum_{k=1}^K( w_{\ast k}+ \sum_{i=1}^{N_\alpha} w_{i0}\beta_{ik} )^2} \nonumber\\
&~~~~\times \left[\prod_{i=1}^{N_{\alpha}}\prod_{k=2}^K \beta_{ik}^{a_k-1}e^{-b_k\beta_{ik}}\frac{b_k^{a_k}}{\Gamma(a_k)}\right]\left[\prod_{i=1}^{N_{\alpha}} (1-e^{-w_{i0}})^{\beta_{i1}}e^{-w_{i0}(1-\beta_{i1})} \right] \nonumber\\
&~~~~\times \left[\prod_{i=1}^{N_{\alpha}}\rho_0(w_{i0};\phi)\right] \alpha^{N_\alpha} p(\phi,\alpha)g_{\ast\alpha}(w_{\ast 1},\ldots,w_{\ast K};\phi)
\label{eq:app:jointposterior}
\end{align}
where $m_{ik}=\sum_{j=1}^{N_\alpha} n_{ijk}+n_{jik}$ and $g_{\ast\alpha}(w_{\ast 1},\ldots,w_{\ast K};\phi)$ is the probability density function of the random vector $\left(W_{1}\left([0,\alpha]\right),\ldots,W_{K}\left([0,\alpha]\right)\right)$.

This is intractable due to the lack of an analytic expression for $g_{\ast\alpha}$. We can however approximate the log-posterior. Noting that $w_{\ast k}=o(\sum_{i=1}^{N_\alpha} w_{ik})$ as $\alpha\rightarrow\infty$, we have $$(w_{\ast k}+\sum_{i=1}^{N_\alpha} w_{ik})^2\simeq (\sum_{i=1}^{N_\alpha} w_{ik})^2+2w_{\ast k}\sum_{i=1}^{N_\alpha} w_{ik}.$$ Using this approximation, one can now integrate out $w_{\ast k}$ and we then obtain the approximation
\begin{align}
&p((w_{1k},\ldots,w_{N_\alpha k})_{k=1,\ldots,K},\phi,\alpha\mid  (z_{ij})_{1\leq i,j\leq N_\alpha})\nonumber\\
&~~~~\simeq \propto \left [\prod_{i=1}^{N_\alpha} \prod_{j=1}^{N_\alpha} \left (\frac{1-e^{-\sum_k w_{i0}\beta_{ik} w_{j0}\beta_{jk}}}{e^{-\sum_k w_{i0}\beta_{ik} w_{j0}\beta_{jk}}}\right )^{z_{ij}}\right ]  e^{-\sum_{k=1}^K( \sum_{i=1}^{N_\alpha} w_{i0}\beta_{ik})^2} \nonumber\\
&~~~~~~~\times\left[\prod_{i=1}^{N_{\alpha}}\prod_{k=2}^K \beta_{ik}^{a_k-1}e^{-b_k\beta_{ik}}\frac{b_k^{a_k}}{\Gamma(a_k)}\right] \left[\prod_{i=1}^{N_{\alpha}} (1-e^{-w_{i0}})^{\beta_{i1}}e^{-w_{i0}(1-\beta_{i1})} \right] \nonumber\\
&~~~~~~~\times \left[\prod_{i=1}^{N_{\alpha}}\rho_0(w_{i0};\phi)\right]\alpha^{N_\alpha}\exp\left [-\alpha \psi\left (2\sum_{i=1}^{N_\alpha} w_{i1},\ldots,2\sum_{i=1}^{N_\alpha} w_{iK};\phi\right ) \right ]p(\phi,\alpha)\label{eq:jointposterioBr}
\end{align}
where $\psi(t_1,\ldots,t_K)$ is the multivariate Laplace exponent, which can be evaluated numerically.
\section{MCMC Diagnostic Plots}\label{app:mcmc_diagnostic}
\subsection{Simulated data}\label{app:simulated data}
We provide here additional trace plots and convergence diagnostics on the simulated data experiments in Section \ref{sec:experiments_simu}.

\subsubsection{Core-periphery structure only}
\label{sec:app:simucore}

\begin{figure}[h]
\centering
\subfloat[$\log \tilde{\alpha}$ trace]{\includegraphics[width=0.3\textwidth]{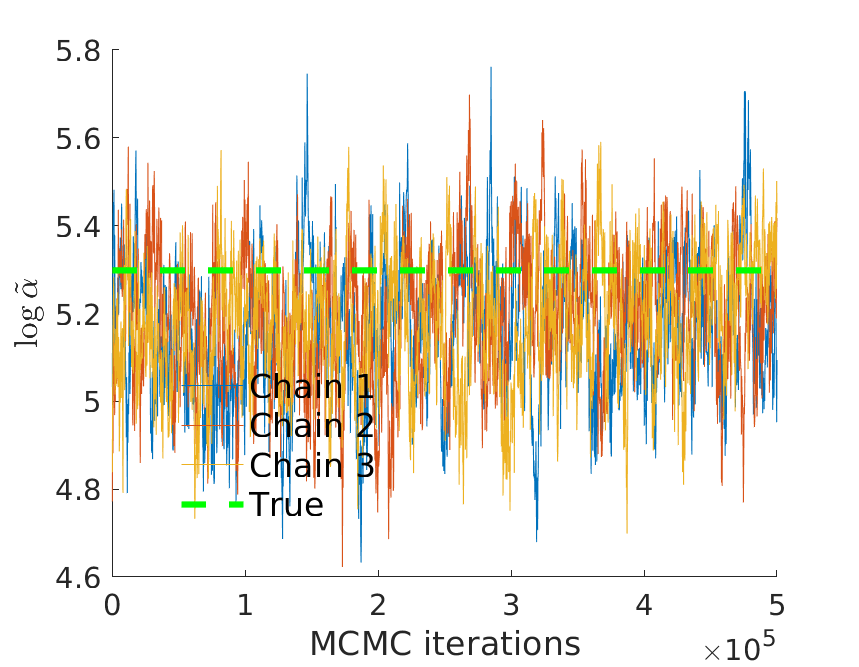}}
\subfloat[$\sigma$ trace]{\includegraphics[width=0.3\textwidth]{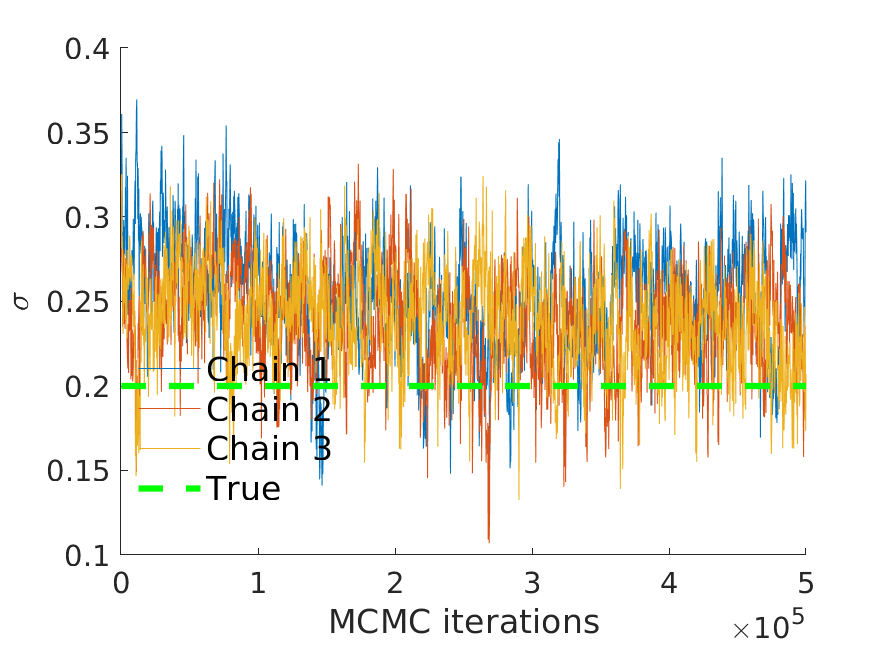}}
\subfloat[$a$ trace]{\includegraphics[width=0.3\textwidth]{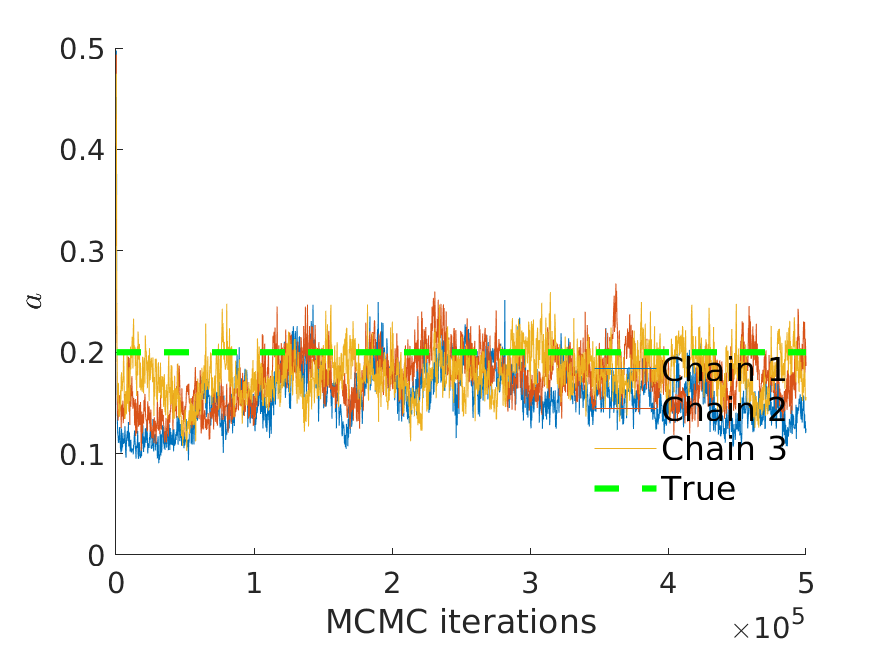}}\\
\subfloat[$\tilde{b}$ trace]{\includegraphics[width=0.3\textwidth]{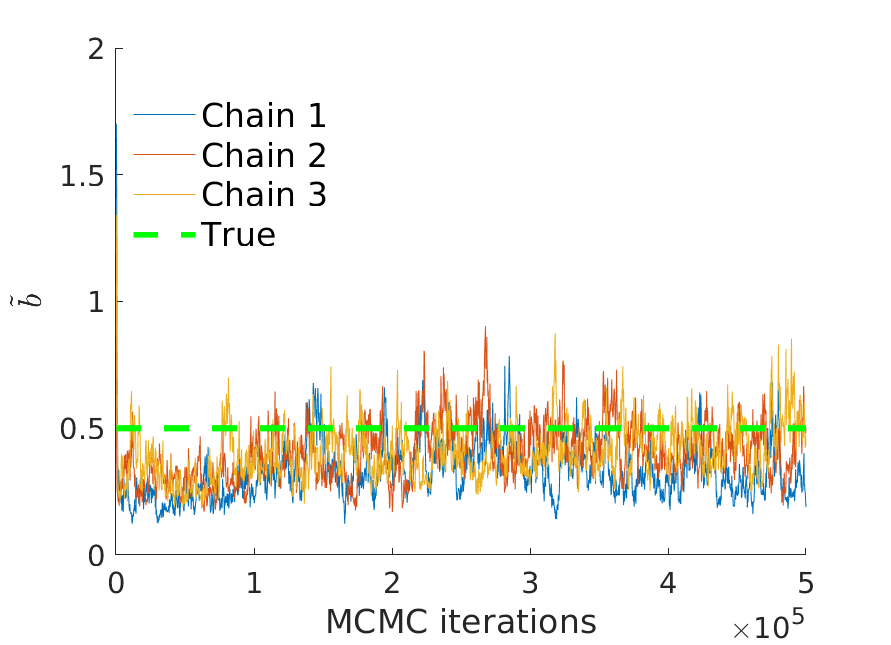}}
\subfloat[$\bar{w}_{\ast}$ trace]{\includegraphics[width=0.3\textwidth]{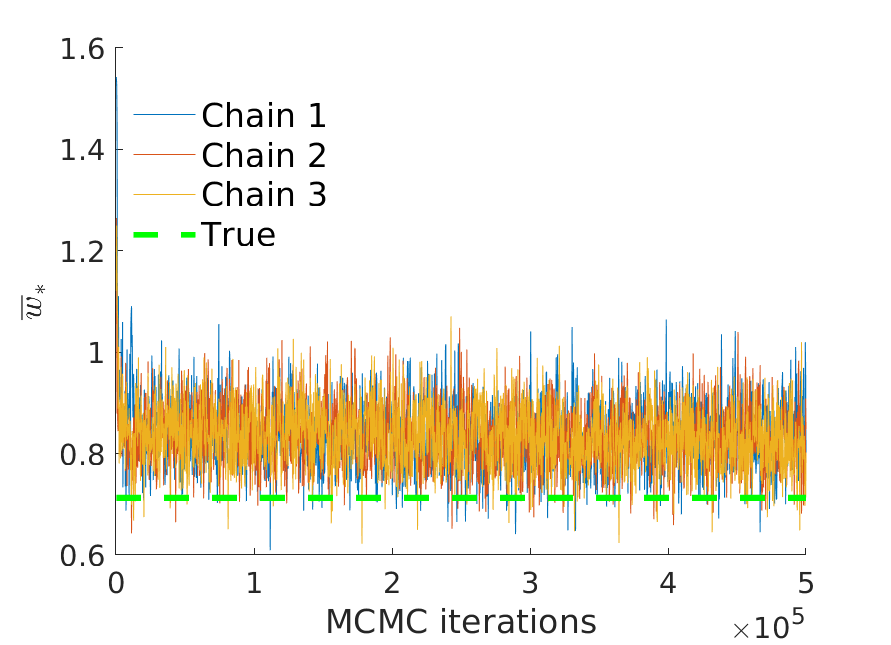}}
\caption{MCMC trace plots for a graph generated with $K=2, \alpha = 200, \sigma= 0.2, \tau = 1, b=\frac{1}{K}, a=0.2$. The green lines correspond to the true values of the parameters}
\label{fig:mcmc_trace_hist}

\end{figure}

\begin{figure}[h]
\centering
\subfloat[$\log \tilde{\alpha}$ histogram]{\includegraphics[width=0.24\textwidth]{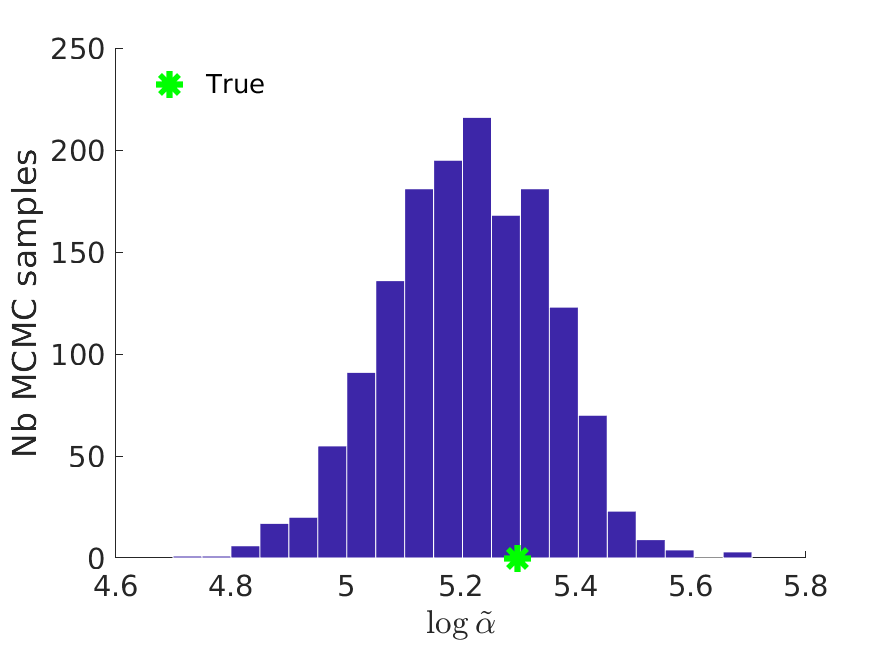}}
\subfloat[$\sigma$ histogram]{\includegraphics[width=0.24\textwidth]{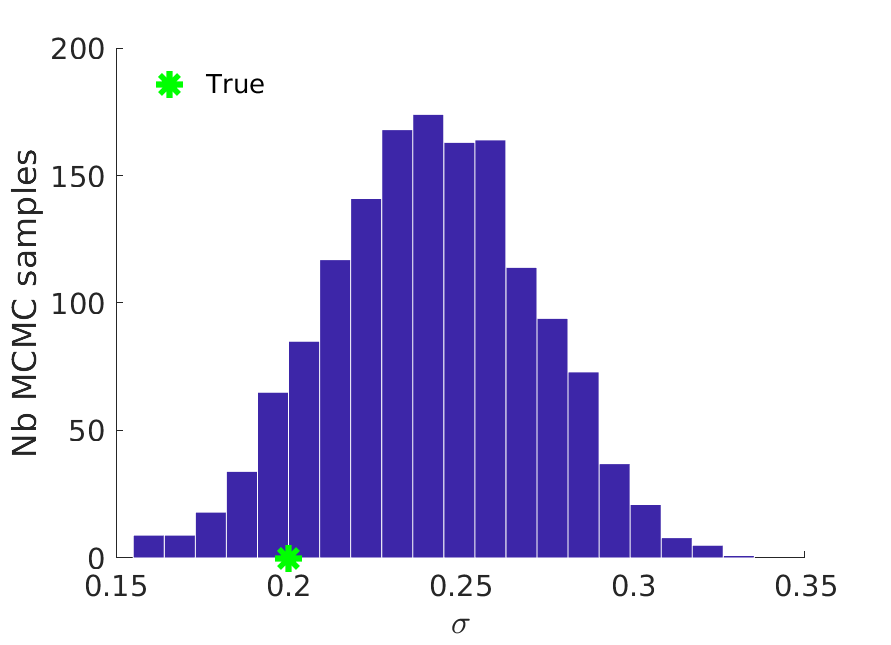}}
\subfloat[$a$ histogram]{\includegraphics[width=0.24\textwidth]{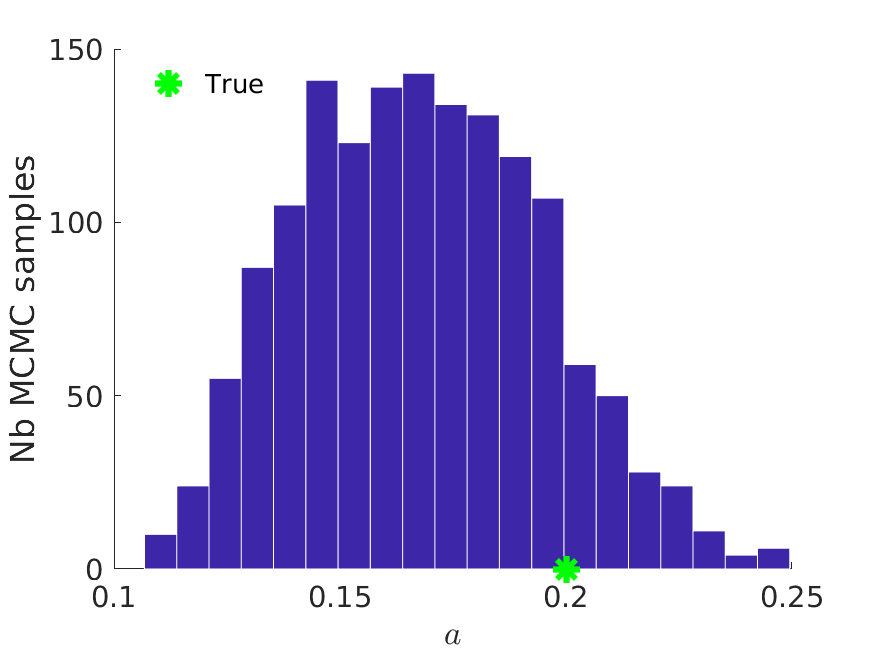}}\\
\subfloat[$\tilde{b}$ histogram]{\includegraphics[width=0.24\textwidth]{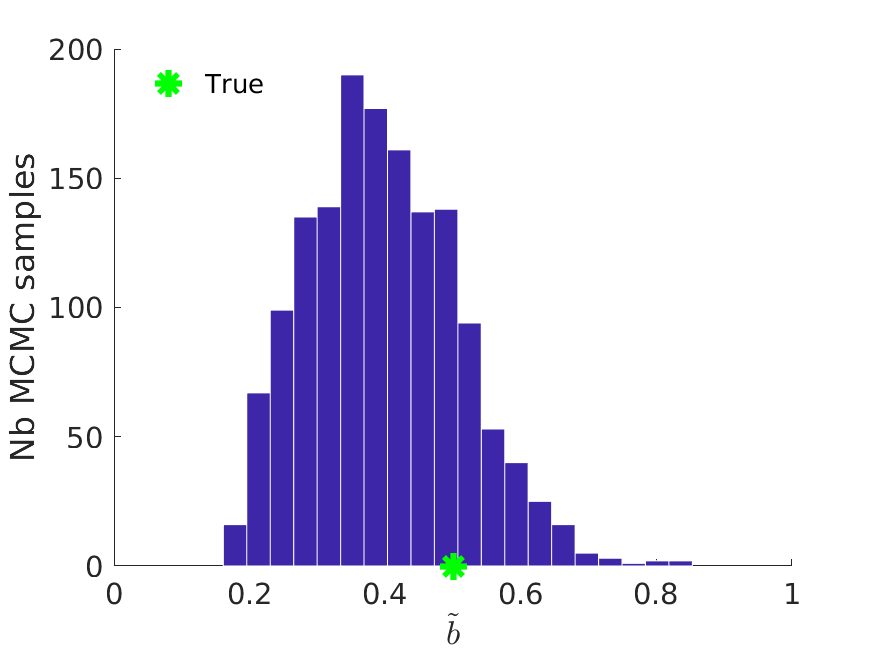}}
\subfloat[$\bar{w}_{\ast}$ histogram]{\includegraphics[width=0.24\textwidth]{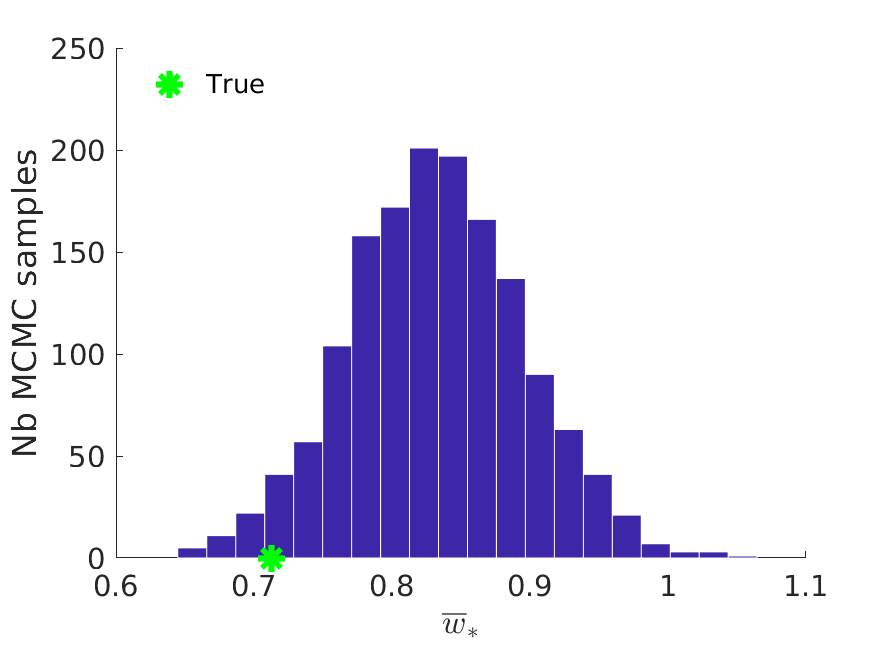}}
\caption{MCMC histograms for a graph generated with $K=2, \alpha = 200, \sigma= 0.2, \tau = 1, b=\frac{1}{K}, a=0.2$. The green stars correspond to the true values of the parameters}
\label{fig:mcmc_hist}

\end{figure}

In Figures \ref{fig:mcmc_trace_hist} and \ref{fig:mcmc_hist} we see the trace plots and histograms for the graph generated with $K=2, \alpha = 200, \sigma= 0.2, \tau = 1, b=\frac{1}{K}, a=0.2$.  The model is overparametrized, and so we plot the identifiable parameters, which are $\log \tilde{\alpha}, \sigma, a, \tilde{b}$ and $\bar{w}_{\ast }$, where $\tilde{\alpha}=\alpha\tau^{\sigma}$ and $\tilde{b}=b\tau$. These correspond to the original parameters if $\tau$ is fixed to be $1$. The green lines and stars respectively correspond to the values of the model parameters used to generate the graphs, and we see that the posterior converges around the true values. 

\begin{figure}[h]
\centering
\subfloat[Log-posterior]{\includegraphics[width=0.4\textwidth]{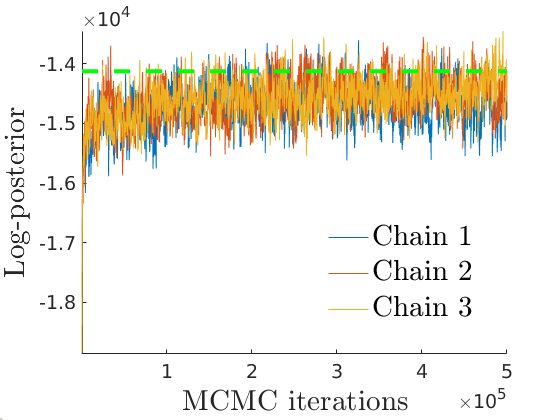}}
\subfloat[Autocorrelation]{\includegraphics[width=0.4\textwidth]{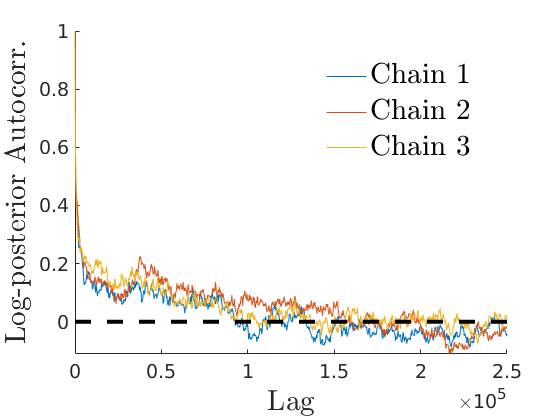}}
\caption{Trace plot and autocorrelation of the log-posterior for a graph generated with $K=2, \alpha = 200, \sigma= 0.2, \tau = 1, b=\frac{1}{K}, a=0.2$. The  green line in $(a)$ corresponds to the value of the approximate log-posterior (up to a constant) under the model parameters used to generate the graph.}
\label{fig:mcmc_logpost}

\end{figure}

In Figure \ref{fig:mcmc_logpost} we give the trace and autocorrelation plots of an approximation of the log-posterior (up to a normalizing constant). The details of the calculation of the approximate log-posterior are given in Section~ \ref{subsec:app:jointposterior}. Here, the green line gives the value of the approximate log-posterior using the true model parameters. The log-posterior trace plot suggests that the chains have converged and the autocorrelation plot shows that the correlation of the samples decreases quickly with increasing lag. 

In order to test the convergence of the chains, we calculate the Gelman-Rubin convergence diagnostic $\hat{R}$ \cite{gelman1992inference}. Due to the high number of parameters in our model, we calculate a univariate statistic using the sampled values of the approximate log-posterior. Recalling that $\hat{R}<1.1$ suggests convergence, we find that in our case $\hat{R}=1.03$. Thus we are satisfied that the chains have indeed converged.

\subsubsection{Core-periphery and community structure}
\label{sec:app:simucorecommunity}

\begin{figure}[h]
\centering
\subfloat[$\log \tilde{\alpha}$ trace]{\includegraphics[width=0.3\textwidth]{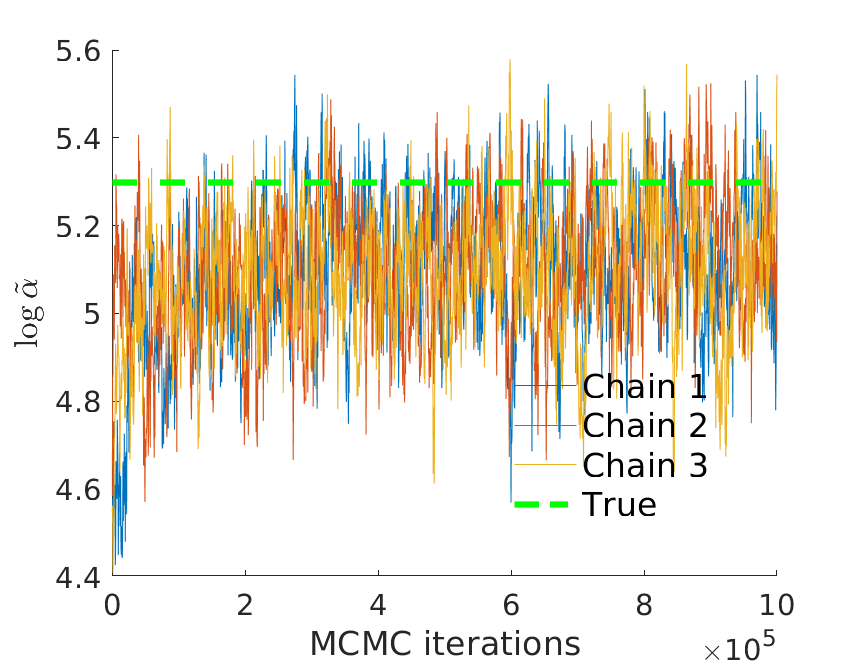}}
\subfloat[$\sigma$ trace]{\includegraphics[width=0.3\textwidth]{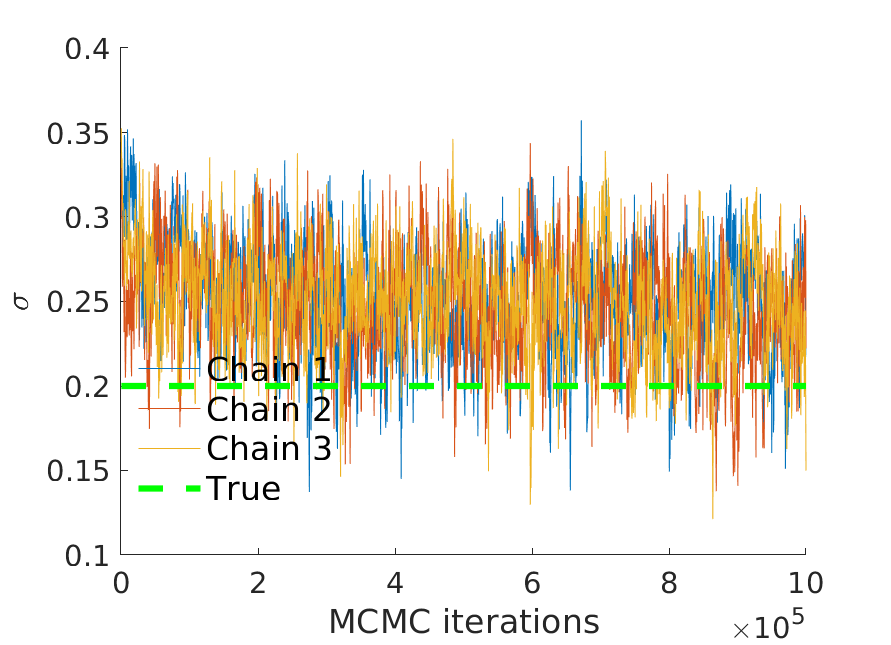}}
\subfloat[$a$ trace]{\includegraphics[width=0.3\textwidth]{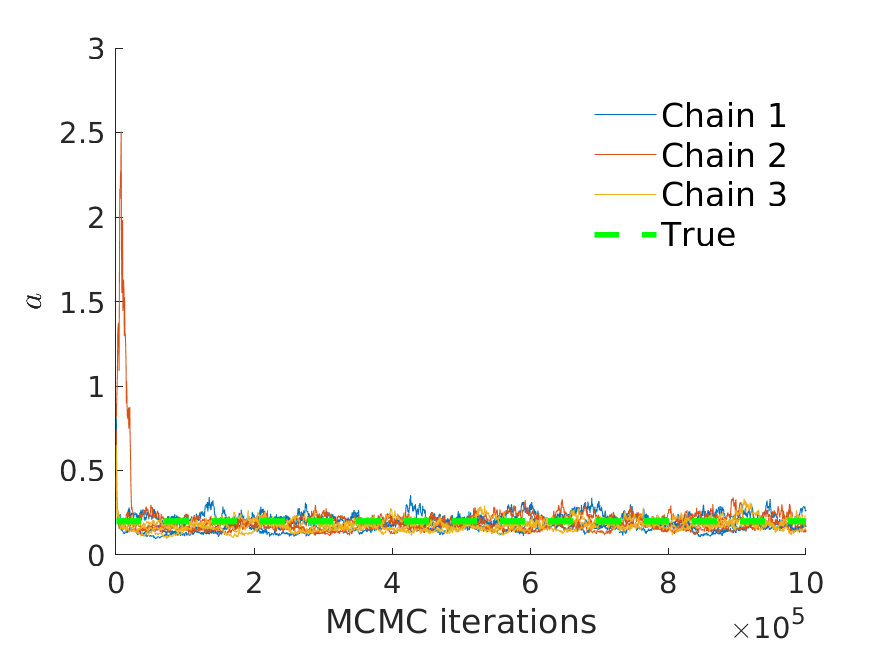}}\\
\subfloat[$\tilde{b}$ trace]{\includegraphics[width=0.3\textwidth]{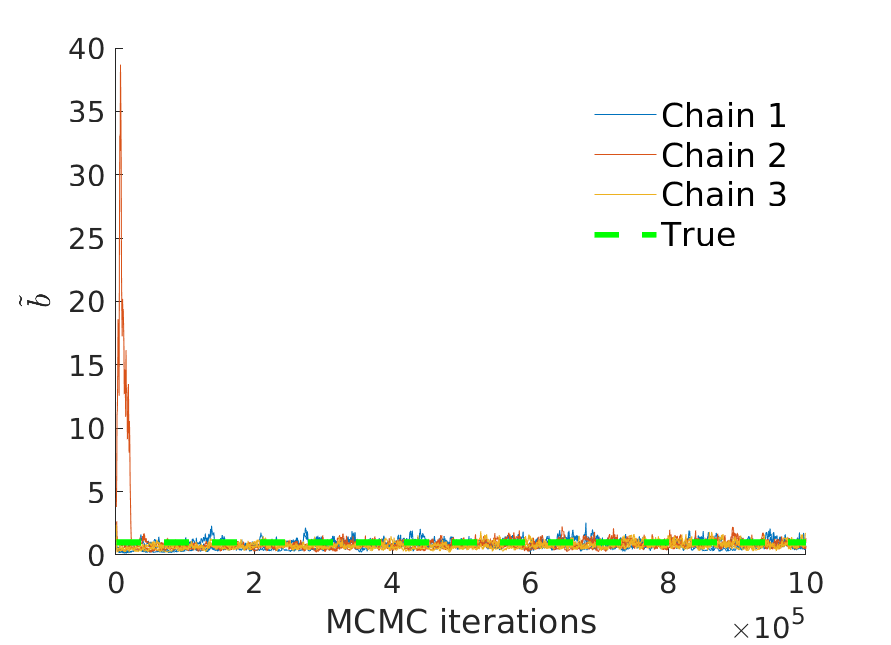}}
\subfloat[$\bar{w}_{\ast}$ trace]{\includegraphics[width=0.3\textwidth]{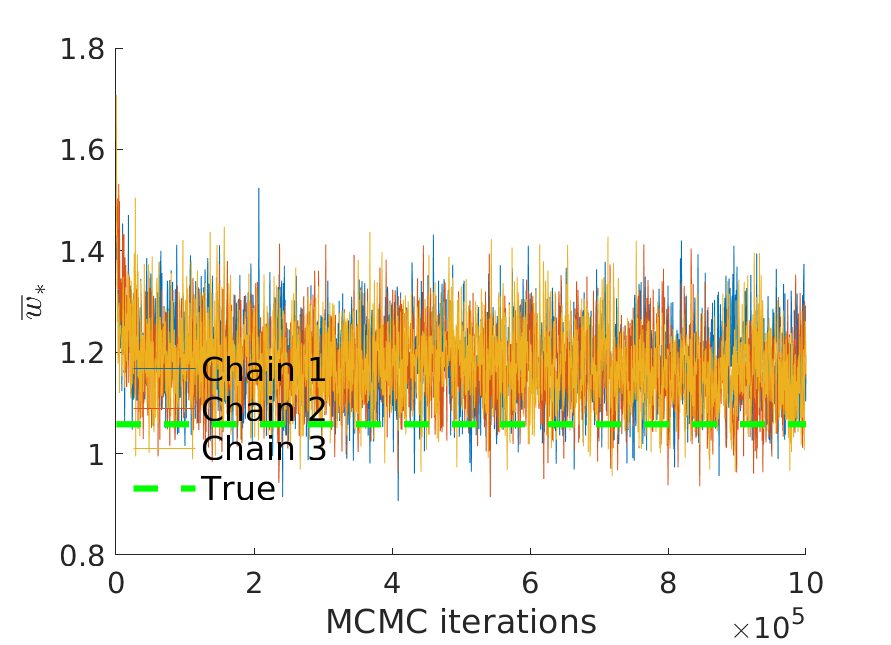}}
\caption{MCMC trace plots for a graph generated with $K=4, \alpha = 200, \sigma= 0.2, \tau = 1, b_i=b=\frac{4}{K}, a_i=a=0.2$. The green lines correspond to the true values of the parameters}
\label{fig:mcmc_trace_hist_4_community}

\end{figure}

\begin{figure}[h]
\centering
\subfloat[$\log \tilde{\alpha}$ histogram]{\includegraphics[width=0.3\textwidth]{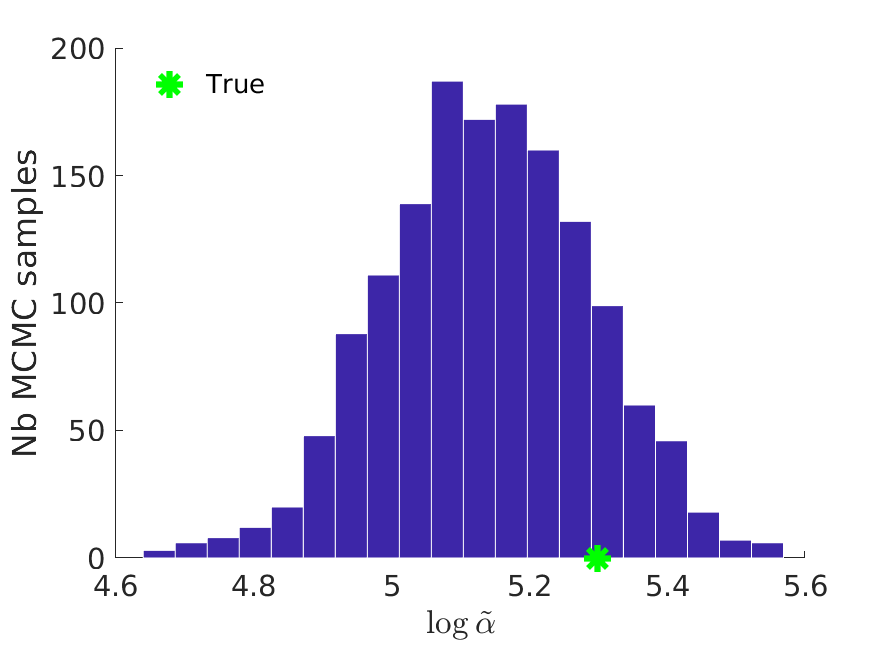}}
\subfloat[$\sigma$ histogram]{\includegraphics[width=0.3\textwidth]{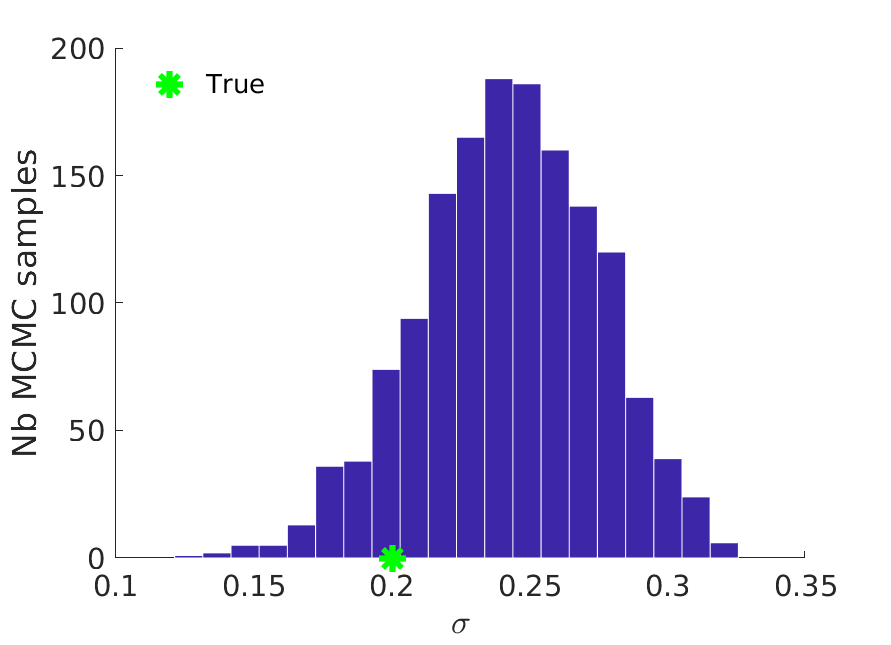}}
\subfloat[$a$ histogram]{\includegraphics[width=0.3\textwidth]{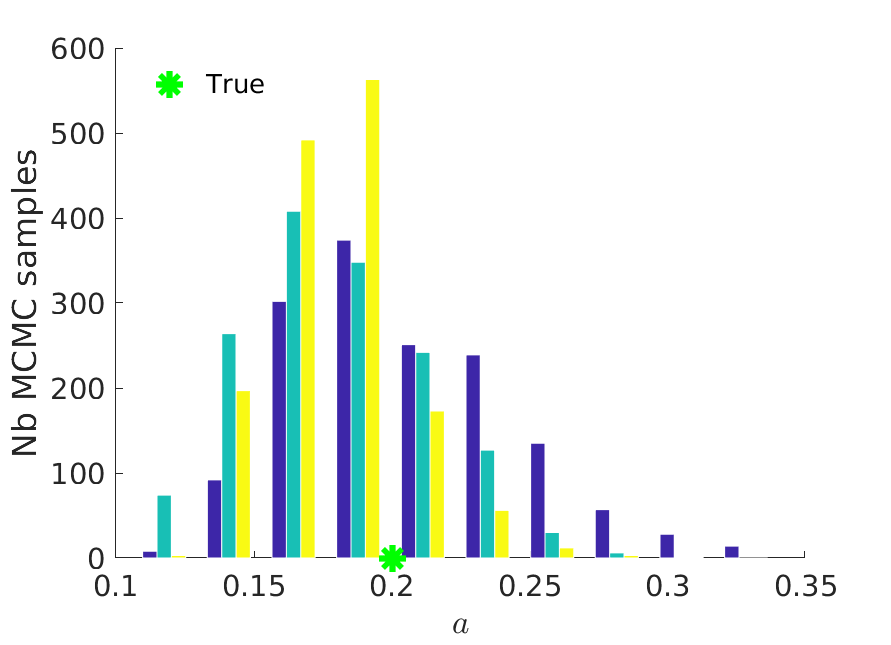}}\\
\subfloat[$\tilde{b}$ histogram]{\includegraphics[width=0.3\textwidth]{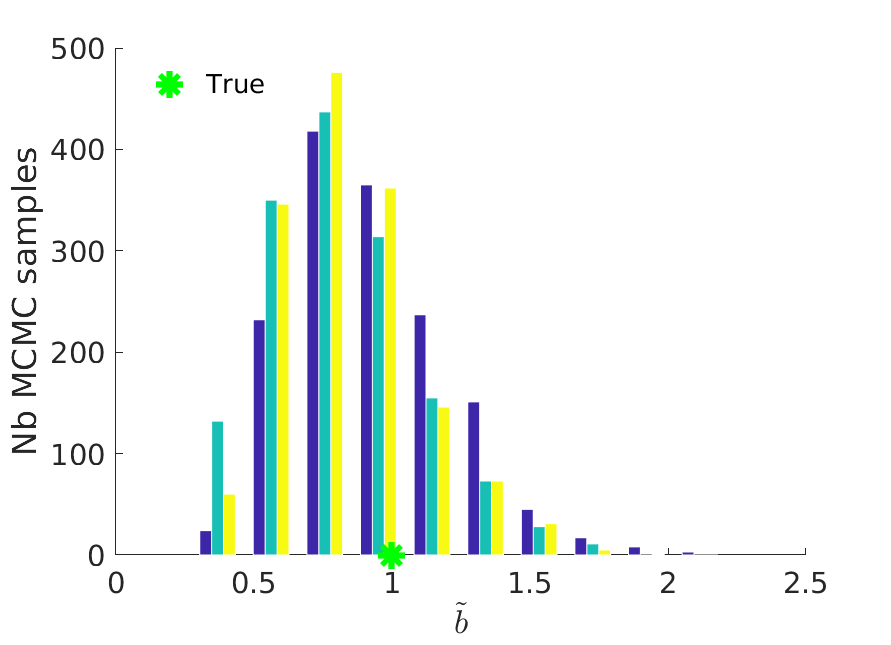}}
\subfloat[$\bar{w}_{\ast}$ histogram]{\includegraphics[width=0.3\textwidth]{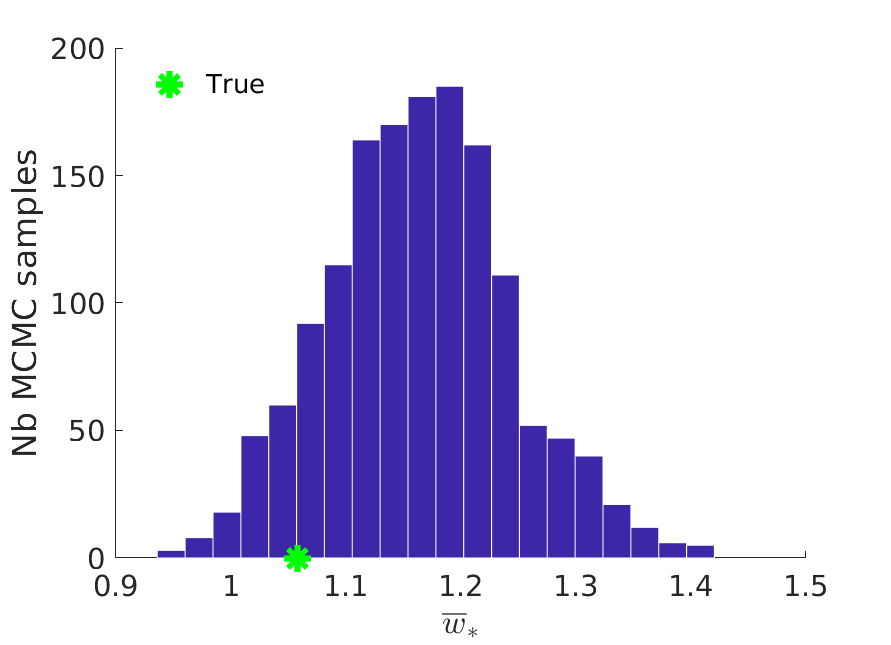}}
\caption{MCMC histograms for a graph generated with $K=4, \alpha = 200, \sigma= 0.2, \tau = 1, b_i=b=\frac{4}{K}, a_i=a=0.2$. The green stars correspond to the true values of the parameters}
\label{fig:mcmc_hist_4_community}

\end{figure}

\begin{figure}[h]
\centering
\subfloat[Core sociability credible intervals for high degree nodes  \label{fig:core_credible_4}]{\includegraphics[width=0.35\textwidth]{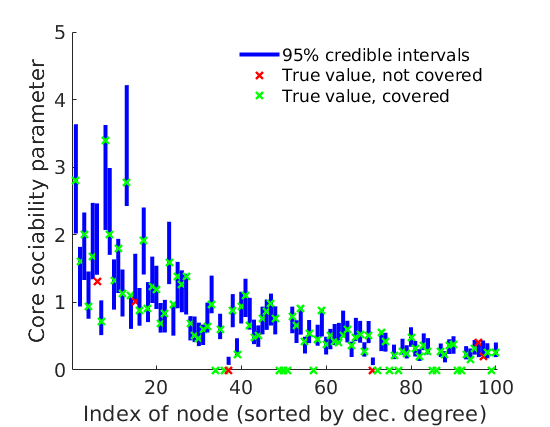}}
\hspace{0.01\textwidth}
\subfloat[Community 1 sociability credible intervals for high degree nodes \label{fig:com1_credible}]{\includegraphics[width=0.35\textwidth]{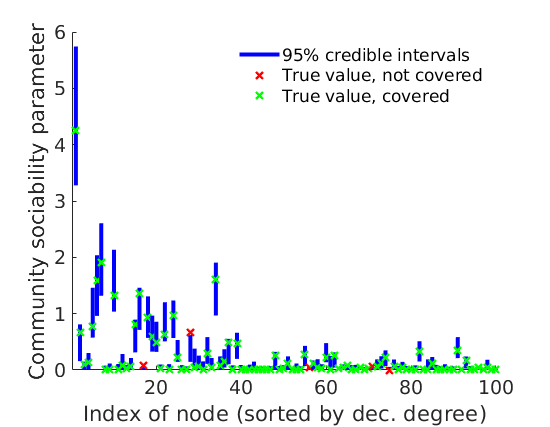}}\\
\subfloat[Community 2 sociability credible intervals for high degree nodes \label{fig:com2_credible}]{\includegraphics[width=0.35\textwidth]{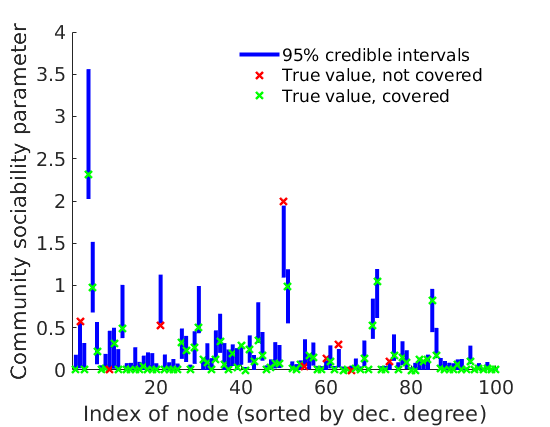}}
\hspace{0.01\textwidth}
\subfloat[Community 3 sociability credible intervals for high degree nodes \label{fig:com3_credible}]{\includegraphics[width=0.35\textwidth]{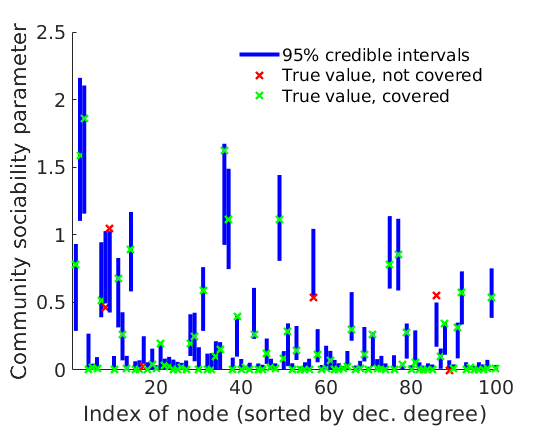}}
\caption{Credible intervals for core and community sociabilities, for the graph generated with $K=4, \alpha = 200, \sigma= 0.2, \tau = 1, b_i=b=\frac{4}{K}, a_i=a=0.2$.}
    \label{fig:all_credible_4}

\end{figure}

In Figures \ref{fig:mcmc_trace_hist_4_community} and \ref{fig:mcmc_hist_4_community} we see the trace plots and histograms for the identifiable parameters for the graph generated with $K=4, \alpha = 200, \sigma= 0.2, \tau = 1, b_i=b=\frac{4}{K}, a_i=a=0.2$. The green lines and stars correspond to the true values of the parameters, and as before we see that the posterior distribution converges around the true values. In Figure \ref{fig:all_credible_4} we plot the credible intervals for each of the sociabilities $w_1,\ldots,w_4$, and see that we are able to recover both the core and community sociabilities. 

\begin{figure}[h]
\centering
\subfloat[Log-posterior]{\includegraphics[width=0.4\textwidth]{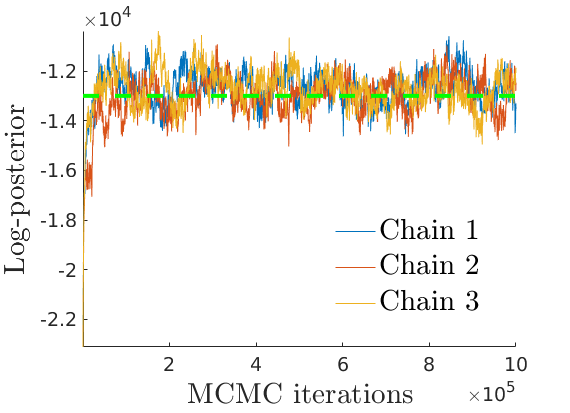}}
\subfloat[Autocorrelation]{\includegraphics[width=0.4\textwidth]{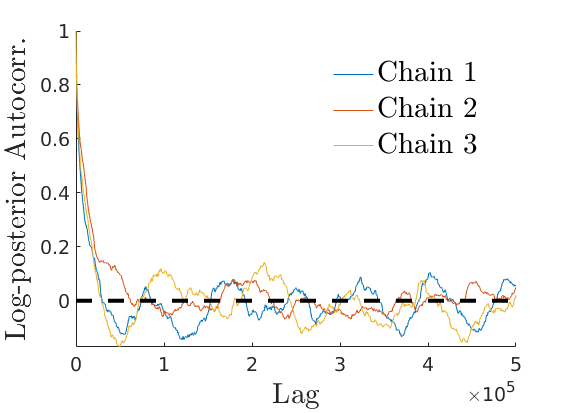}}
\caption{Trace plot and autocorrelation of the log-posterior for a graph generated with $K=4, \alpha = 200, \sigma= 0.2, \tau = 1, b_i=b=\frac{4}{K}, a_i=a=0.2$. The green line in $(a)$ corresponds to the value of the approximate log-posterior (up to a constant) under the model parameters used to generate the graph.}
\label{fig:mcmc_logpost_4_community}

\end{figure}

As before, to test convergence we calculate the approximate log-posterior probability density function (up to a normalizing constant). We see from Figure~\ref{fig:mcmc_logpost_4_community} that the chain has converged to the true value. This is confirmed by the Gelman-Rubin diagnostic, which in this case comes out as $\hat{R}=1.03$. However, from the autocorrelation plot we see that the mixing is not as good as in the case without community structure, with dependencies not vanishing as quickly as we would like for increasing lag. This indicates that estimation in the setting with core-periphery and community structure can be more difficult, as we might expect

\subsubsection{No core-periphery structure}

\begin{figure}[h]
\centering
\subfloat[Posterior predictive degree for network with no core \label{fig:post_pred_zero}]{\includegraphics[width=0.4\textwidth]{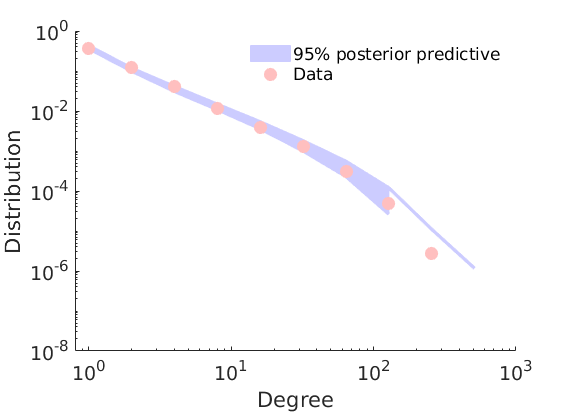}}
\subfloat[Posterior predictive degree for dense network \label{fig:post_pred_dense}]{\includegraphics[width=0.4\textwidth]{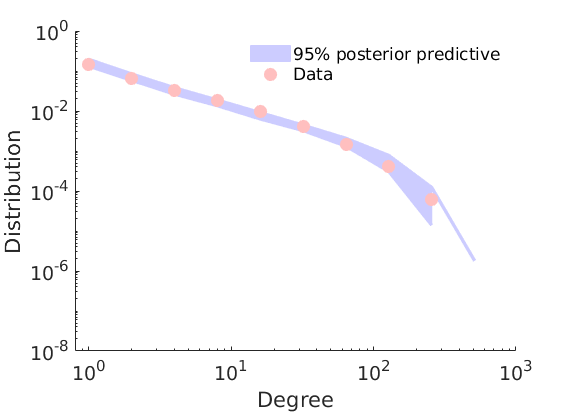}}
\caption{Fitting our model to networks which (a) have no core, and (b) are dense}
\label{fig:post_pred_cf}
\end{figure}

We also test our model in two situations in which we do not expect it to work, in order to check that the behaviour is still sensible. Recalling that in our model, the overall generated graph is still sparse, we first simulate data from the model of \citep{Caron2017}, with $\sigma>0$. This model generates a sparse graph without a core-periphery structure. In this case, the size of the core estimated by our model tend to zero. Furthermore, we see from Figure \ref{fig:post_pred_cf}\subref{fig:post_pred_zero} that we still accurately recover the degree distribution in this case.

Secondly, we test our model on a graph generated from the model of \citep{Caron2017}, but with $\sigma<0$. In the construction of our model, the core and periphery are distinguished by the fact that the core nodes form a dense subgraph in an otherwise sparse graph. Thus, when the overall graph is dense, we do not have this distinction and our model struggles to identify any latent core-periphery structure. However, in this case this structure is not present. We see two different types of behaviour, depending on the parameters of the model and the initial conditions.
\begin{enumerate}
    \item $\sigma$ is (correctly) estimated to be negative, and the size of the core goes to zero. This is the behaviour we see in Figure \ref{fig:post_pred_cf}\subref{fig:post_pred_dense} and is the behaviour we would expect.
    \item $\sigma$ is (incorrectly) estimated to be positive, and the whole graph is estimated to be in the core. Although this may not seem intuitive, it still fits with our definition of the core being a dense subgraph within a sparse network.
\end{enumerate}

\subsection{Real data}
In this section we provide additional trace plots and convergence diagnostics on the experiments in Section \ref{sec:experiments_real}.
\subsubsection{US Airport network}
\label{sec:app:airports}

\begin{figure}[h]
\centering
\subfloat{\includegraphics[width=0.4\textwidth]{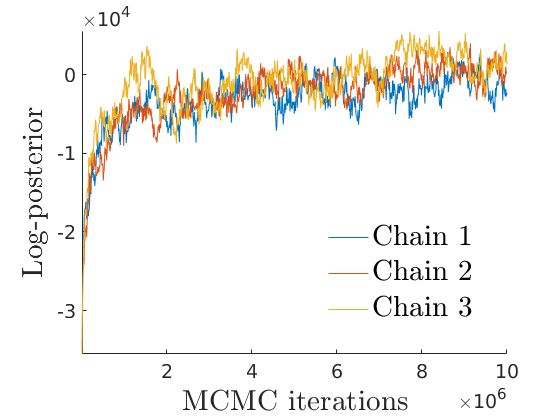}}
\subfloat{\includegraphics[width=0.4\textwidth]{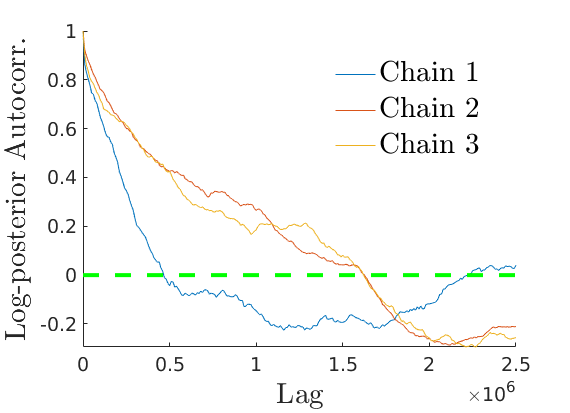}}
\caption{Trace plot (left) and autocorrelation plot (right) of the approximated log-posterior probability density (up to a constant) of the US Airport network}
\label{fig:airports_diagnostics}
\end{figure}

\begin{figure}[h]
\centering
\subfloat[$\log \tilde{\alpha}$ trace]{\includegraphics[width=0.3\textwidth]{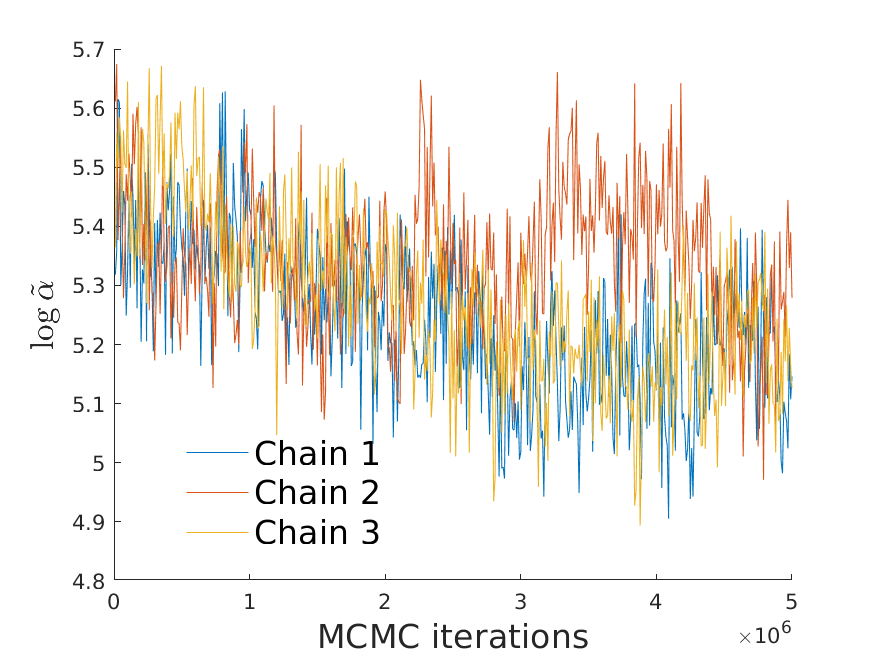}}
\subfloat[$\sigma$ trace]{\includegraphics[width=0.3\textwidth]{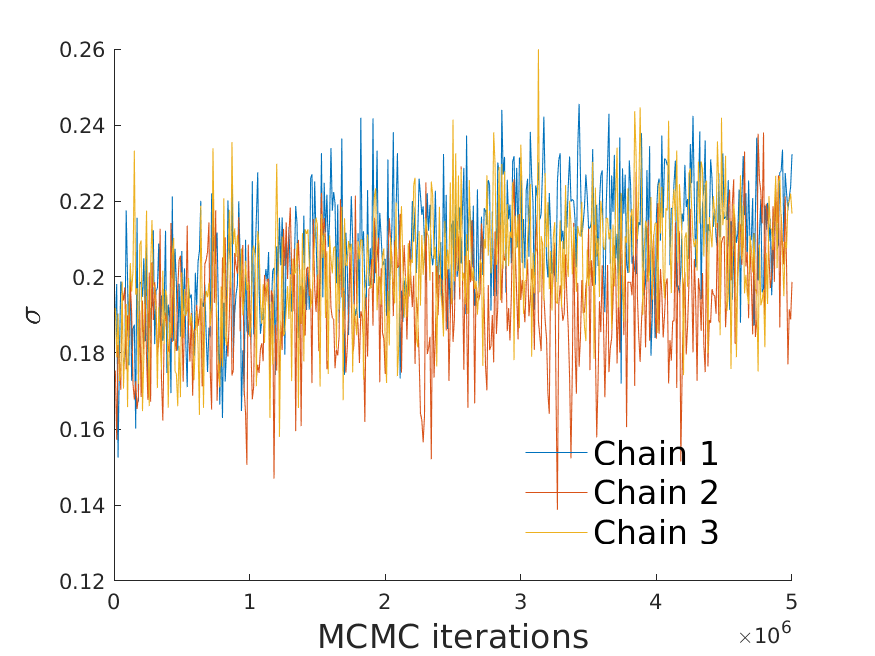}}
\subfloat[$a$ trace]{\includegraphics[width=0.3\textwidth]{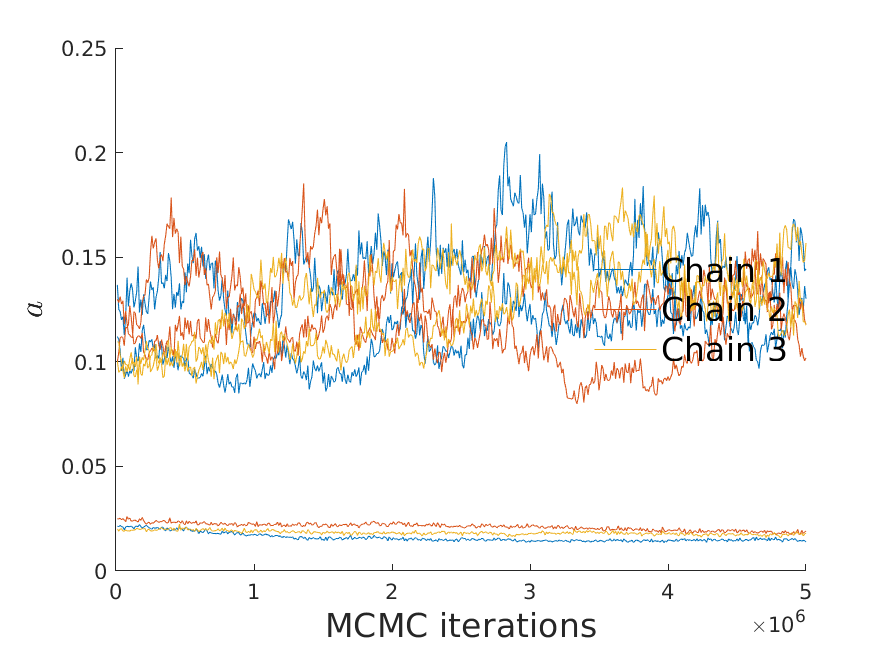}}\\
\subfloat[$\tilde{b}$ trace]{\includegraphics[width=0.3\textwidth]{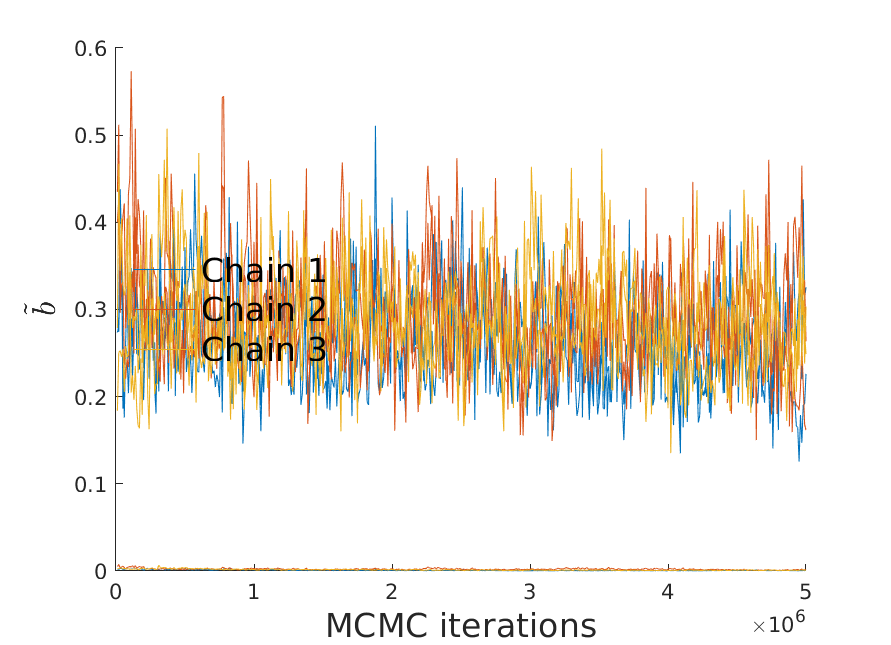}}
\subfloat[$\bar{w}_{\ast}$ trace]{\includegraphics[width=0.3\textwidth]{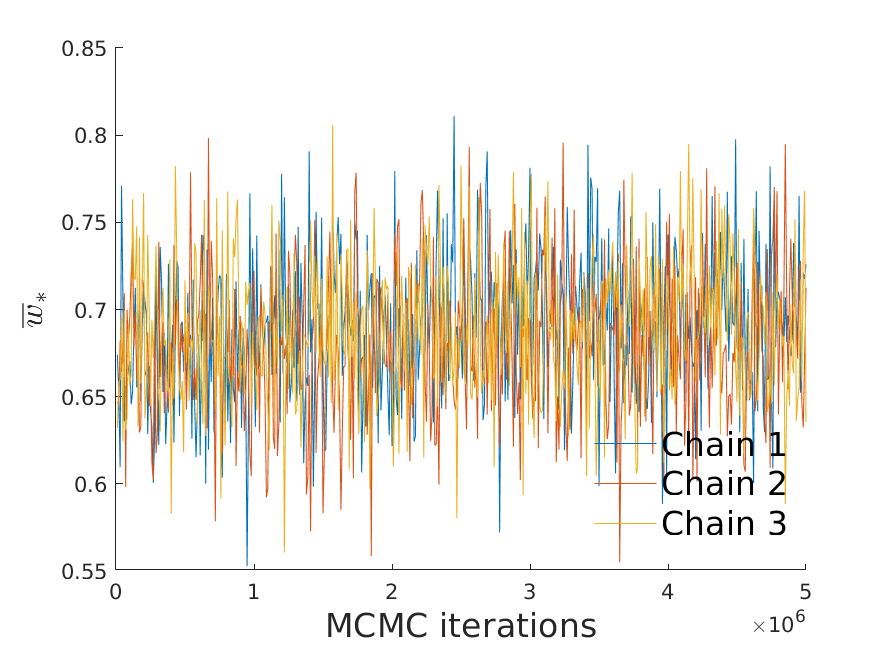}}

\caption{MCMC trace plots for the US Airport network}
\label{fig:mcmc_trace_airports}
\end{figure}

In Figure \ref{fig:airports_diagnostics} we give the trace plot and autocorrelation of the approximate logposterior for the US Airport network. The Gelman-Rubin statistic comes out to be $\hat{R}=1.25$ in this case, suggesting that the chain has not converged despite the large number of iterations. However, we observed that increasing the number of iterations does not change significantly the value of the core-periphery parameters. In Figure \ref{fig:mcmc_trace_airports} we report the trace plots for the identifiable parameters of the US Airport network
 
\subsubsection{World Trade network}
\label{sec:app:WT}

\begin{figure}[h]
\centering
\subfloat{\includegraphics[width=0.4\textwidth]{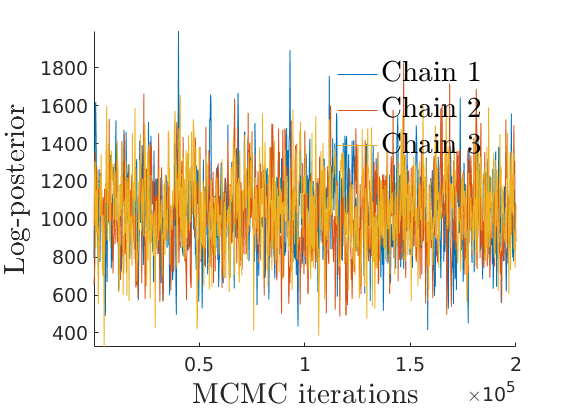}}
\subfloat{\includegraphics[width=0.4\textwidth]{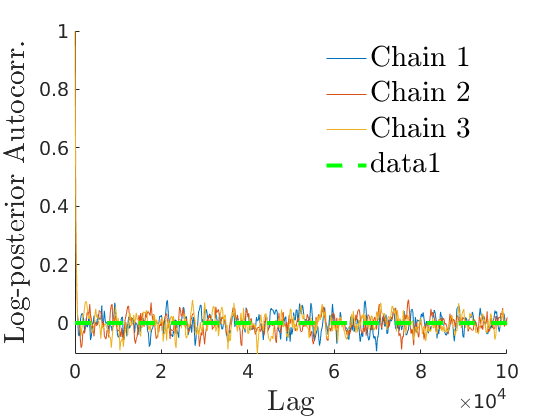}}
\caption{Trace plot (left) and autocorrelation plot (right) of the approximated log-posterior probability density (up to a constant) of the World Trade network}
\label{fig:trade_diagnostics}
\end{figure}

In Figure \ref{fig:trade_diagnostics} we give the trace plot and autocorrelation of the approximate logposterior for the World Trade network. In this case, the Gelman-Rubin statistic of $\hat{R}=1.009$ suggests convergence of the Markov chain. 

\begin{figure}[h]
\centering
\subfloat[$\log \tilde{\alpha}$ trace]{\includegraphics[width=0.3\textwidth]{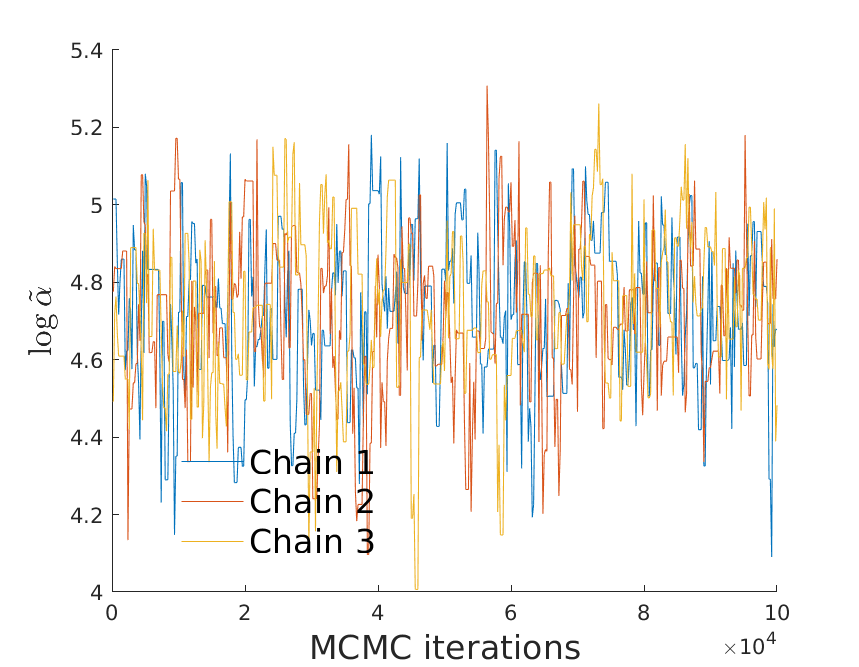}}
\subfloat[$\sigma$ trace]{\includegraphics[width=0.3\textwidth]{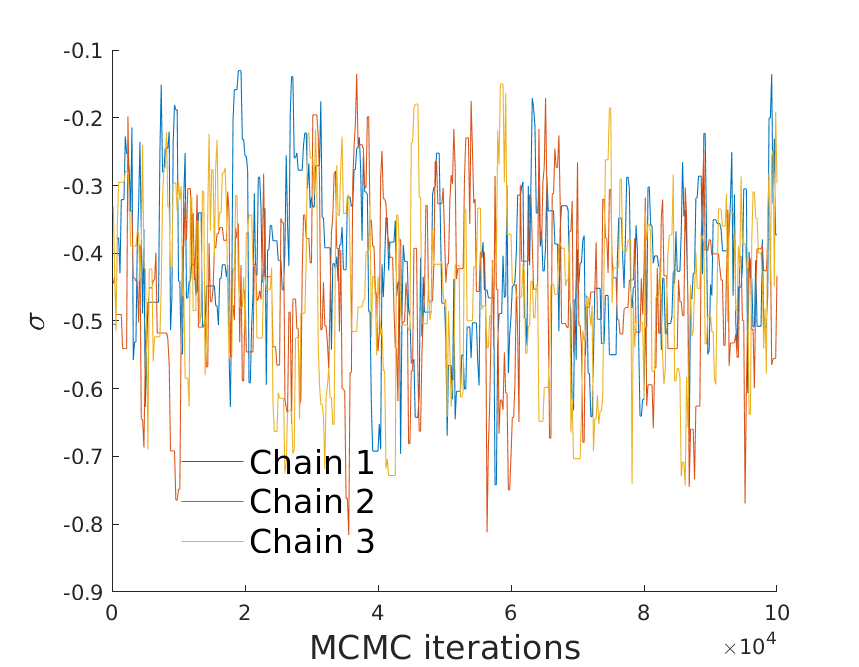}}
\subfloat[$a$ trace]{\includegraphics[width=0.3\textwidth]{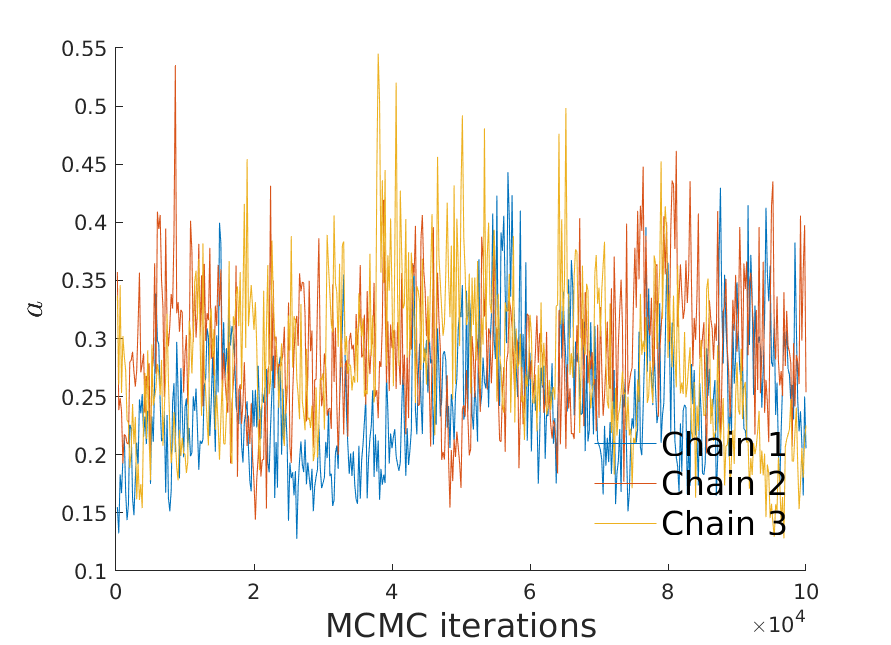}}\\
\subfloat[$\tilde{b}$ trace]{\includegraphics[width=0.3\textwidth]{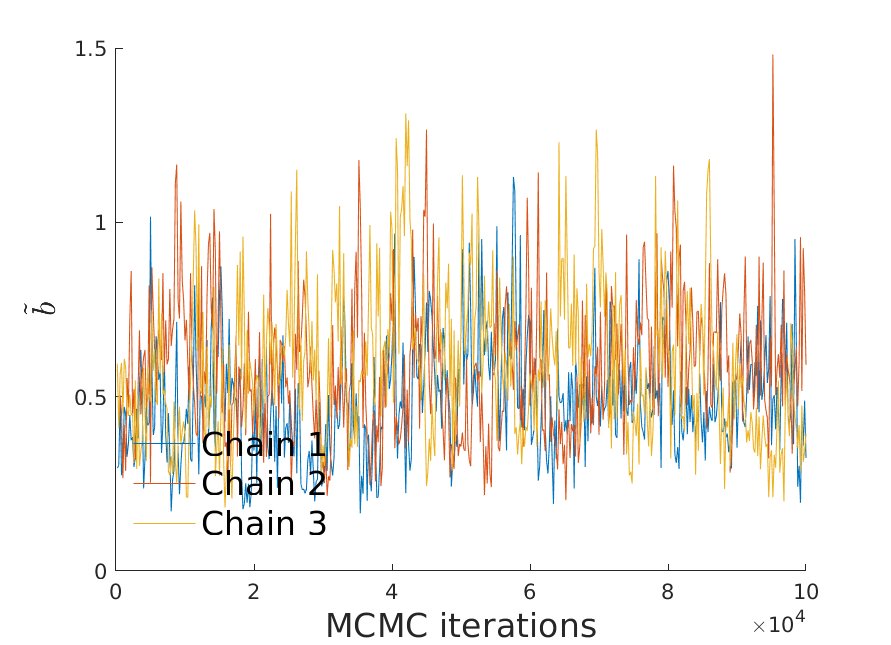}}
\subfloat[$\bar{w}_{\ast}$ trace]{\includegraphics[width=0.3\textwidth]{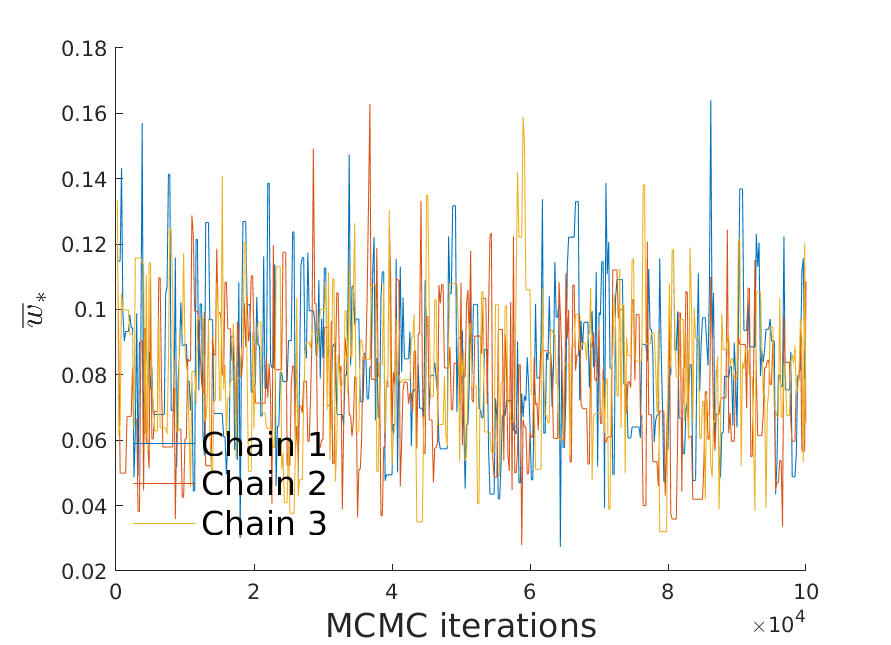}}

\caption{MCMC trace plots for the World Trade network}
\label{fig:mcmc_trace_trade}

\end{figure}

In Figure \ref{fig:mcmc_trace_trade} we see the trace plots for the identifiable parameters of the World Trade network. 
\section{US Airport Network Case Study Repeated Without Alaskan Airports}\label{app:us_airport_no_alaska}
\begin{figure}[h]
\centering
\subfloat[$\log \tilde{\alpha}$ trace]{\includegraphics[width=0.3\textwidth]{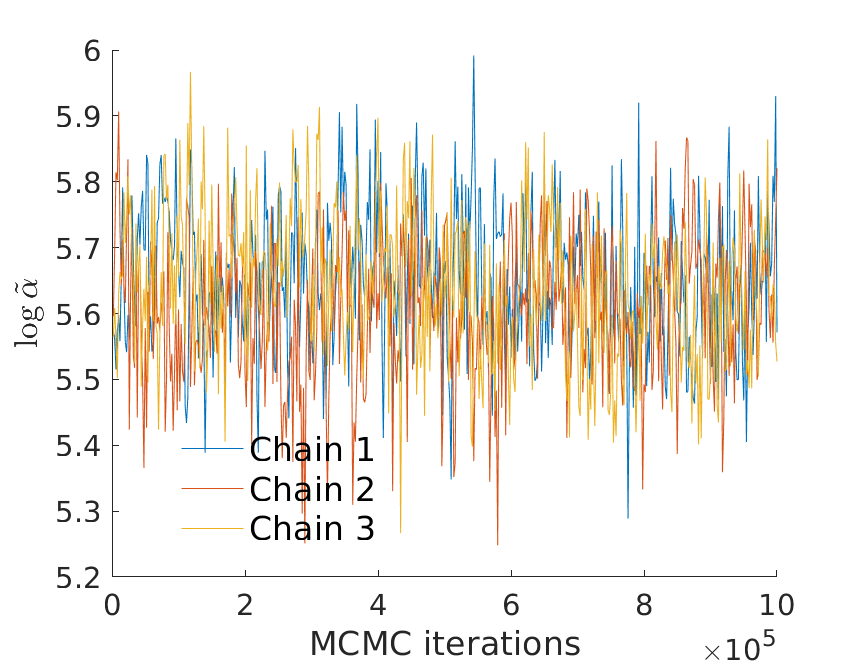}}
\subfloat[$\sigma$ trace]{\includegraphics[width=0.3\textwidth]{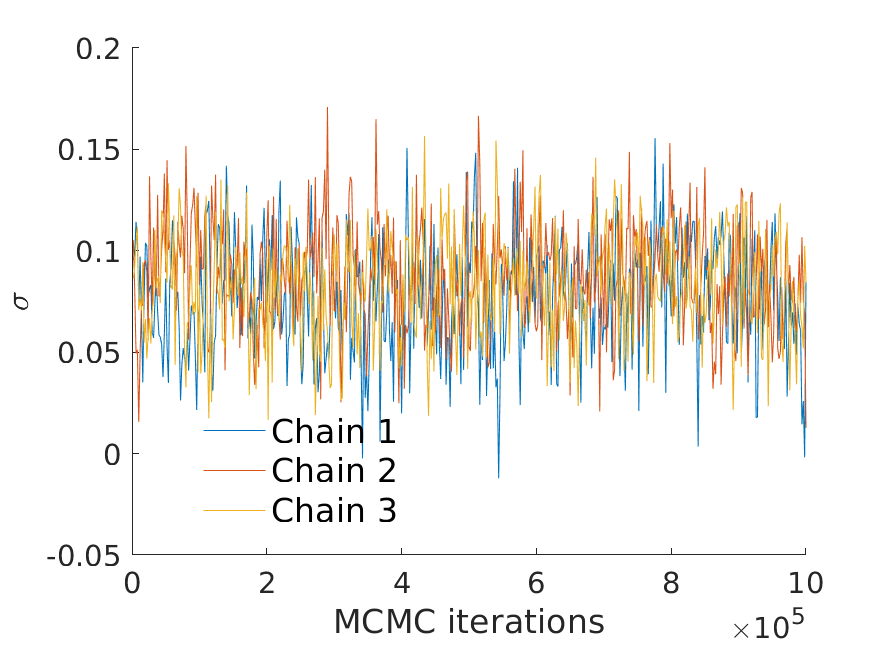}}
\subfloat[$a$ trace]{\includegraphics[width=0.3\textwidth]{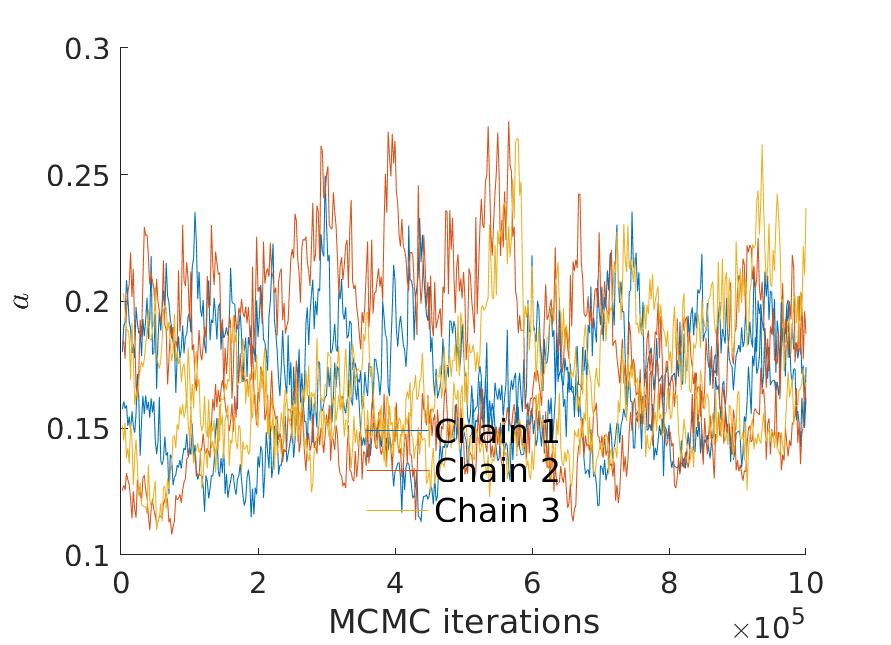}}\\
\subfloat[$\tilde{b}$ trace]{\includegraphics[width=0.3\textwidth]{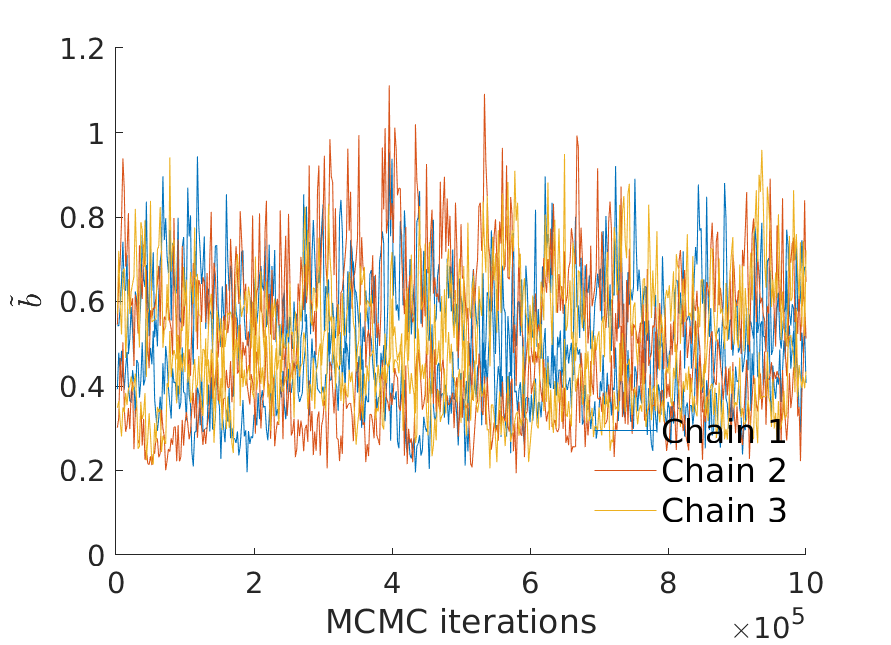}}
\subfloat[$\bar{w}_{\ast}$ trace]{\includegraphics[width=0.3\textwidth]{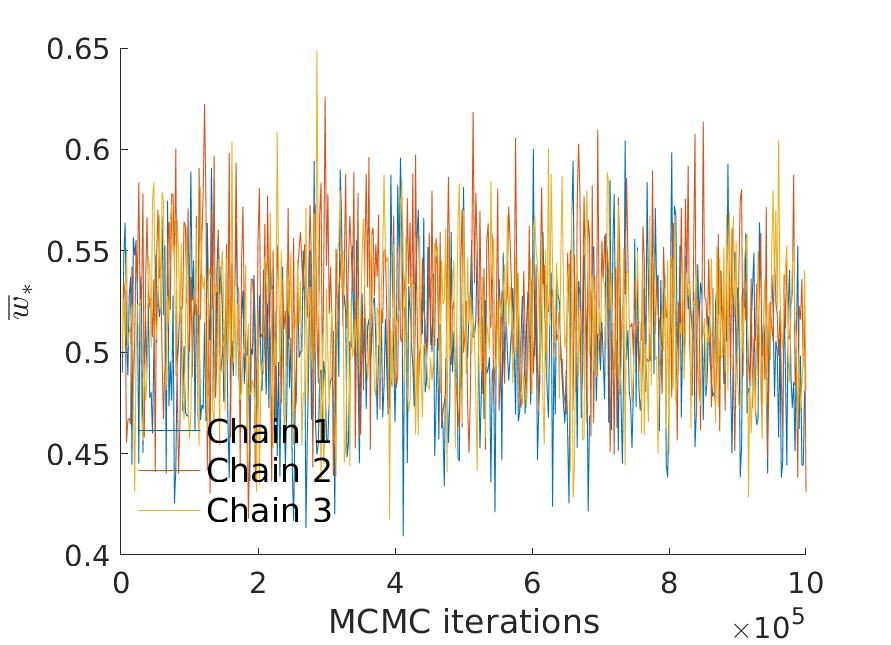}}

\caption{MCMC trace plots for the US Airport network without Alaskan airports}
\label{fig:mcmc_trace_airports_no_alaska}
\end{figure}

In Section \ref{sec:us_airport}, we noted a problem that was occurring due to the sociability parameter for the Alaskan community. Here, we repeat our analysis having taken out all the Alaskan airports, as well as any airport that was only connected to Alaskan airports. In Figure \ref{fig:mcmc_trace_airports_no_alaska} we see the trace plots for the identifiable parameters. From this we see that we no longer have the same problem of an estimate of one of the $b_i$ being very close to 0.

\begin{figure}[h]
\centering
\subfloat[Core-periphery posterior  degree \label{fig:airports_cp_logpost_no_alaska}]{\includegraphics[width=0.4\textwidth]{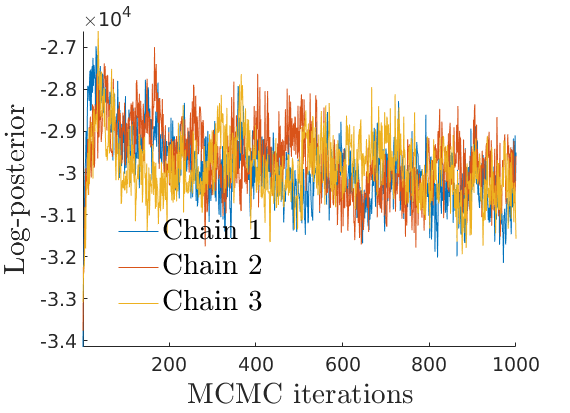}}
\subfloat[Two-community posterior  degree\label{fig:airports_autocorr_no_alaska}]{\includegraphics[width=0.4\textwidth]{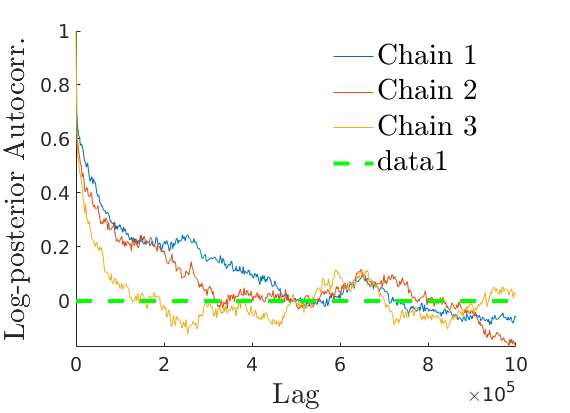}}

\caption{Trace plot (left) and autocorrelation plot (right) of the approximated log-posterior probability density (up to a constant) of the US Airport network without Alaskan airports}
\label{fig:airports_post_logpost_no_alaska}
\end{figure}

In Figure \ref{fig:airports_post_logpost_no_alaska} we see the autocorrelation plot of the approximate log-posterior density. This seems as though it is converging to a stable value. If we look at the Gelman-Rubin convergence diagnostic, this comes out to be $\hat{R}=1.05$ in this case, indicating convergence.

\begin{figure}[h]
\centering
\subfloat[Core-periphery posterior  degree \label{fig:airports_cp_post_no_alaska}]{\includegraphics[width=0.4\textwidth]{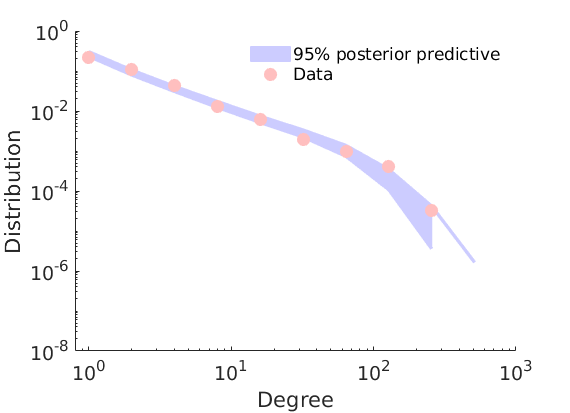}}
\subfloat[Three-community posterior  degree\label{fig:airports_gamma_post_no_alaska}]{\includegraphics[width=0.4\textwidth]{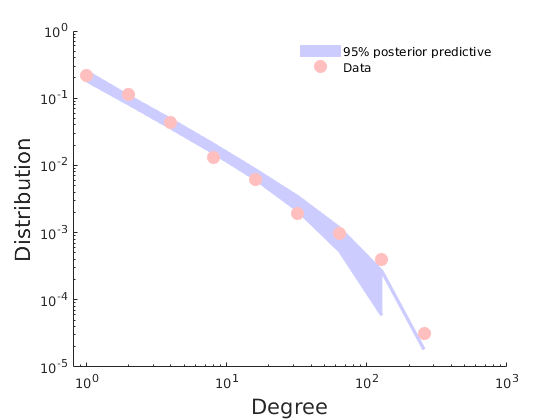}}

\caption{Posterior Predictive Degree Distribution for the US Airport network without Alaskan airports}
\label{fig:airports_post_pred_no_alaska}
\end{figure}

In Figure \ref{fig:airports_post_pred_no_alaska} we see the posterior predictive degree plot on the left, compared to the corresponding plot for the model of \citep{todeschini2016exchangeable} with three communities on the right. We see that we are no longer overestimating the high degree nodes. We also see that the overlapping communities model has estimated $\sigma<0$ in this case, whereas for our model we still estimate $\sigma>0$.

If we look at the reweighted KS statistics as before in Table \ref{table:airports_gof_no_alaska}, we see that our model is still providing a better fit to the data, but not necessarily any better than the fit before excluding the Alaskan airports. In this case we report the R\'enyi statistics $L_2$ and $U_1$, because we see that our model is overestimating the number of low degree nodes, and underestimating the number of high degree nodes.

\begin{table}[htb]
\centering
\begin{tabular}{ |p{2cm}||p{2cm}|p{2cm}|p{1.9cm}|p{1.9cm}|  }
 \hline
\multicolumn{5}{|c|}{Distance Measures} \\
 \hline
\multirow{2}{4em}{Method} & Reweighted KS& Unweighted KS & \multicolumn{2}{c|}{R\'enyi Statistics} \\
\cline{4-5}
 &  $D$&  $K$&  $L_{2}$&  $U_{1}$\\
 \hline
 Core-periphery& $\mathbf{0.213 \pm 0.089}$  &   $\mathbf{1.947 \pm 0.427}$ & $0.032 \pm 0.002$    &$\mathbf{0.939 \pm 0.728}$  \\ 
 SNetOC &$0.385 \pm 0.169 $   &$2.409 \pm 0.527 $ &   $\mathbf{0.08 \pm 0.001}$   & $2.665 \pm 1.587$      \\
  \hline
\end{tabular}
\caption{Distance Measures for the US Airport Network}
\label{table:airports_gof_no_alaska}
\end{table}

\begin{figure}[h]
\centering
\subfloat[Core-periphery posterior  degree \label{fig:airports_cp_adj_no_alaska}]{\includegraphics[width=0.4\textwidth]{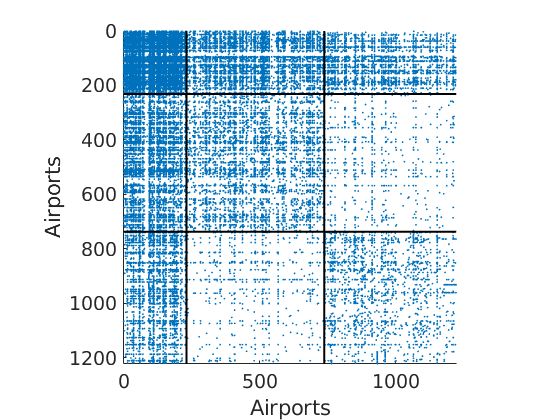}}
\subfloat[Two-community posterior  degree\label{fig:airports_filled_adj_no_alaska}]{\includegraphics[width=0.4\textwidth]{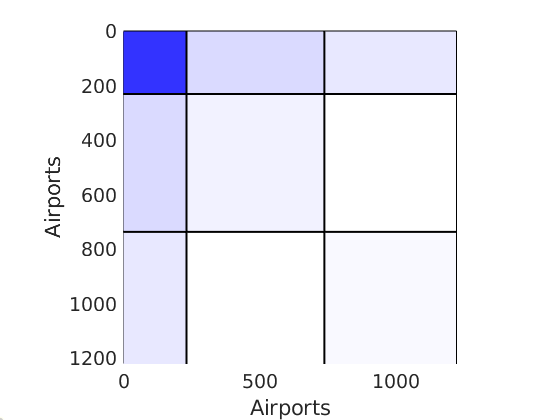}}

\caption{Adjacency matrix for the US Airport network without Alaskan airports, ordered into core and periphery, and then by highest community weight}
\label{fig:airports_adj_both_no_alaska}
\end{figure}

Furthermore, we see that the core and communities identified here are largely the same as before (without the ``Alaska'' community). In Figure \ref{fig:airports_adj_both_no_alaska} we plot the adjacency matrix in this case, again ordered into core and periphery, and then by highest community weight.

\end{document}